\documentclass[pdflatex,sn-mathphys-num]{sn-jnl}

\usepackage{graphicx}%
\usepackage{multirow}%
\usepackage{amsmath,amssymb,amsfonts}%
\usepackage{amsthm}%
\usepackage{mathrsfs}%
\usepackage[title]{appendix}%
\usepackage{xcolor}%
\usepackage{textcomp}%
\usepackage{manyfoot}%
\usepackage{booktabs}%
\usepackage{algorithm}%
\usepackage{algorithmicx}%
\usepackage{algpseudocode}%
\usepackage{listings}%

\theoremstyle{thmstyleone}%
\newtheorem{theorem}{Theorem}
\newtheorem{proposition}[theorem]{Proposition}%
\newtheorem{property}{Property}

\theoremstyle{thmstyletwo}%
\newtheorem{remark}{Remark}%

\theoremstyle{thmstylethree}%
\newtheorem{definition}{Definition}%

\newlength{\mycolumnwidth}
\newif\ifcompile
\compiletrue 

\newif\ifdraft 

\usepackage{bm}

\usepackage{mathrsfs}
\usepackage{scalerel}

\newcommand\scale[2]{\vstretch{#1}{\hstretch{#1}{#2}}}
	
\newcommand{\LP}{
		\mathbin{\mathpalette\LIPcls+}
	}
\newcommand{\LM}{
		\mathbin{\mathpalette\LIPcls-}
		}
\newcommand{\LT}{
		\mathbin{\mathpalette\LIPcls\times}
	}

\newcommand{\LIPcls}[2]{%
		\ooalign{$#1\bigtriangleup$\crcr  \hidewidth\raisefix{#1}\hbox{$#1\scale{0.45}{\bm{#2}}$}\hidewidth}}

\def\raisefix#1{%
  \ifx#1\displaystyle
    \raise.14em
  \else
    \ifx#1\textstyle
      \raise.14em
    \else
      \ifx#1\scriptstyle
        \raise.112em
      \else
        \raise.0933em
      \fi
    \fi
  \fi
}

\newcommand{\LIPplus}{\LP}
\newcommand{\LIPminus}{\LM}
\newcommand{\LIPtimes}{\LT}

\makeatletter
\DeclareRobustCommand\bigop[1]{%
  \mathop{\vphantom{\sum}\mathpalette\bigop@{#1}}\slimits@
}
\newcommand{\bigop@}[2]{%
  \vcenter{%
    \sbox\z@{$#1\sum$}%
    \hbox{\resizebox{\ifx#1\displaystyle1.1\fi\dimexpr\ht\z@+\dp\z@}{!}{$\m@th#2$}}%
  }%
}
\makeatother

\newcommand{\oplusLIP}{%
  \mathbin{\mathpalette\oplusLIPcls{+}}%
}
\newcommand{\ominusLIP}{%
  \mathbin{\mathpalette\oplusLIPcls{-}}%
}
\newcommand{\oplusLIPcls}[2]{%
  \ooalign{%
    $#1\bigcirc$\cr
    \hidewidth$#1\triangle$\hidewidth\cr
    \hidewidth\raisefix{#1}\hbox{$#1\scale{0.45}{\bm{#2}}$}\hidewidth%
  }%
}

\newcommand{\Real}{\mathbb R}
\newcommand{\Realb}{\overline{\Real}}

\newcommand{\Zint}{\mathbb Z}

\newcommand{\la}{\lambda}

\newcommand{\I}{\mathcal{I}}

\newcommand{\Fcurv}{\mathcal{F}}
\newcommand{\Fcurvb}{\overline{\Fcurv}}

\newcommand{\Lscr}{\mathscr{L}}
\usepackage{siunitx}

\usepackage{etoolbox}
\newrobustcmd\Bfs{\DeclareFontSeriesDefault[rm]{bf}{b}\bfseries} 
\robustify\bfseries
\newcommand\mcc[1]{\multicolumn{1}{c}{#1}} 

\newcommand{\LIPPref}[1]{\hyperref[#1]{(\textbf{P\ref{#1}})}}

\usepackage{paralist}

\graphicspath{{./Fig/}}
\DeclareGraphicsExtensions{.pdf,.jpeg,.png,.jpg}

\usepackage[caption=false,font=footnotesize]{subfig}

\usepackage{url}

\usepackage{placeins} 

\begin{document}

\title[Logarithmic Mathematical Morphology]{Logarithmic Mathematical Morphology: theory and applications}


\author*[1]{\fnm{Guillaume} \sur{Noyel}}\email{guillaume.noyel@mines-paris.org}

\affil*[1]{\orgdiv{CRESTIC}, \orgname{Universit\'e de Reims-Champagne-Ardenne}, \city{Reims}, \country{France}}


\abstract{In Mathematical Morphology for grey-level functions, an image is analysed by another image named the structuring function. This structuring function is translated over the image domain and summed to the image. However, in an image presenting lighting variations, when the structuring function is added to this image, the amplitude of this translated structuring function should vary according to the image intensity. Such a property is not verified in Mathematical Morphology for grey level functions, when the structuring function is summed to the image with the usual additive law. This issue has been addressed by defining a framework that uses an additive law, for which the amplitude of the structuring function added to the image varies according to the amplitude of the image itself. This additive law is chosen within the Logarithmic Image Processing framework and models the lighting variations with a physical cause such as a change of light intensity. 
The framework is named Logarithmic Mathematical Morphology (LMM) and allows the definition of operators which are robust to such lighting variations.} 




\maketitle

%
%
\section{Introduction}
\label{sec:intro}

Mathematical Morphology (MM) was originally defined by Matheron \cite{Matheron1975} for sets 
and then extended to functions with real values by Serra \cite{Serra1982}, Sternberg \cite{Sternberg1979,Sternberg1986} and Maragos \cite{Maragos_1989}. In this latter case, a function is analysed by another function named structuring element or structuring function. 
This extension includes grey level images whose values are within the bounded interval $[0,M[$ of the real space $\Real$. For example, for 8 bit-digitised images, $M$ is equal to 256.
Generally, structuring functions and morphological operators are invariant under horizontal translations (i.e. in space) and under vertical translations (i.e. in intensity) \cite{Heijmans1991}. 

However, the application of MM to grey level images presents two limitations.
\begin{inparaenum}[(1)]
\item Firstly, albeit grey level images have bounded values, MM was defined for functions with values within the unbounded real space $\Real$ \cite{Heijmans1990,Heijmans1994}. 
Indeed, given an image $f$ and a structuring function $b$ both defined on the same domain $D \subset \Real^n$ 
and whose values are within the range $\left[0,M\right[$\>, their sum $f+b$ does not lie within the interval $\left[0,M\right[$\>. Practical solutions to this issue consist of using either 
\begin{inparaenum}[(i)]
\item a structuring function whose supremum is equal to zero,
\item or a flat structuring element whose values are equal to zero 
\item or to truncate the values of the resulting image to the maximal possible value, $M$ \cite{Heijmans1990}. 
\end{inparaenum}
\item Secondly, adding a structuring function to an image without taking into account the image intensity into the amplitude of the translated structuring function is not physically justified. 
As in human vision, the eye response to light intensity variations is known to be logarithmic \cite{Jourlin1988,Brailean1991,Sun2012,Varshney2013,Jourlin2016_chap1}, 
it follows that in images the contrast variations are also logarithmic and the darkest variations are more attenuated than the brightest ones \cite{Land1977,Kimmel2003}.
The amplitude of the structuring function translated by the image intensity must therefore vary according to the image intensity, i.e. the grey value.\label{review:corr:transl_sf} Such a translation of the structuring function enables the definition of morphological operators with an invariance under horizontal translation but not under vertical translation, i.e. in the intensity domain $\left[0,M\right[$\>, with the usual additive law $+$ \cite{Heijmans1994,Roerdink2009}. 
\end{inparaenum}

\begin{figure}[!t]
\centering
\subfloat[Input]{\includegraphics[width=0.32\mycolumnwidth]{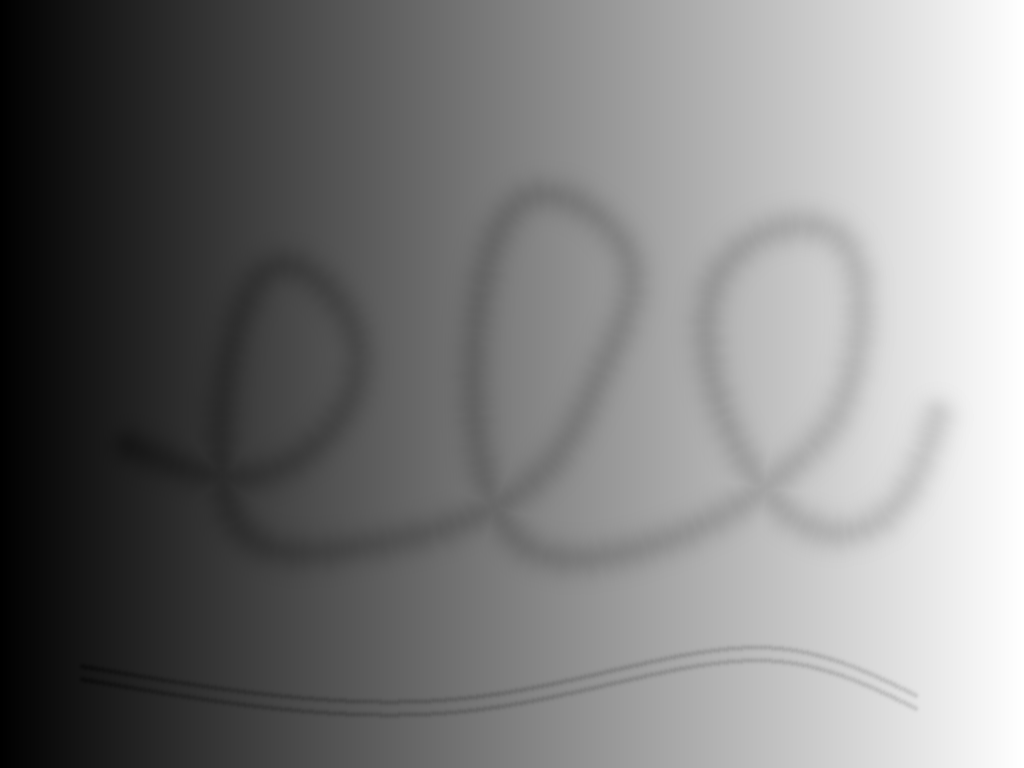}%
\label{fig_intro:input}}
\hfil
\subfloat[Input (LIP-greyscale)]{\includegraphics[width=0.32\mycolumnwidth]{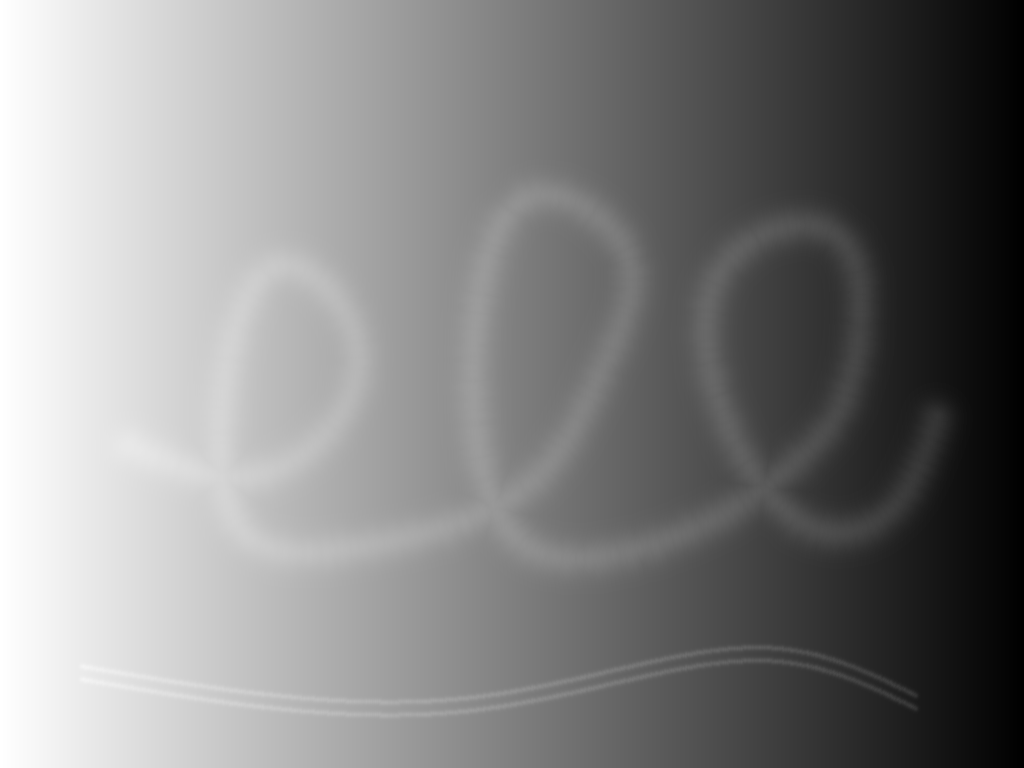}%
\label{fig_intro:input_LIP_scale}}
\hspace{0.03\mycolumnwidth}
\subfloat[Probe]{\includegraphics[width=0.24\mycolumnwidth]{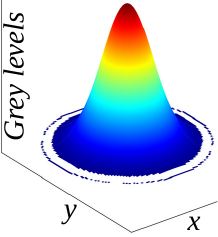}%
\label{fig_intro:probe}}
\hspace{0.05\mycolumnwidth}
\hfil
\\
\subfloat[Top-hat]{\includegraphics[width=0.32\mycolumnwidth]{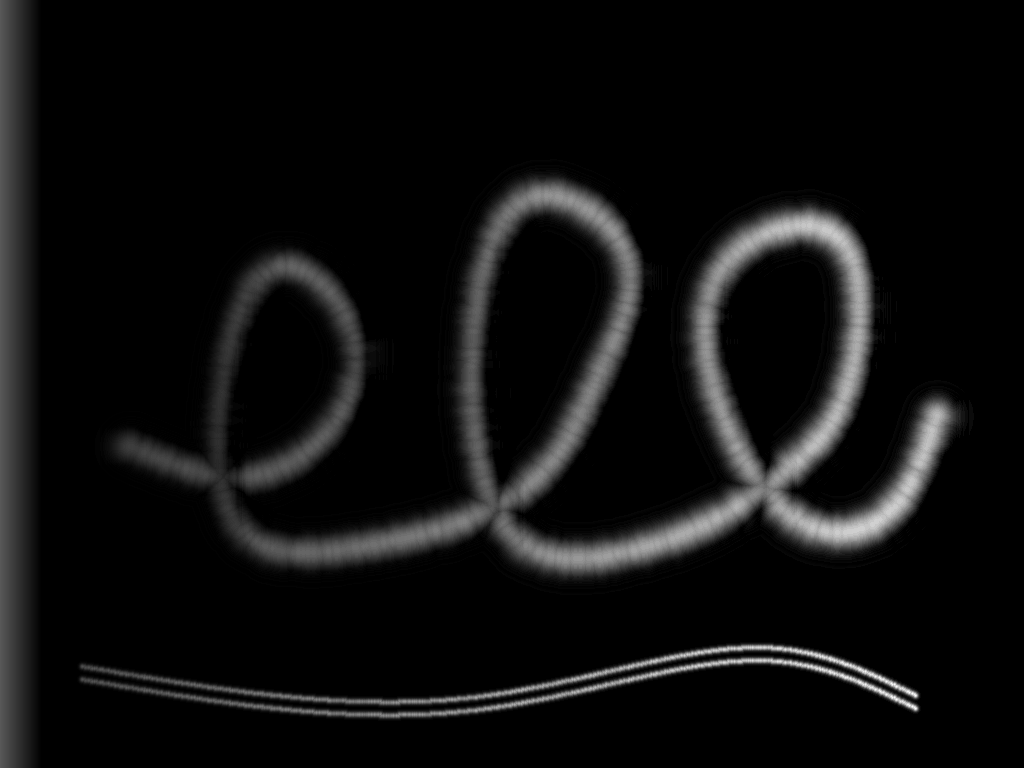}%
\label{fig_intro:tophat}}
\hfil
\subfloat[LIP top-hat]{\includegraphics[width=0.32\mycolumnwidth]{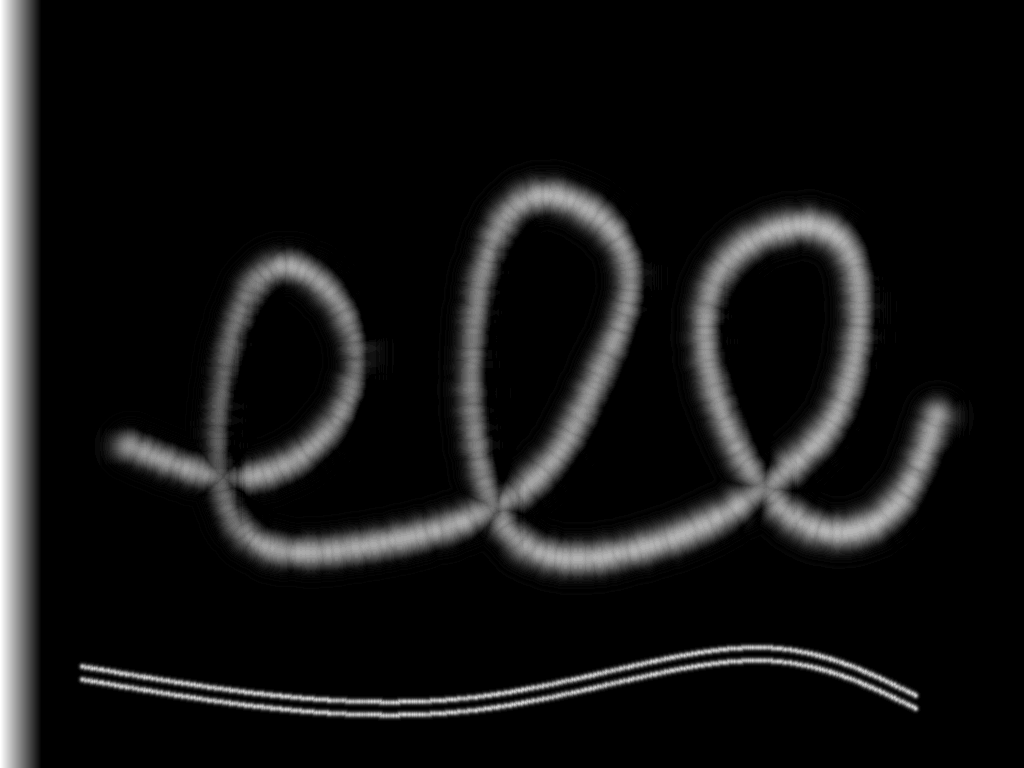}%
\label{fig_intro:Ltophat}}
\hfil
\subfloat[LMM]{\includegraphics[width=0.32\mycolumnwidth]{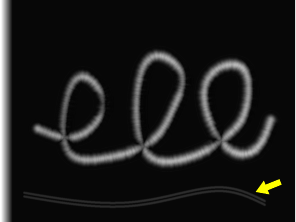}%
\label{fig_intro:Ldiff_Lopen}}

\caption{Comparison between usual methods and LMM to detect a spiral in (a) a simulated image also containing confounders : two close curves and a lighting drift. (b) This image is converted in the LIP-greyscale (i.e. the inverted greyscale). (d) The top-hat with a flat disk extracts the spiral and both curves. However, it is sensitive to the lighting drift. (e) The LIP-top-hat with a flat disk extracts the spiral and both curves without the lighting drift. (f) The LMM operator is a LIP-difference between two LIP-openings: the first LIP-opening is done by (c) a probe composed of a ring and a Gaussian and the second by a probe composed of the same ring. The LMM operator successfully extracts the spiral and strongly attenuates both close curves.}
\label{fig_intro}
\end{figure}

In parallel to the genesis of grey-level MM, Jourlin has developed the Logarithmic Image Processing (LIP) model which is adapted to human vision \cite{Jourlin1988,Jourlin2001,Jourlin2016} and which allows to process images as a human eye would do. It is not only rigorously defined from a mathematical point of view, but it also possesses strong physical properties. In particular, the LIP-addition $\LP$ of two images results in an image (with bounded values within the interval $\left[0,M\right[\>$). 
It also allows to simulate the variations of light intensity or of camera exposure-time in images.

The aim of this paper is to address both previously listed limitations of the MM application to grey level images by presenting a framework named \textit{Logarithmic Mathematical Morphology} (LMM)  that was recently introduced \cite{Noyel2019a,Noyel2021,Noyel2022}. 
Such a framework enables the amplitude of the translated structuring function by the LIP-addition to the image, to be adjusted according to the image intensity, thanks to the LIP-additive law $\LP$. LMM extends the theory of MM for images and functions by introducing operations of Logarithmic Image Processing.


In this article and beyond the prior work, the theory of LMM will be detailed. New theoretical results will be added at both following levels.
\begin{inparaenum}[(1)]
\item A link will be established between LMM and the functional Asplund metric defined with the LIP-additive law $\LP$. Such a metric is robust to lighting variations caused by a change of the light intensity or the camera exposure-time \cite{Jourlin2016}.
\item New morphological operators with the same robustness to lighting variations will be introduced for non flat structuring functions. 
\end{inparaenum}
LMM will also be validated with experiments and compared to state-of the art methods.



%
%

\section{Background}
\label{sec:full_back}


\subsection{Related work: detection operators robust to lighting variations}
\label{ssec:rel_wrk}

Previously in the literature, there have been some attempts to create operators theoretically robust to lighting changes. 
However, such operators generally do not take into account the physical causes of these lighting changes such as variations of light intensity or equivalently variations of camera exposure-time. In LMM, these causes are modelled by the LIP-additive law $\LP$ \cite{Jourlin2016}.
\figurename \ref{fig_intro}, shows an image composed of a spiral, a lighting drift and two confounding curves (Fig.~\ref{fig_intro:input}). In this example, the LMM operator better detects the spiral without the confounding curves (Fig.~\ref{fig_intro:Ldiff_Lopen}) than the usual methods based on a top-hat (Fig.~\ref{fig_intro:tophat}) or a logarithmic top-hat (Fig.~\ref{fig_intro:Ltophat}).

Let $f : D \mapsto \Realb$ be a function defined on a domain $D \subset \Real^n$ whose values are lying in $\Realb = \Real \cup \{-\infty , +\infty\}$. 
Let $b : D \mapsto \Realb$ be a structuring function whose values are equal to $-\infty$ outside a domain $D_b \subset D$, $\forall x \in D \setminus D_b$, $b(x)=-\infty$. In addition, if its values in $D_b$ are equal to zero, the structuring function is called a (flat) structuring element and it is represented by the upper-case letter $B$.

%
%
%



\subsubsection{Top-hat operators}
\label{sssec:rel_wrk:top_hat}

Meyer \cite{Meyer1979,Serra1982} created the \textit{top-hat} operator $TH_B(f)$ to detect peaks in a function $f$. It is equal to the difference between the function and its morphological opening $\gamma_B(f)$ by a flat structuring element $B$, $TH_B(f) = f - \gamma_B(f)$. The complementary operation for the detection of valleys is the \textit{bottom-hat} $BTH_B(f)$ defined as the difference between a morphological closing $\phi_B(f)$ of the function and the function itself, $BTH_B(f) = \phi_B(f)-f$.
Both top-hats are invariant to artificial variations of intensity caused by the addition or the subtraction of any real constant $c$ to a function $f$, $TH_B(f+c) = TH_B(f)$ and $BTH_B(f+c) = BTH_B(f)$. However, they are not invariant to any lighting variation with a physical cause and modelled by a LIP-addition $\LP$ or a LIP-subtraction $\LM$ of a constant to an image $f : D \mapsto \left[O,M\right[$\>. To address this issue, Jourlin et al. \cite{Jourlin1997,Jourlin2001} have introduced LIP top-hats where the LIP-difference $\LM$ replaces the usual difference ``$-$''. Zaharescu \cite{Zaharescu2007} proposed variants of LIP top-hats.
However, all these top-hats are still defined with a flat structuring element. They constitute a particular case of the extended tops-hats that will be presented in this paper (in section~\ref{ssec:RobOp:Extensions_top_hats}). 

\subsubsection{Morphological gradients}
\label{sssec:rel_wrk:morpho_grad}

Beucher \cite{Beucher1993} defined the \textit{morphological gradient} $\varrho_B(f)$ as the difference between the dilation $\delta_B(f)$
and the erosion $\varepsilon_B(f)$ 
of a function $f : D \mapsto \Real$ by a flat structuring element $B$, $\varrho_B(f) = \delta_B(f) - \varepsilon_B(f)$. 
The so-called ``morphological gradient'' corresponds in fact to the norm of the usual gradient of a function \cite{Beucher1990}. In order to be the norm of a gradient, it must be defined with a flat structuring element. It is invariant to the addition or subtraction of any real constant $c$ to a function, $\varrho_B(f+c) = \varrho_B(f)$. However, it is not invariant to a lighting variation with a physical cause and modelled by a LIP-addition $\LP$ or a LIP-subtraction $\LM$ of a constant to an image. Jourlin \cite{Jourlin1997} addressed this issue by defining a LIP-morphological gradient where the LIP-difference $\LM$ replaces the usual difference ``$-$'',\linebreak $\varrho_B^{LIP}(f) = \delta_B(f) \LM \varepsilon_B(f)$. 

\subsubsection{Scale Invariant Feature Transform (SIFT)}
\label{sssec:rel_wrk:SIFT}

Lowe introduced the \textit{Scale Invariant Feature Transform} to detect image features which are ``partially invariant to change in illumination'' \cite{Lowe2004}. In SIFT, salient points are first detected as some extrema of a function scale-space obtained by differences between Gaussian filtering of the image. A local image descriptor is then associated to each salient point. This image descriptor is based on an orientation histogram of the gradient of the Gaussian filtered image. As the gradient is computed by differences between image values, this makes it insensitive to illumination changes caused by the addition of a constant. The orientation is weighted by the norm of the gradient which is equal to the so-called ``morphological gradient'' and has therefore the same invariance. In addition, a normalisation between 0 and 1 of the orientation histograms makes the SIFT descriptor invariant to the multiplication of the image by a constant. However, such invariances do not have any physical causes.

\subsubsection{Trees of connected components}
\label{sssec:rel_wrk:CCTree}

Trees of connected components are based on \textit{level sets}. An image \textit{level set} is the set of these pixels whose values are greater or equal to a given threshold value. By increasing the threshold value, the connected components of a higher level set are included within the connected components of a lower level set. These inclusion relations can be represented by trees of connected components. Various types of trees can be built according to the inclusion relation between the level sets. One can cite e.g., the component-tree, also named max-tree \cite{Salembier1998}, or the tree of shapes, also named inclusion tree \cite{Monasse2000}. Recent segmentation methods by trees of connected components have been presented in \cite{Passat2011,Xu2016}. Trees of connected components of a real function are theoretically invariant to intensity changes caused by applying to the function a continuous and increasing transformation. However, as the intensity of an image is quantised in discrete values in the range $[\![0,1, \ldots, M-1]\!]$ with a constant step, the lower intensities are poorly represented by these discrete values. For this reason, in low-lighted images, the connected components trees may present a limited robustness to lighting variations \cite{Noyel2019c}.

\subsubsection{Intensity variant Mathematical Morphology}
\label{sssec:rel_wrk:IntVarMM}

Heijmans \cite{Heijmans1991} named \textit{t-operators} the usual morphological operators for functions  because of their invariance under horizontal translation (i.e. in space) and under vertical translation (i.e. in intensity). He also defined a class of morphological operators, the \textit{h-operators}, which are invariant under \textit{horizontal} translation but adaptive in the vertical domain (i.e. in intensity). Although the examples of \textit{h-operators} given in \cite{Heijmans1991} were mathematically well-defined for real functions, they were not physically justified. LMM operators form a particular case of \textit{h-operators} which are in addition defined for functions with bounded values such as images and are physically justified.


\subsection{Mathematical foundations}
\label{ssec:back}


\subsubsection{Mathematical Morphology}
\label{sssec:back:MM}

MM is defined on \textit{complete lattices} \cite{Birkhoff1967,Heijmans1990,Banon1993}. 
A set $\Lscr$ on which a partial order relation is defined is called a \textit{complete lattice} if every subset $\mathscr{X}$ of $\Lscr$ has an infimum (i.e., a greatest lower bound), $\wedge \mathscr{X}$, and a supremum (i.e., a least upper bound), $\vee \mathscr{X}$.
In the case of MM for functions, the set of functions $\Realb^D$ defined on a domain $D$ with values in $\Realb$ is a complete lattice with the order $\leq$.
Details about this complete lattice are in Appendix \ref{ssec:app:MM:real_fct}.
Between any two complete lattices $\Lscr_1$ and $\Lscr_2$, the fundamental morphological operations of \textit{erosion} and \textit{dilation} are defined as follows \cite{Serra1988,Heijmans1990,Banon1993}.
\begin{definition}
\label{def:erosion_dilation}
A mapping or an operator $\varepsilon : \Lscr_1 \mapsto \Lscr_2$ is an \textup{erosion}, if and only if (iff) it distributes over infima, that is $\varepsilon( \wedge \mathscr{X} ) = \wedge \varepsilon( \mathscr{X} )$, for any family $\mathscr{X} \subset \Lscr_1$.
The operator $\delta : \Lscr_2 \mapsto \Lscr_1$ is a \textup{dilation}, iff it distributes over suprema, that is $\psi( \vee \mathscr{X} ) = \vee \psi( \mathscr{X} )$, for any $\mathscr{X} \subset \Lscr_2$.\label{pre:def_dilation_erosion}
\end{definition}
Details about Definition \ref{pre:def_dilation_erosion} are in Appendix~\ref{ssec:app:MM:def_ero_dil}.
Moreover, the pair $(\varepsilon,\delta)$ forms an \textit{adjunction} between $\Lscr_1$ and $\Lscr_2$ if for all $X \in \Lscr_1$, $Y \in \Lscr_2$ the following equivalence holds
\begin{equation}
\delta(Y) \leq X \Leftrightarrow Y \leq \varepsilon(X). \label{eq:adjunction}
\end{equation}  
In an \textit{adjunction} $(\varepsilon,\delta)$, $\varepsilon$ is an \textit{erosion} and $\delta$ a \textit{dilation} \cite{Heijmans1990}.
If one reverses the ordering of both lattices $\Lscr_1$ and $\Lscr_2$, the dilation becomes an erosion and vice versa. The erosion and the dilation are called \textit{adjunct operators}. The adjunction constitutes a bijection between the erosion and the dilation. For every dilation $\delta$, there is a unique erosion $\varepsilon$ such that $(\varepsilon,\delta)$ is an adjunction and vice-versa.
Moreover, if $(\varepsilon,\delta)$ is an \textit{adjunction}, then the combination $\delta \varepsilon$ is an \textit{opening} on $\Lscr_1$ and the combination $\varepsilon \delta$ is a \textit{closing} on $\Lscr_2$ \cite{Ronse1991}. \textit{Opening} and \textit{closing} are morphological filters defined as follows \cite{Matheron1975,Serra1982,Ronse1991}.

\begin{definition}
An operator $\psi : \Lscr \mapsto \Lscr$ on the complete lattice $\Lscr$ is called an \textup{opening} if $\psi$ is increasing ($\forall X, Y \in \Lscr$, if $X \leq Y$ then $\psi(X) \leq \psi(Y)$), anti-extensive ($\forall X \in \Lscr$, $\psi(X) \leq X$) and idempotent ($\psi \circ \psi = \psi$). $\psi$ is a \textup{closing} if it is increasing, extensive ($\forall X \in \Lscr$, $X \leq \psi(X)$) and idempotent.
\label{pre:def_opening_closing}
\end{definition}

In the lattice $\Realb^D$ of real functions, let $b : D \mapsto \Realb$ be a structuring function. The functional dilation $\delta_b : \Realb^D \mapsto \Realb^D$ and erosion $\varepsilon_b : \Realb^D \mapsto \Realb^D$ are \textit{t-operators} which are invariant under horizontal and vertical translations and which are usually expressed \cite{Serra1982} by:
\begin{align}
\delta_b(f)(x)		 &=\vee 	\left\{ f(x - h) + b(h), h \in D \right\} = (f \oplus b) (x) \label{eq:dilate_funct}\\
\varepsilon_b(f)(x)&=\wedge \left\{ f(x + h) - b(h), h \in D \right\} = (f \ominus b) (x). \label{eq:erode_funct}%
\end{align}
In the case of ambiguous expressions, the following conventions are used: $f(x - h) + b(h) = -\infty$ when $f(x - h) = -\infty$ or $b(h) = -\infty$, and $f(x + h) - b(h) = +\infty$ when $f(x + h) = +\infty$ or $b(h) = -\infty$ \cite{Heijmans1990}. The symbols $\oplus$ and $\ominus$ represent the extension to functions of Minkowski operations between sets \cite{Serra1982}. 
Overviews of MM are available in \cite{Serra1988,Heijmans1990,Soille2003,Najman2013,Meyer2019_1,Meyer2019_2} and some recent advances in the field can be found in \cite{Schonfeld2008,Angulo2011,vanDeGronde2014,Merveille2018}.

\subsubsection{Logarithmic Image Processing}
\label{sssec:back:LIP}

The LIP model is a mathematical framework which allows to process images in a way which is compatible with the human visual system \cite{Brailean1991,Jourlin2016}. This makes it valid not only for images acquired with transmitted light but also with reflected light. The LIP model is based on the \textit{physical law of transmittances}, 
\begin{align}
T_{f \LP g} (x) &= T_f(x) \cdot T_g(x)\text{,} \label{eq:LIP:transmittance_law}%
\end{align}
where the transmittance $T_{f \LP g}$ of the superimposition of two semitransparent objects generating the images $f \in \I = [0,M[^D$ and $g \in \I$ is equal to the point-wise product of their respective transmittances $T_f$ and $T_g$. The transmittance $T_f(x)$ at point $x \in D$ is also related to the image grey value $f(x)$ by the equation 
\begin{align}
T_f(x) &= 1 - f(x)/M\text{,} \label{eq:LIP:transmittance}%
\end{align} where $M$ is the upper-bound of the grey value interval $[0,M[$\>. Due to this relation, the LIP-greyscale is inverted compared to the usual greyscale (\figurename~\ref{fig_intro:input_LIP_scale}). This means that $0$ corresponds to the white extremity, when there is no obstacle between the light source and the camera, whereas $M$ corresponds to the black extremity when no light is passing. By replacing the transmittances $T_f$ and $T_g$ by their expressions in the transmittance law, the addition $\LP$ of two images $f$ and $g$ is deduced:
\begin{equation}
	f \LIPplus g = f + g - \frac{f \cdot g}{M}\>. \label{eq:LIP:plus}%
\end{equation}
In transmitted light, the addition of two images corresponds to the superimposition of two semitransparent objects generating the images $f$ and $g$. From \eqref{eq:LIP:plus}, the LIP-multiplication $\LT$ of an image by a scalar $\lambda \in \Real$ is deduced, $\lambda \LIPtimes f = M - M \left( 1 - f/M \right)^{\lambda}$. It is equivalent to a variation of thickness or opacity of the object by a factor $\la$. If $\la > 1$, the image becomes darker, whereas if $\la \in [0,1]$ it becomes brighter.
The opposite function $\LM f = -1 \LT f$ is then deduced: 
\begin{equation}
	\LM f = \frac{-f}{1-f/M}. \label{eq:LIP:neg}%
\end{equation}
One can notice that $\LM f$, where $f \geq 0$, is not an image as it takes negative values. It belongs to the set $\Fcurv_M = \left]-\infty,M\right[^D$ of real functions $f$ whose values are bounded by $M$, $f : D \mapsto \left]-\infty,M\right[$\>. From a physical point of view, the negative values $\LM f$,  where $f \geq 0$, are light intensifiers that can be used to compensate the attenuation of the semi-transparent object generating the image $f$. Their superimposition with the image $f$ is equal to zero (i.e. the white intensity),\linebreak $f \LP (\LM f) = f \LM f= 0$. This is an important physical property that will be used in this paper. In particular, the LIP-addition of a negative constant will allow to compensate the light attenuation due to a variation of camera exposure-time or of light intensity \cite{Jourlin2016}. From \eqref{eq:LIP:neg}, the difference between two functions with bounded values $f$ and $g \in \Fcurv_M$ is deduced: 
\begin{equation}
	f \LM g = \frac{f-g}{1-g/M}\>. \label{eq:LIP:minus}%
\end{equation}
$f \LM g$ is an image iff $f \geq g$. The space of functions whose values are bounded by $M$, $(\Fcurv_M,\LP,\LT)$, is a \textit{real vector space} and the space of images, $(\I,\LP,\LT)$, represents the \textit{positive cone} of this vector space \cite{Jourlin2001,Jourlin2016}. $\Fcurv_M$ and $\I$ are both ordered by the usual order $\leq$ \cite{Jourlin2001}. 
As the LIP addition and subtraction $\LP$ and $\LM$ are defined between functions of $\Fcurv_M$, they are also defined between function values of $\left]\infty,M\right[$\>.
Let $a, b, c$ be function values, i.e., elements of $\left]-\infty,M\right[$\>, the following properties hold:
\begin{enumerate}[P1 :]
  \item $\LM (\LM a) = a$ \label{LIP:P1}
	\item $a \LP b = b \LP a$ \label{LIP:P2}
	\item $(a \LP b) \LP c = a \LP (b \LP c)$ \label{LIP:P3}
	\item $(a \LM b) \LM c = a \LM (b \LP c)$ \label{LIP:P4}
	\item $a \LP (\LM b) = a \LM b$ \label{LIP:P5}
	\item $(a \LP b) \LM c =  (a \LM c) \LP b$ \label{LIP:P6}
	\item $(a \LP c) \LM (b \LP c) = a \LM b$ \label{LIP:P7}
	\item $(\LM a) \LM b = \LM (a \LP b)$ \label{LIP:P8}
	\item $a \LP b \leq c \Leftrightarrow a\leq c \LM b$. \label{LIP:P9}
	\item $a \leq c \Leftrightarrow a \LP c \leq b \LP c$. \label{LIP:P10}
\end{enumerate}
The LIP operators, $\LP$ and $\LM$, behave in a similar way as the usual operators $+$ and $-$ on the interval $\left]-\infty,M\right[$\>.

%
%

\subsection{Logarithmic Mathematical Morphology}
\label{sec:LMM}


%
%
\subsubsection{The framework}
\label{ssec:LMM:framework}

LMM is defined in the lattice $\Fcurvb_M=[-\infty,M]^D$ of functions with values in $[-\infty,M]$ \cite{Noyel2019a,Noyel2021,Noyel2022}, whereas MM for functions is defined in the lattice $\Realb^D$ of functions with values in $\left[-\infty,+\infty\right]$. An isomorphism will be used to link both lattices and to deduce the properties of LMM from the properties of MM for functions. Before defining this isomorphism, let us recall some properties of lattice isomorphisms.

A lattice isomorphism between complete lattices commutes with arbitrary supremum and infimum. For totally ordered lattices (in particular subsets of the completed real line), a map is a lattice
isomorphism iff it is a strictly increasing bijection, it is a dual isomorphism iff it is a strictly decreasing
bijection. For lattices that are also vector spaces where addition and multiplication by a positive scalar are
increasing, see \cite{Birkhoff1967}.

This isomorphism will be defined by analogy with the transmittance
 $T_f=1-f/M$ of \eqref{eq:LIP:transmittance}, the law of transmittances $T_{f\LP g} = T_f.T_g$ in \eqref{eq:LIP:transmittance_law} where the combination of $f$ and $g \in \left[0,M\right[^D$ leads to multiplying their transmittances and defines the LIP-addition $f \LP g = f+g-f.g/M$ in \eqref{eq:LIP:plus}.
Let us define the map:
\begin{align*}
     \tau: \begin{cases}
			\left[0,M\right] &\mapsto \left[0,1\right]\\
		  x &\mapsto 1-x/M
		 \end{cases}.
\end{align*}	
It is a strictly decreasing bijection; thus it is a dual automorphism (i.e. anti-isomorphism) between the two complete lattices $\left[0,M\right]$ and $\left[0,1\right]$. The binary operation $\LP$ defined by $x \LP y = x+y-xy/M$ lies on $\left[0,M\right]$ and satisfies $\tau(x \LP y) = \tau(x).\tau(y)$. In order to consider the inverse $\LM$, $\tau$ is extended to $[-\infty,M]$, it is then a dual isomorphism $\left[-\infty,M\right] \mapsto \left[0,+\infty\right]$, with $\tau(M)=0$, $\tau(0)=1$ and $ \tau(-\infty)=+\infty$. The definition of the operation $\LP$ is then extended to $\left]-\infty,M\right]$ (with $\tau(\left]-\infty,M\right]) = \left[0,+\infty\right[$). For $x$ in $\left]-\infty,M\right[$ (that is, $\tau(x) \in \left]0,+\infty\right[$), we define the inverse $\LM x$ by $x \LP (\LM x) =0$, which gives $\tau(x \LP (\LM x)) = \tau(0)$. This leads to the following expression of $\LM x$
\begin{align*}
	\tau(\LM x) = 1/\tau(x),\> \text{ giving }\> 
	(\LM x) = -x/(1-x/M).
\end{align*}	
Then, the LIP-difference $x \LM y$ is defined by
\begin{align*}
     x \LM y &= x \LP (\LM y),\> \text{ in other words }\> \tau(x \LM y) = \tau(x)/\tau(y).
\end{align*}
For a scalar $\lambda$, $\lambda \LT x$ is defined by
\begin{align*}
     \tau(\lambda \LT x) = \tau(x)^{\lambda},\>\text{ therefore }\>\lambda \LT x = M-M(1-x/M)^{\lambda}.
\end{align*}
Hence $\tau$ is an isomorphism between the vector spaces $\left]-\infty,M\right[$ (with $\LP$ and $\LT$) and $\left]0,+\infty\right[$ (with multiplication $.$ and scalar exponentiation \textasciicircum). By applying pointwise the isomorphisms to functions, $\tau$ is a lattice dual isomorphism $\left[0,M\right]^D \mapsto \left[0,1\right]^D$ which is extended to $\left[-\infty,M\right]^D \mapsto \left[0,+\infty\right]^D$. $\tau$ is therefore a vector space isomorphism $(\left]-\infty,M\right[^D,\LP,\LT) \mapsto (\left]0,+\infty\right[^D,.,\text{\textasciicircum} )$.

In order to have a lattice isomorphism and a range that is a vector space with addition and scalar multiplication, $\tau$ is composed with a negative logarithm function, so $\xi$ is defined on $\left[-\infty,M\right]$ by $\xi(x) = -M.\ln(\tau(x))$. Then $\xi$ is a strictly increasing bijection $\left[-\infty,M\right] \mapsto \left[-\infty,+\infty\right]$, hence a lattice isomorphism, with $\xi(-\infty)=-\infty$, $\xi(0)=0$ and $\xi(M)=+\infty$. By restricting $\xi$ to $\left]-\infty,M\right[$\>, it is a vector space isomorphism $(]-\infty,M[,\LP,\LT) \mapsto (]-\infty,+\infty[,+,.)$.

By applying pointwise the isomorphisms to functions, $\xi$ is a lattice isomorphism $\left[0,M\right]^D \mapsto \left[0,+\infty\right]^D$ which is extended to $\left[-\infty,M\right]^D \mapsto \left[-\infty,+\infty\right]^D$,
 and a vector isomorphism $\left]-\infty,M\right[^D \mapsto \left]-\infty,+\infty\right[^D$.
Here $\left[0,M\right[^D$ is the positive cone of the vector space. As an isomorphism, $\xi$ satisfies $\xi(f \LP g) = \xi(f) + \xi(g)$, $\xi(\LM f) = -\xi(f)$, $\xi(f \LM g) = \xi(f) - \xi(g)$ and $\xi(\lambda \LT f) = \lambda . \xi(f)$, for $f,g \in \left]-\infty,M\right[^D$\> and $\lambda \in \Real$.


Now, the LIP-Minkowski operations are defined by isomorphism with the original ones $\oplus$ and $\ominus$ through $\xi$. The LIP-Minkowski addition $\oplusLIP$ and subtraction $\ominusLIP$ of functions $f$ and $g$ in $\left[-\infty,M\right]^D$ are defined by
\begin{align*}
     \xi(f \oplusLIP g) &= \xi(f) \oplus \xi(g)\>  \text{ and }\>  \xi(f \ominusLIP g) = \xi(f) \ominus \xi(g).
\end{align*}
The expression for the value of $f \oplusLIP g$ at a point $x \in D$ is easily derived:
\begin{align*}
	(f \oplusLIP g)(x) &= \xi^{-1} \left( \sup_h \left\{\xi\left(f(x-h)\right) + \xi\left(g(h)\right)\right\} \right) &\\
	&= \sup_h \left\{ \xi^{-1} \left[\xi\left(f(x-h)\right) + \xi\left(g(h)\right)\right] \right\} &\text{ $\xi^{-1}$  is a lattice isomorphism}\\
	&= \sup_h \left\{ \xi^{-1} \left[\xi\left(f(x-h)\right)\right] \LP \xi^{-1} \left[\xi\left(g(h)\right)\right] \right\} &\text{ $\xi^{-1}$  is a vector isomorphism}\\
	&= \sup_h \left\{ f(x-h) \LP g(h) \right\}.&
\end{align*}
\begin{remark}\label{rem:xi_commutes_with_sup}
The isomorphism $\xi$ commutes with the supremum or the infimum because it is an increasing bijection.
\end{remark}
Here, the rule that the $-\infty$ value for $f(x-h)$ or $g(h)$ is an absorbant follows by isomorphism from the same rule with the ordinary Minkowski addition.
A similar derivation is made for $(f \ominusLIP g)(x)$
\begin{align*}
	(f \ominusLIP g)(x) &= \xi^{-1} \left( \inf_h \left\{\xi\left(f(x+h)\right) - \xi\left(g(h)\right)\right\} \right) &\\
	&= \inf_h \left\{ f(x+h) \LM g(h) \right\},&
\end{align*}\label{eq:LIP_ominus_minkowski}
where the rule that the $+\infty$ value for $f(x+h)$ or $g(h)$ is an absorbant, is obtained by isomorphism.
Then, the LIP dilation and erosion which are defined from the LIP Minkowski operations in the same way as for the usual operations. The dilation $\delta_b^{\LP} : \Fcurvb_M \mapsto \Fcurvb_M$ by the structuring function $b$ is $f \mapsto f \oplusLIP b$ and the erosion $\varepsilon_b^{\LP} : \Fcurvb_M \mapsto \Fcurvb_M$ by $b$ is $f \mapsto f \ominusLIP b$.
For each point $x\in D$, both mappings are expressed by:
\begin{align}
\delta_b^{\LP}(f)(x)		 &= \left(f \oplusLIP b\right)(x) = \vee 	\left\{ f(x - h) \LP b(h), h \in D \right\} \label{eq:LIP-dilation}\\
\varepsilon_b^{\LP}(f)(x)&= \left(f \ominusLIP b \right)(x) = \wedge \left\{ f(x + h) \LM b(h), h \in D \right\}. \label{eq:LIP-erosion}%
\end{align}
In the case of ambiguous expressions, the following conventions will be used: $f(x - h) \LP b(h) = -\infty$ when $f(x - h) = -\infty$ or $b(h) = -\infty$, and $f(x + h) \LM b(h) = M$ when $f(x + h) = M$ or $b(h) = -\infty$.
The LIP-additive law $\LP$ and the LIP-negative law $\LM$ perform morphological transformations that are compatible with the human visual system.

By isomorphism, the LIP dilation and erosion form an adjunction, therefore, they are a dilation and erosion in the lattice theoretical sense.
\begin{proposition} \label{prop:LMM:base_operators}
The pair of mappings $(\varepsilon_b^{\LP}, \delta_b^{\LP})$ forms an \textup{adjunction}, where $\varepsilon_b^{\LP}$ is an \textup{erosion} and $\delta_b^{\LP}$ is a \textup{dilation}.
\end{proposition}


\begin{definition} \label{def:LMM:base_operators}
$\varepsilon_b^{\LP}$ is called a \textup{logarithmic-erosion} and $\delta_b^{\LP}$ a \textup{logarithmic-dilation}. 
\end{definition}
As $(\varepsilon_b^{\LP} , \delta_b^{\LP})$ forms an adjunction, an \textit{opening} and a \textit{closing} can be defined by combination of both operators of logarithmic-erosion $\varepsilon_b^{\LP}$ and logarithmic-dilation $\delta_b^{\LP}$ (see section \ref{sssec:back:MM}). The operators $\gamma_{b}^{\LP}$ and $\varphi_{b}^{\LP}$ defined by
\begin{align}
\gamma_{b}^{\LP} &= \delta_b^{\LP} \varepsilon_b^{\LP}, \label{eq:LIP-opening}\\
\varphi_{b}^{\LP} &= \varepsilon_b^{\LP} \delta_b^{\LP} \label{eq:LIP-closing}%
\end{align}
are an \textit{opening} and a \textit{closing} (by adjunction), respectively.

\begin{definition} \label{def:LMM:opening_closing}
 $\gamma_{b}^{\LP}$ is called a \textup{logarithmic-opening} and $\varphi_{b}^{\LP}$ a \textup{logarithmic-closing}.
\end{definition}

The isomorphism  $\xi: \Fcurvb_M \mapsto \Realb^D$ and its inverse  $\xi^{-1}: \Realb^D \mapsto \Fcurvb_M$ were both defined in \cite{Jourlin1995} by $\xi(f) = -M \ln{(1-f/M)}=-M\ln{(\tau(f))}$ and $\xi^{-1}(f) = M(1-\exp{(-f/M)})=\tau^{-1}\left(\exp{(-f/M)}\right)$, where $\tau^{-1}=M(1-f)$. With this isomorphism $\xi$, the following proposition is immediate.
\begin{proposition}	\label{prop:link_MM_LMM}
	Let $f \in \Fcurvb_M$ be a function and $b \in \Fcurvb_M$ a structuring function. The logarithmic-dilation $\delta_b^{\LP}$ and the logarithmic-erosion $\varepsilon_b^{\LP}$ are related to the functional dilation $\delta_b$, or $\oplus b$, and erosion $\varepsilon_b$, or $\ominus b$, respectively, by the equations:
	\begin{align}
		\delta_b^{\LP} (f) &= f \oplusLIP b =  \xi^{-1} \left[ \xi(f) \oplus \xi(b) \right] 
		= \xi^{-1} \left( \delta_{\xi(b)}\left[\xi(f)\right] \right),\label{eq:LIP_dil_prop}\\
		\varepsilon_b^{\LP} (f) &= f \ominusLIP b = \xi^{-1} \left[ \xi(f) \ominus \xi(b) \right]
		= \xi^{-1} \left( \varepsilon_{\xi(b)} \left[\xi(f)\right] \right). \label{eq:LIP_ero_prop}
	\end{align}
\end{proposition}
These relations facilitate the implementation of the LMM operators as those of usual MM already exist in numerous image analysis software.

Another useful property in MM is that 
the erosion of a function is equal to the negative of the dilation of its \textit{negative function}, and vice versa. Such a property is the duality by \textit{negative function} of both operators and, for LMM, it is established in proposition \ref{prop:LMM_duality}. The \textit{negative function} $f^{Lneg}$ of $f$ is equal to $f^{Lneg} = \LM f$ because of  $\xi(\LM f) = -\xi(f)$, it corresponds by isomorphism to the usual negative \cite{Heijmans1994}.

\begin{proposition}[\cite{Noyel2019a}]
	\label{prop:LMM_duality}
Let  $\overline{b} \in \Fcurvb_M$ be the reflected structuring function of $b$, where $\forall x \in D$, $\overline{b}(x) = b(-x)$. The logarithmic-erosion $\varepsilon_{b}^{\LP}$ and dilation $\delta_b^{\LP}$ are dual by their \textup{negative functions}:
\begin{align}
	(\delta_b^{\LP} (f^{Lneg}))^{Lneg} = \varepsilon_{\overline{b}}^{\LP} (f) \quad \text{and} \quad (\varepsilon_{b}^{\LP} (f^{Lneg}))^{Lneg} = \delta_{\overline{b}}^{\LP} (f)\>. \label{eq:duality_LMM}
\end{align}
\end{proposition}

\begin{proof}[Proof of proposition \ref{prop:LMM_duality}]
It follows immediately by isomorphism from the corresponding property for the usual dilation and erosion based on Minkowski operations in the lattice $[-\infty,+\infty]^D$, namely $-[-f \oplus g] = f \ominus \overline{g}$ and $-[-f \ominus g] = f \oplus \overline{g}$ (here $\overline{g}$ is the symmetrical or reflection of $g$, where $\overline{g}(x) = g(-x)$).
In the lattice $[-\infty,M]^D$, we have, $\forall f, b \in [-\infty,M]^D$, 
	\begin{align*}
	(\delta_b^{\LP} (f^{Lneg}))^{Lneg} &= \LM \left[ (\LM f) \oplusLIP b\right]
	= f \ominusLIP \overline{b}
	= \varepsilon_{\overline{b}}^{\LP} (f).
	\end{align*}
	Similarly, we have $(\varepsilon_{b}^{\LP} (f^{Lneg}))^{Lneg} = \delta_{\overline{b}}^{\LP} (f)$.
\end{proof}

\begin{figure}[!t]
\centering
\subfloat[Erosions]{\includegraphics[width=0.49\mycolumnwidth]{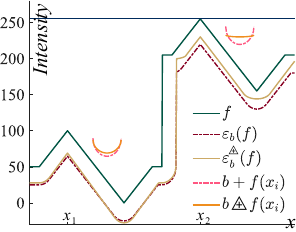}%
\label{fig:comp_MM_vs_LMM_signal:ero}}
\hfil
\subfloat[Dilations]{\includegraphics[width=0.49\mycolumnwidth]{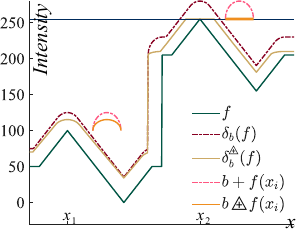}%
\label{fig:comp_MM_vs_LMM_signal:dil}}
\\
\subfloat[Openings]{\includegraphics[width=0.49\mycolumnwidth]{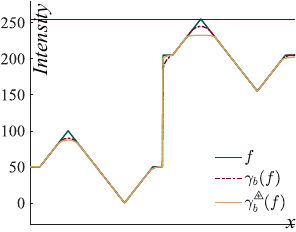}%
\label{fig:comp_MM_vs_LMM_signal:open}}
\hfil
\subfloat[Closings]{\includegraphics[width=0.49\mycolumnwidth]{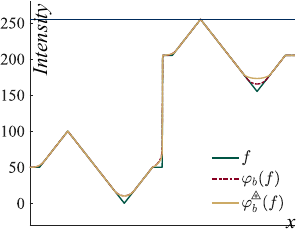}%
\label{fig:comp_MM_vs_LMM_signal:close}}
\caption{In an image $f$ (represented in the LIP-greyscale), functional MM is compared to LMM for: 
(a) the erosions $\varepsilon_b(f)$, $\varepsilon_b^{\protect \LP}(f)$, 
(b) the dilations $\delta_b(f)$, $\delta_b^{\protect \LP}(f)$, 
(c) the openings $\gamma_b(f)$, $\gamma_b^{\protect \LP}(f)$ and 
(d) the closings $\varphi_b(f)$, $\varphi_b^{\protect \LP}(f)$. 
(a) and (b) For both image peaks $f(x_1)$ and $f(x_2)$, the vertically translated structuring functions $b+f(x_i)$ and  $b {\protect \LP} f(x_i)$, where $i\in\left\{1,2\right\}$, are represented (after an horizontal translation) for the erosions $\varepsilon_b(f)$, $\varepsilon_b^{\protect \LP}(f)$ and for the dilations $\delta_b(f)$, $\delta_b^{\protect \LP}(f)$. The structuring function $b$ is a semi-circle with a diameter of 50 pixels.}
\label{fig:comp_MM_vs_LMM_signal}
\end{figure}

In the unidimensional image of \figurename \ref{fig:comp_MM_vs_LMM_signal}, operators of functional MM are compared to those of LMM. A semi-circle with a diameter of 50 pixels acts as a structuring function (sf) $b$. In LMM, the amplitude of the sf translated by the LIP-addition or LIP-subtraction of an image value changes according to the image values because of the LIP-laws, $\LP$ or $\LM$, used in \eqref{eq:LIP-dilation} and \eqref{eq:LIP-erosion}. LMM operators are therefore \textit{h-operators} which are only invariant under horizontal translation (see section \ref{sssec:rel_wrk:IntVarMM}). 
However in functional MM, the amplitude of the sf translated by the addition of the image value remains the same. This generates \textit{t-operators} which are invariant under horizontal and vertical translations. 
Moreover, in \figurename \ref{fig:comp_MM_vs_LMM_signal:dil}, the logarithmic-dilation $\delta_b^{\LP}(f)$ of $f$ is always below the upper bound $M = 256$, whereas the functional dilation $\delta_b(f)$ of $f$ may exceed this bound. Such a property is due to the LIP-addition law $\LP$. 
In \figurename \ref{fig:comp_MM_vs_LMM_signal:ero}, the negative values of the functional erosion $\varepsilon_b(f)$ have no physical justification, whereas those of the erosion $\varepsilon_b^{\LP}(f)$ correspond to light intensifiers. 
In \figurename \ref{fig:comp_MM_vs_LMM_signal:open}, the difference between the functional opening $\gamma_b (f)$ and the logarithmic-opening $\gamma_b^{\LP} (f)$ is greater for the grey-levels close to $M$ than for those close to zero. Indeed, in LMM, the amplitude of the sf that is LIP-added to an image is greater for higher image intensities than for lower ones due to the nonlinearity of the LIP laws $\LP$ and $\LM$. 
The same observation exists between the functional closing $\varphi_b(f)$ and the logarithmic-closing $\varphi_b^{\LP}(f)$ (\figurename \ref{fig:comp_MM_vs_LMM_signal:close}).

\begin{remark}\label{rem:LIP_translated_probe}
Let $b : D_b \mapsto \left]-\infty,M\right[$ be a structuring function whose supremum value is $s=\vee_{x\in D_b}{\left\{b(x)\right\}}$ and whose infimum value is $i=\wedge_{x\in D_b}{\left\{b(x)\right\}}$. The amplitude of the structuring function $b$ is equal to 
\begin{align*}
A_b &= s - i,
\end{align*}
where $A_b \in \Real$ is a real number. Clearly, the amplitude of the structuring function $b$ does not vary according to the image amplitude. However, if a constant is LIP-added to the structuring function, let us study its effect on the translated structuring function.
Let $c$ be a constant function $c : D_b \mapsto \left]-\infty,M\right[$, such that for any $x \in D_b$, $c(x) = C$, where $C \in \left]-\infty,M\right[$\>. 
The amplitude of the translated structuring function $b \LP c$ by the LIP-addition of the constant function $c$ is equal to:
\begin{align*}
A_{b\LP c} &= \vee_{x\in D_b}{\left\{b(x) \LP c(x)\right\}} - \wedge_{x\in D_b}{\left\{b(x) \LP c(x)\right\}}&\nonumber\\
					&= \left[\vee_{x\in D_b}{\left\{b(x)\right\}} \LP C\right] - \left[\wedge_{x\in D_b}{\left\{b(x)\right\}} \LP C\right] \nonumber &\quad \LIPPref{LIP:P10}\\
					&= \left[s \left(1-C/M\right) + C\right] - \left[i \left(1-C/M\right) + C\right]&\nonumber\\
					&= \left(1-C/M\right)\left(s - i\right) &\nonumber\\
					&= \left(1-C/M\right)A_b. &\nonumber
\end{align*} 
The factor $(1 - C/M)$ lies within the interval $]0,+\infty[$\>.
Therefore, the amplitude $A_{b\LP c}$ of the translated structuring function $b\LP c$ by the LIP-addition of a constant function $c$ is modified by a factor $(1-C/M)$ with respect to the amplitude $A_b$ of the structuring function $b$. When the value $C$ depends on an image intensity, the amplitude of the translated structuring function varies according to the image intensity.
\end{remark}

\begin{remark}\label{rem:link_mult_add_morpho}
The idea of a logarithmic mapping of grey values between morphological adjunctions
using additive versus multiplicative structuring functions can be found in \cite{Heijmans1990} and \cite[Section 5.7-B]{Heijmans1994}. 
However, this mapping relies on a single logarithmic function which, unlike the LIP isomorphism $\xi$, lacks a physical justification for images whose grey values lie in the interval $\left[0,M\right[$\>. For other image modalities or perceptual models, however, such a logarithmic function may be justified. Further details are provided in Appendix \ref{sec:app:detailed_remark_log_iso}.
\end{remark}

\subsubsection{Rank filters}
\label{ssec:LMM:rankfilt}

Functional dilations $\delta_b$ and erosions $\varepsilon_b$ are based on supremum and infimum operations. As supremum and infimum are very sensitive to noise such as speckle \cite{Ronse1991}, they can be replaced by \textit{rank filters} \cite{Soille2003} also named \textit{order statistics filters} \cite{Maragos1987}, \textit{percentile filters} or \textit{rank order filters}. The filter of rank $k$ selects the $k^{th}$ smallest element of a set. It corresponds to the $k^{th}$ \textit{minimum} represented by $\wedge^k$. Its dual, the $k^{th}$ \textit{maximum} $\vee^k$ selects the $k^{th}$ greatest element of a set. 
In \eqref{eq:erode_funct}, a $k^{th}$ \textit{minimum filter} $\zeta_{b,k} : \Realb \mapsto \Realb$ can be defined  by replacing the infimum $\wedge$ by the the $k^{th}$ minimum $\wedge^k$. The filter $\zeta_{b,k}$ is called the \textit{minimum-rank filter of rank $k$} by the structuring function $b: D \mapsto \Realb$. Similarly, in \eqref{eq:dilate_funct}, a $k^{th}$ \textit{maximum filter} $\vartheta_{b,k}$ can be defined with the $k^{th}$ maximum $\vee^k$. The filter $\vartheta_{b,k}$ is the \textit{maximum-rank filter of rank $k$}. Both operations are defined by:
\begin{align}
	\zeta_{b,k}(f)(x) &= \wedge^k \{ f(x+h) - b(h), h \in D_b \}, \label{eq:_erotol}\\
	\vartheta_{b,k}(f)(x) &= \vee^k   \{ f(x-h) + b(h), h \in D_b \}. \label{eq:_diltol}
\end{align}
If $k=1$, the \textit{minimum-rank filter of rank $1$}, $\zeta_{b,1}$, is equal to the functional \textit{erosion} $\varepsilon_b$ and the \textit{maximum-rank filter of rank $1$}, $\vartheta_{b,1}$, is equal to the functional \textit{dilation} $\delta_b$.

In LMM, the $k^{th}$ \textit{minimum logarithmic filter} $\zeta^{\LP}_{b,k} : \Fcurvb_M \mapsto \Fcurvb_M$ and the $k^{th}$ \textit{maximum logarithmic filter} $\vartheta^{\LP}_{b,k}$ can also be defined by using the $k^{th}$ minimum $\wedge^k$ and the $k^{th}$ maximum $\vee^k$ in place of the minimum $\wedge$ and the maximum $\vee$ 
in \eqref{eq:LIP-erosion} and \eqref{eq:LIP-dilation}, respectively. The structuring function is denoted $b \in \Fcurvb_M$.
Both filters $\zeta_{b,k}^{\LP}$ and $\vartheta_{b,k}^{\LP}$ are called \textit{LIP-minimum-rank filter} and \textit{LIP-maximum-rank filter of rank $k$}. They both are defined as follows:
\begin{align}
	\zeta_{b,k}^{\LP}(f)(x) &= \wedge^k \{ f(x+h) \LM b(h), h \in D_b \}, \label{eq:_LIP_erotol}\\
	\vartheta_{b,k}^{\LP}(f)(x) &= \vee^k   \{ f(x-h) \LP b(h), h \in D_b \}. \label{eq:_LIP_diltol}
\end{align}
The $k^{th}$ \textit{LIP-minimum-rank filter} $\zeta^{\LP}_{b,k}$ and \textit{LIP-maximum-rank filter} $\vartheta^{\LP}_{b,k}$ are also related to the $k^{th}$ \textit{minimum-rank filter} $\zeta_{b,k}$ and \textit{maximum-rank filter} $\vartheta_{b,k}$ by replacing the erosions, $\varepsilon_b$ and $\varepsilon^{\LP}_b$, and dilations, $\delta_b$ and $\delta^{\LP}_b$, by their corresponding rank filters in \eqref{eq:LIP_ero_prop} and \eqref{eq:LIP_dil_prop}. The relations are as follows:
\begin{align}
		\vartheta_{b,k}^{\LP} (f) &= \xi^{-1} \left( \vartheta_{\xi(b),k}\left[\xi(f)\right] \right) \nonumber\\
		&= M \left[1 - \exp{\left(- \vartheta_{\xi(\acute{b}),k}(\acute{f}) \right)}\right], \label{eq:LIP_diltol_prop}\\
		\zeta^{\LP}_{b,k} (f) &= \xi^{-1} \left( \zeta_{\xi(b),k} \left[\xi(f)\right] \right)  \nonumber\\
		&= M \left[1 - \exp{\left(-\zeta_{\xi(\acute{b}),k}(\acute{f}) \right)}\right], \label{eq:LIP_erotol_prop}
\end{align}
where $\acute{f}: D \mapsto \Realb$ is equal to $\acute{f}= -\ln{\left( 1 - f/M \right)}$.

%
%

\section{Operators robust to lighting variations}
\label{sec:RobOp}

Examples of operators robust to lighting changes caused by variations of the camera exposure-time or of the light intensity will be given. These lighting changes are modelled by the LIP-addition of a constant.


\subsection{Map of LIP-additive Asplund distances}
\label{ssec:RobOp:Aspl}


\subsubsection{Link with LMM}
\label{sssec:RobOp:Aspl:linkLMM}


The functional Asplund metric with the LIP-additive law $\LP$ was defined by Jourlin \cite{Jourlin2016}.
Let $f$ and $g \in \Fcurv_M $ be two functions. One of them, e.g. $g$, is chosen as a probing function and both following numbers are defined: $c_1 = \inf{ \{c, f \leq c \LP g \}}$ and $c_2 = \sup{ \{c, c \LP g \leq f \} }$, where $c$ lies within the interval $]-\infty,M[$\>. $c_1$ and $c_2$ are the constants to be LIP-added to the probe $b$ such that it is in contact with the function $f$ from above or from below, respectively. The LIP-additive Asplund metric $d_{asp}^{\protect \LP}$ is defined by $d_{asp}^{\LP}(f,g) = c_1 \LM c_2$.
Importantly, this metric is theoretically invariant under lighting changes modelled by a LIP-addition of a constant:
$\forall k \in ]-\infty,M[$\>, $d_{asp}^{\LP}(f,b) = d_{asp}^{\LP}(f \LP k , b)$ and $d_{asp}^{\LP}(f,f \LP k) = 0$ \cite{Jourlin2016}.

The map of Asplund distances of a function $f$ of $\Fcurv_M$ by a probe $b : D_b \mapsto ]-\infty,M[$\>, where $D_b$  is a subset of $D$, is the mapping  $Asp_{b}^{\LP}: \Fcurv_M \mapsto \I$. Such a mapping is obtained by computing the distance between the function and the probe for each point $x$ of the function domain $D$. It is defined by $Asp_{b}^{\LP} f(x) = d_{asp}^{\LP} (f_{\left|D_b(x)\right.},b)$,
where $f_{\left|D_b(x)\right.}$ is the restriction of $f$ to the neighbourhood $D_b(x)$ centred on $x \in D$.
The map of Asplund distances $Asp_{b}^{\LP}$ which was related to MM in \cite{Noyel2017a,Noyel2019b,Noyel2019d}, is equal to 
\begin{align}
	Asp_{b}^{\LIPplus}(f) &= c_{1_{b}} (f) \LM c_{2_{b}} (f), \label{eq:map_As_c1_c2_add}%
\end{align}
where $c_{1_b}: \Fcurvb_M  \mapsto \Fcurvb_M$ is the \textit{map of the least upper bounds} (\textit{mlub}) and $c_{2_{b}}: \Fcurvb_M \mapsto \Fcurvb_M$ is the \textit{map of the greatest lower bounds} (\textit{mglb}). The \textit{mlub} $c_{1_b}$ is a dilation and the \textit{mglb} $c_{2_b}$ is an erosion which are both equal, for all $x \in D$, to:
\begin{align}
	c_{1_{b}} (f)(x) &=  \inf_{h \in D_b}{ \{c, f(x+h) \leq c \LIPplus b(h) \} }  \nonumber \\
	&= \vee \left\{ f(x+h) \LM b(h) , h \in D_b \right\},	\label{eq:upper_map_add_2}\\
	c_{2_{b}} (f)(x) &=  \sup_{h \in D_b}{ \{c, c \LIPplus b(h)	\leq f(x+h) \} }\nonumber \\
	&= \wedge \left\{ f(x+h) \LM b(h) , h \in D_b \right\}.\label{eq:lower_map_add_2}%
\end{align}


As shown in the next proposition, 
there is a strong link between the map of Asplund distances and LMM, as can be seen by comparing \eqref{eq:upper_map_add_2} with \eqref{eq:LIP-dilation} and \eqref{eq:lower_map_add_2} with \eqref{eq:LIP-erosion}.

\begin{proposition}	\label{prop:link_AsAdd_LMM}
	Let $f \in \Fcurvb_M$ be a function and $b \in \Fcurvb_M$ be a structuring function, where for all $x \in D_b$, $D_b \subset D$, $b(x) > -\infty$. The map of Asplund distances between the function $f$ and the structuring function $b$ is equal to:
\begin{align}
	Asp_{b}^{\LP} (f) &= \delta_{\LM \overline{b}}^{\LP} (f) \LM \varepsilon_b^{\LP}(f). \label{eq:map_AsAdd_LMM}%
\end{align}
	For the \textit{mlub} and the \textit{mglb} of $f$, $c_{1_b} (f)$ and $c_{2_b} (f)$, we have:
\begin{align}	
c_{1_b} (f) &= \delta_{\LM \overline{b}}^{\LP} (f),\label{eq:mlub_LMM}\\
c_{2_b} (f) &= \varepsilon_b^{\LP}(f). \label{eq:mglb_LMM}%
\end{align}
In the case of ambiguous expressions, the following conventions are used: $Asp_{b}^{\LP} (f)(x) = M$ when $\delta_{\LM \overline{b}}^{\LP} (f)(x) = M$ or $\varepsilon_b^{\LP}(f)(x) = -\infty$, and $Asp_{b}^{\LP} (f)(x) = 0$ when $\delta_{\LM \overline{b}}^{\LP} (f)(x) = \varepsilon_b^{\LP}(f)(x)$. 
\end{proposition}

\begin{proof}[Proof of proposition \ref{prop:link_AsAdd_LMM}]
From \eqref{eq:upper_map_add_2}, we have $\forall x \in D$:
\begin{align*}
	c_{1_{b}} (f)(x) &=  \vee \left\{ f(x+h) \LM b(h) , h \in D_b \right\} \\
	&=  \vee \left\{ f(x-h) \LM b(-h) , -h \in D_b \right\} \\
	&=  \vee \left\{ f(x-h) \LP (\LM \overline{b}(h)) , h \in D_{\overline{b}} \right\} \qquad \LIPPref{LIP:P5}\\
	&= \delta_{\LM \overline{b}}^{\LP} (f)(x).
\end{align*}

By comparing \eqref{eq:LIP-erosion} with \eqref{eq:lower_map_add_2}, we have for all $x \in D$:

\begin{equation*}
	c_{2_{b}}(f)(x) = \varepsilon_b^{\LP}(f)(x). \label{eq:mglb_Lero}
\end{equation*}
From \eqref{eq:map_As_c1_c2_add}, we deduce that $Asp_{b}^{\LIPplus}(f) =  \delta_{\LM \overline{b}}^{\LP} (f) \LM \varepsilon_b^{\LP}(f)$.
\end{proof}

\begin{figure}[!t]
\centering
\subfloat[Image and probe]{\includegraphics[width=0.49\mycolumnwidth]{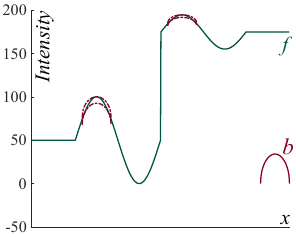}%
\label{fig:signal_mapAspAdd:sgn}}
\hfil
\subfloat[Maps]{\includegraphics[width=0.49\mycolumnwidth]{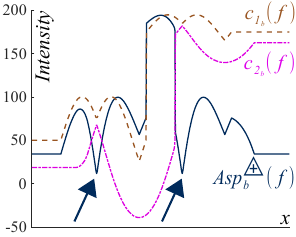}%
\label{fig:signal_mapAspAdd:map}}
\caption{In the LIP-greyscale, (a)~an image $f$ is analysed by a semi-circular probe $b$ with a diameter of 50 pixels from above and below. 
(b)~The \textit{mlub} $c_{1_b} (f)$, the \textit{mglb} $c_{2_b} (f)$ and the map of Asplund distances $Asp_b^{\protect \LP} (f)$ between the image and the probe. Both arrows point towards the minima of the map of Asplund distances.}
\label{fig:signal_mapAspAdd}
\end{figure}

As illustrated in \figurename \ref{fig:signal_mapAspAdd}, the map of Asplund distances consists of a double probing of a function $f$ by the same probe $b$ from above and from below. The \textit{mlub} $c_{1_b} (f)$ and the \textit{mglb} $c_{2_b} (f)$ correspond to a dilation $\delta_{\LM \overline{b}}^{\LP} (f)$ and an erosion $\varepsilon_b^{\LP}(f)$ of the image $f$, respectively, by their respective structuring functions $\LM \overline{b}$ or $b$ (\figurename \ref{fig:signal_mapAspAdd:map}). When the probe $b$ is similar to the image $f$ (according to the Asplund metric), the map of Asplund distances, $Asp_b^{\protect \LP} (f)$, of the image presents a minimum. In \figurename \ref{fig:signal_mapAspAdd:map}, both minima of the map of distances correspond to both bumps of the image $f$ (\figurename \ref{fig:signal_mapAspAdd:sgn}).
A link also exists between the double probing used in the map of Asplund distances and the logarithmic opening and closing, as shown in Appendix~\ref{app:link_double_prob_LIPopen_close}.
\label{link_double_prob_LIPopen_close}

\subsubsection{Link with the LIP-morphological gradient}
\label{sssec:RobOp:Aspl:properties}
Let $b_0 : D \mapsto \Fcurvb_M$ be a symmetric and constant structuring element which is defined for all $x \in D_{b_0}$, where $D_{b_0} \subset D$, by $b_0(x) = b_0$ and $b_0(-x) = b_0(x)$.
In the case of a symmetric and constant structuring element $b_0$, the map of LIP-additive Asplund distances $Asp_{b_0}^{\LP}$ is equal to the LIP-morphological gradient $\varrho_{B_0}^{LIP}$.  
For all $x \in D$, we have:
\begin{align}
	Asp_{b_0}^{\LP} (f)(x) &= \varrho_{B_0}^{LIP}(f)(x). \label{eq:link_masAsp_grad}
\end{align}
$B_0$ is a flat structuring element with the same domain $D_{b_0}$ as the one of the constant structuring element $b_0$.

\begin{proof}[Proof of equation \eqref{eq:link_masAsp_grad}]
According to the previous definition of $b_0$, we have therefore $b_0 = \overline{b}_0$. For all $x \in D$, we have:
\begin{align*}
	Asp_{b_0}^{\LP} (f)(x) &=  \delta_{\LM \overline{b}_0}^{\LP} (f)(x) \LM \varepsilon_{b_0}^{\LP}(f)(x)  \nonumber\\
	&=  \vee_{h \in D_{\overline{b}_0}} 	\left\{ f(x - h) \LP (\LM \overline{b}_0) \right\}  
	\LM \wedge_{h \in D_{b_0}} \left\{ f(x + h) \LM b_0 \right\} \nonumber\\
	&=  \left( \vee_{h \in D_{b_0}} 	\left\{ f(x - h) \right\} \LP (\LM b_0)  \right) 
	\LM \left( \wedge_{h \in D_{b_0}} \left\{ f(x + h) \right\} \LP (\LM b_0)  \right)     \qquad \LIPPref{LIP:P5} \nonumber\\
	&=  \vee_{h \in D_{b_0}} 	\left\{ f(x - h) \right\} \LM \wedge_{h \in D_{b_0}} \left\{ f(x + h) \right\}  \qquad \LIPPref{LIP:P7}\nonumber\\
	&= \delta_{B_0} (f)(x) \LM \varepsilon_{B_0}(f)(x)  \nonumber\\
	&= \varrho_{B_0}^{LIP}(f)(x). 
\end{align*}
\end{proof}



When the structuring function $b : D_b \mapsto \Fcurvb_M$ is non flat, the map of LIP-additive Asplund distances $Asp_{b}^{\LP}$ is an extension of the \textit{LIP-morphological gradient} $\varrho_{B}^{LIP}$. 
However, contrary to the morphological gradient, the map of Asplund distances is no more the norm of a gradient. For example, let $Y \subset D$ be a constant (i.e., a flat) zone of a function $f : D \mapsto \Fcurvb_M$ and let $X  = Y \ominus D_b$ be the eroded flat zone by the domain $D_b$ of the structuring function $b$. $\ominus$ represents the binary erosion \cite{Serra1982,Serra1988}. In the eroded flat zone $X$, the map of Asplund distances is equal to a constant, whereas a gradient and its norm should be equal to zero.
We have, for all $x \in X$:
\begin{align}
	Asp_{b}^{\LP} (f)(x) &= b_{sup} \LM b_{inf}. \label{eq:mapAsp_flatzone}
\end{align}
$b_{sup}$ and $b_{inf}$ are the supremum and the infimum, respectively, of the structuring function $b$.
\begin{proof}[Proof of equation \eqref{eq:mapAsp_flatzone}]\label{proof:eq:mapAsp_flatzone}
We have, for all $x \in X$:
\begin{align*}
	Asp_{b}^{\LP} (f)(x) &=  c_{1_{b}} f(x) \LM c_{2_{b}} f(x) \nonumber\\
	&= \vee_{h \in D_b} \left\{ f(x+h) \LM b(h) \right\} 
	\LM \wedge_{h \in D_b} \left\{ f(x+h) \LM b(h) \right\} \nonumber\\
	&= \vee_{h \in D_b} \left\{ f(x) \LM b(h) \right\} 
	\LM \wedge_{h \in D_b} \left\{ f(x) \LM b(h) \right\} \nonumber\\
	&= \left[ f(x) \LM \wedge_{h \in D_b} \left\{ b(h) \right\} \right]
	\LM \left[ f(x) \LM \vee_{h \in D_b} \left\{ b(h) \right\} \right]  \nonumber\\
		&= \LM \wedge_{h \in D_b} \left\{ b(h) \right\} \LP \vee_{h \in D_b} \left\{ b(h) \right\}  \qquad \LIPPref{LIP:P2}, \LIPPref{LIP:P7}, \LIPPref{LIP:P1}\nonumber\\
		&= \vee_{h \in D_b} \left\{ b(h) \right\} \LM \wedge_{h \in D_b} \left\{ b(h) \right\}  \qquad \LIPPref{LIP:P2}\nonumber\\
		&= b_{sup} \LM b_{inf}. 
\end{align*}
In the eroded flat zone $X$, the Asplund metric is therefore equal to the constant $b_{sup} \LM b_{inf}$.
\end{proof}

\begin{figure}[!t]
\centering
\subfloat{\includegraphics[width=0.65\mycolumnwidth]{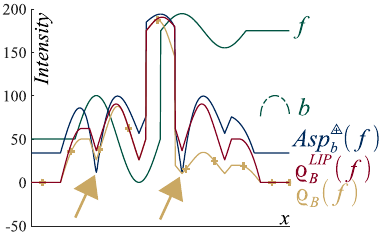}%
\label{fig:signal_comp_mapAspAdd_morphograd:sgn}}
\caption{In the LIP-greyscale, comparison between the morphological gradient $\varrho_B(f)$, the LIP-morphological gradient $\varrho^{{\protect LIP}}_B(f)$ and the map of Asplund distances $Asp_b^{{\protect \LP}}(f)$ of the image $f$. The flat structuring element $B$ has the same domain $D_B$ as the one of the structuring function $b$ in \figurename \ref{fig:signal_mapAspAdd:sgn}, which is a semi-circle with a diameter of 69 pixels. Both arrows point towards the regional minima of the gradients which correspond to the minima on the map of Asplund distances.}
\label{fig:signal_comp_mapAspAdd_morphograd}
\end{figure}

In \figurename \ref{fig:signal_comp_mapAspAdd_morphograd}, the morphological gradient $\varrho_B(f)$ of the image $f$ and its LIP-morphological gradient $\varrho^{{\protect LIP}}_B(f)$, both with a flat structuring element $B$, are compared to the map of Asplund distances $Asp_b^{{\protect \LP}}(f)$  with a structuring function $b$. This structuring function has a bump shape which was designed to detect the bumps of the image $f$. For both image bumps, the map of Asplund distances $Asp_b^{\LP}(f)$ presents two deep minima with the same dynamic range, whereas the LIP-morphological gradient $\varrho_B^{LIP}(f)$ and the morphological gradient $\varrho_B(f)$ have two regional minima which have a lower dynamic range. In addition, the regional minima of the morphological gradient $\varrho_B(f)$ have not the same depth between each other. For the flat zones of the image $f$, both gradients $\varrho_B(f)$ and $\varrho_B^{LIP}(f)$ are equal to zero, whereas the map of Asplund distances $Asp_b^{{\protect \LP}}(f)$ is equal to a positive constant defined by \eqref{eq:mapAsp_flatzone}.

The map of LIP-additive Asplund distances $Asp_b^{\LP}$ is therefore the extension of the LIP-morphological gradient $\varrho_B^{LIP}$ for non flat structuring functions. It gives to this gradient the properties of a metric which is robust to lighting variations. The LIP-morphological gradient is thereby a double probing of an image by a flat structuring element.


\subsubsection{Map of Asplund distances with a tolerance to extrema}
\label{sssec:RobOp:Aspl:AspTol}

In the case of discrete images, a map of LIP-additive Asplund distances with a tolerance (to extrema) can be defined as in \cite{Noyel2019b} and related to Mathematical Morphology. The relation is similar to the one in \eqref{eq:map_As_c1_c2_add}, with a \textit{map of the least upper bounds} (\textit{mlub}) and \textit{map of the greatest lower bounds} (\textit{mglb}).

The $k^{th}$ LIP-minimum-rank filter $\zeta_{b,k}^{\LP}: \Fcurvb_M^D \mapsto \Fcurvb_M^D$ and the $k^{th}$ LIP-maximum-rank filter $\vartheta_{b,k}^{\LP}: \Fcurvb_M^D \mapsto \Fcurvb_M^D$ are defined in \eqref{eq:_LIP_erotol} and in \eqref{eq:_LIP_diltol}, respectively. The proof of proposition \ref{prop:link_AsAddtol_LMM} is in Appendix \ref{app:prop:link_AsAddtol_LMM}.
\begin{proposition}\label{prop:link_AsAddtol_LMM}
	Let $f \in \Fcurvb_M$ be a function defined on a discrete grid, e.g. $D\subset \Zint^n$. Let $b \in \Fcurvb_M$ be a structuring function, where for all $x \in D_b$, $D_b \subset D$, $b(x) > -\infty$. Let $(1-p)$ be a percentage of points of $D_b$ to be discarded. The map of LIP-additive Asplund distances with a tolerance $p$ between the function $f$ and the structuring function $b$ is equal to:
\begin{align}
	Asp_{b,p}^{\LP} (f) &= \vartheta_{\LM \overline{b},n_1}^{\LP} (f) \LM \zeta_{b,n_2}^{\LP}(f). \label{eq:map_AsAddtol_LMM}%
\end{align}
The number of points to be suppressed, $n_1$ and $n_2$, for the \textup{mlub} $\vartheta_{\LM \overline{b},n_1}^{\LP}$ and for the \textup{mglb} $\zeta_{b,n_2}^{\LP}$ are equal to
$n_1 = round( n_{suppr} /2)$ and $n_2 = n_{suppr} - n_1$, respectively, where $n_{suppr} = round[(1-p)\#D_b]$ and $\#D_b$ is the cardinal of $D_b$. For the \textup{mlub}, $c_{1_b,p} (f)$, and the \textup{mglb} of $f$, $c_{2_b,p} (f)$, we have:
\begin{align}	
c_{1_b,p} (f) &= \vartheta_{\LM \overline{b},n_1}^{\LP} (f),\label{eq:mlub_AsAddtol_LMM}\\
c_{2_b,p} (f) &= \zeta_{b,n_2}^{\LP}(f). \label{eq:mglb_AsAddtol_LMM}%
\end{align}
\end{proposition}

\begin{figure}[!t]
\centering
\subfloat[Image, mlub, mglb]{\includegraphics[width=0.49\mycolumnwidth]{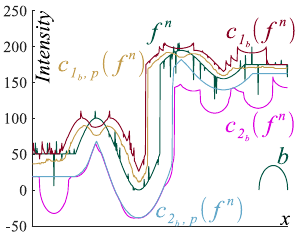}%
\label{fig:signal_mapAspAdd_noised_im:1}}
\hfil
\subfloat[Map of distances]{\includegraphics[width=0.49\mycolumnwidth]{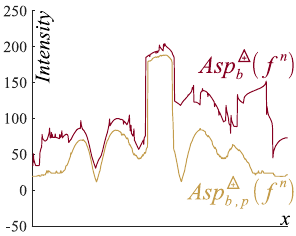}%
\label{fig:signal_mapAspAdd_noised_im:2}}
\caption{(a) $f^n$ is an image with a Gaussian white noise, with a standard deviation of \num{20} grey levels and a density of \num{0.08}. 
The mlub $c_{1_b}(f^n)$ and the mglb $c_{2_b}(f^n)$ (without tolerance) are compared to the mlub $c_{1_b,p}(f^n)$ and the mglb $c_{2_b,p}(f^n)$ with a tolerance of $p= \SI{85}{\percent}$. $b$ is a semi-circular probe with a diameter of 69 pixels.
(b) The map of Asplund distances $Asp^{\protect \LP}_{b}(f^n)$ (without tolerance) is compared to the map of Asplund distances with tolerance $Asp^{\protect \LP}_{b,p}(f^n)$. The LIP-greyscale is used to represent grey-levels.}
\label{fig:signal_mapAspAdd_noised_im}
\end{figure}

In \figurename~\ref{fig:signal_mapAspAdd_noised_im:1}, a Gaussian white noise is added to the image $f$ of \figurename~\ref{fig:signal_mapAspAdd:sgn} in order to obtain a noised image $f^n$. The mlub $c_{1_b}(f^n)$ and the mglb $c_{2_b}(f^n)$ of $f^n$ (without tolerance) are compared to the mlub $c_{1_b,p}(f^n)$ and the mglb $c_{2_b,p}(f^n)$ of $f^n$ with a tolerance $p$. One can notice that these latter are less sensitive to the local extrema caused by peaks of noise. In \figurename~\ref{fig:signal_mapAspAdd_noised_im:2}, a similar observation can be made between the map of Asplund distances $Asp^{\protect \LP}_{b}(f^n)$ (without tolerance) and the map of Asplund distances of $f^n$ with tolerance $Asp^{\protect \LP}_{b,p}(f^n)$. This latter map has a similar shape as the map of Asplund distances $Asp^{\protect \LP}_{b}(f)$ of the image $f$ without noise (\figurename~\ref{fig:signal_mapAspAdd:map}).

\subsection{A novel operator: the LIP-difference between LIP-erosions}
\label{ssec:RobOp:LipDiff_LipEro}

\begin{figure}[!t]
\centering
\subfloat[Bump and probe]{\includegraphics[width=0.49\mycolumnwidth]{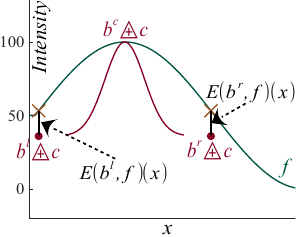}%
\label{fig:signal_LdiffEro:bump}}
\hfil
\subfloat[Inflection point neighbourhood and probe]{\includegraphics[width=0.49\mycolumnwidth]{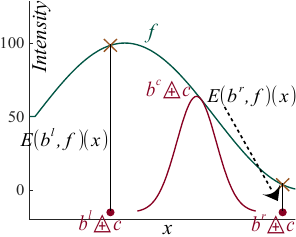}%
\label{fig:signal_LdiffEro:trans}}
\\
\subfloat[Detector]{\includegraphics[width=0.49\mycolumnwidth]{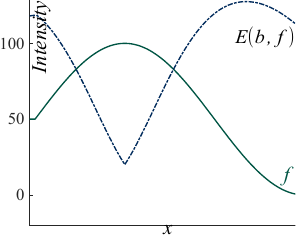}%
\label{fig:signal_LdiffEro:detect}}
\caption{In the LIP-greyscale, an image $f$ is analysed by a probe $b$ from below. The objective is to detect (a) a bump but not (b) an inflection point and its neighbourhood between two extrema. 
After the LIP-addition of a constant $c$ to the probe, $b \protect \LP c$, the left $E(b^l,f)(x)$ and right detectors $E(b^r,f)(x)$ are the LIP-differences between $f$ and the left element $b^l$ of the probe $b$ and its right element $b^r$, respectively. 
(c)~Bump detector $E(b,f)$.}
\label{fig:signal_LdiffEro}
\end{figure}

In \figurename \ref{fig:signal_LdiffEro} representing a unidimensional image $f$, a probe $b$ was designed to detect a bump (\figurename \ref{fig:signal_LdiffEro:bump}) but not an inflection point and its neighbourhood between two local extrema (\figurename \ref{fig:signal_LdiffEro:trans}). The probe $b$ is composed of three elements: 
\begin{inparaenum}[(i)]
\item the left point $b^l$, 
\item the right point $b^r$ and 
\item a central bump $b^c(x)=a \exp{\left(\frac{-x^2}{2\sigma^2}\right)} + h$,  if $\left|x\right|\leq 3\sigma$, with approximately the same dynamic range as the bump to be detected but with a smaller width.
\end{inparaenum}
Both points $b^l$ and $b^r$ are located at a radius $r=48$ pixels from the probe origin and a height of $h_p=50$ grey levels. The central bump has the following parameters: an amplitude $a=60$ grey levels, a standard deviation $\sigma=11$ pixels, a minimal height $h=50$ grey levels. 
For each point $x$ of the domain $D$, the probe is set in contact with the image $f$ from below by LIP-adding a constant $c$. This latter one is equal to the value of the \textit{mglb}, $c_{2_b}f(x)$, which has been defined in \eqref{eq:lower_map_add_2}. The LIP-difference 
is computed between the image $f$ and the left and right points, $b^l$ and $b^r$, of the probe, $b\LP c_{2_b}f(x)$, which is in contact with the image $f$. The left and right detectors, $E (b^l, f)$ and $E (b^r, f) : D \mapsto \left]-\infty,M\right[$\>, are defined as follows: 
\begin{align}
	E (b^l, f)(x) &= \wedge_{h \in D_b}{ \{ f(x+h) \LM [ b^l(h) \LP c_{2_b}(f)(x) ] \} } \nonumber\\
	&= \wedge_{h \in D_b}{ \{ f(x+h) \LM b^l(h) \} } \LM c_{2_b}(f)(x), \quad \LIPPref{LIP:P4} \label{eq:LAC_er_left}\\
	E (b^r, f)(x) &= \wedge_{h \in D_b}{ \{ f(x+h) \LM [ b^r(h) \LP c_{2_b}(f)(x) ] \} } \nonumber\\
	&= \wedge_{h \in D_b}{ \{ f(x+h) \LM b^r(h) \} } \LM c_{2_b}(f)(x). \quad \LIPPref{LIP:P4} \label{eq:LAC_er_right}
\end{align}
In the event of a bump similar to the probe, the left and right detectors, $E (b^l, f)$ and $E (b^r, f)$, will have close values (\figurename~\ref{fig:signal_LdiffEro:bump}), whereas in the event of an inflection point, one of the detectors will have a value much higher than the other one (\figurename~\ref{fig:signal_LdiffEro:trans}). Such a property allows to separate bumps (which are similar to the probe) from the inflection points and their neighbourhoods between two extrema. The bump detector $E (b, f) : D \mapsto \left]-\infty,M\right[$ is therefore defined as the point-wise supremum $\bigvee$ between the left and right detectors:
\begin{align}
	E (b,f) &= \bigvee{\left\{ E (b^l, f) , E (b^r, f) \right\}}. \label{eq:LIP:LMM:detector_one_dir}
\end{align}
As illustrated in \figurename~\ref{fig:signal_LdiffEro:detect}, in the event of a bump, the detector presents a deep minimum, whereas in the event of an inflection point and its neighbourhood, this minimum disappears.
 Both the left and right detectors are related to LMM via the following two properties, the proofs of which can be found in Appendix~\ref{app:property:lr_detectors_LMM_property:lr_detectors_rob_LIPadd}.

\begin{property} 
\label{property:lr_detectors_LMM}
The left and right detectors, $E (b^l, f)$ and $E (b^r, f)$, are equal to LIP-differences between logarithmic-erosions: 
\begin{align}
	E (b^l, f) &= \varepsilon_{b^l}^{\LP}(f) \LM \varepsilon_b^{\LP}(f), \label{eq:LIP:LMM:LAC_er_left2}\\
	E (b^r, f) &= \varepsilon_{b^r}^{\LP}(f) \LM \varepsilon_b^{\LP}(f). \label{eq:LIP:LMM:LAC_er_right2}
\end{align}
\end{property}

\begin{property} 
\label{property:lr_detectors_rob_LIPadd}
The left and right detectors, $E (b^l, f)$ and $E (b^r, f)$, and the bump detector, $E (b, f)$, are insensitive to the LIP-addition (or the LIP-subtraction) of any constant $c \in \left]-\infty,M\right[$ to (or from) the image $f$:
\begin{align}
	E (b^l, f \LP c) &= E (b^l, f), \nonumber\\
	E (b^r, f \LP c) &= E (b^r, f), \nonumber\\
	E (b, f \LP c) 	 &= E (b, f). \nonumber
\end{align}
\end{property}

\begin{figure}[!t]
\centering
\subfloat[Image and detector]{\includegraphics[width=0.49\mycolumnwidth]{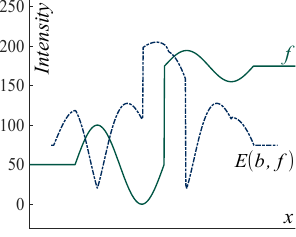}%
\label{fig:signal_LdiffEro_full:1}}
\hfil
\subfloat[Image and probe at two places]{\includegraphics[width=0.49\mycolumnwidth]{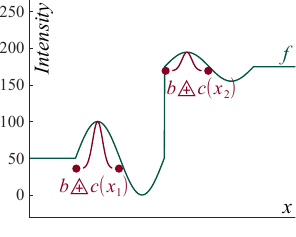}%
\label{fig:signal_LdiffEro_full:2}}
\caption{In the LIP-greyscale, (a)~unidimensional image $f$ and bump detector $E(b,f)$.
(b) Image and probe $b$ for two points $x_1, x_2 \in D$. The probe $b$ is the same as the one of \figurename~\ref{fig:signal_LdiffEro}.}
\label{fig:LIP:LMM:signal_LdiffEro_full}
\end{figure}

\figurename~\ref{fig:signal_LdiffEro_full:1} illustrates the application of the bump detector to a unidimensional image $f$. The probe $b$ is the same as the one of \figurename~\ref{fig:signal_LdiffEro}. The detector presents two minima  of the same depth, one for each bump of the image. In the image $f$, the bump amplitudes are related by the LIP-addition of a constant $c$ which models a change in the image intensity caused by a variation of the light intensity or of the exposure time of the camera. The bump detector $E(b,f)$ is therefore robust to lighting variations modelled by the LIP-addition of a constant. Due to the LIP-addition of the image \textit{mglb}, $c_{2_b}(f)$, to the probe $b$, the translated probe $b \LP c_{2_b}(f)$ has an amplitude which changes according to the image intensity (\figurename~\ref{fig:signal_LdiffEro_full:2}).

\subsection{Other operators: LIP-differences between LIP-morphological operations}
\label{ssec:RobOp:LipDiff_LipOp}

\begin{figure}[hbt!]
\centering
\subfloat[Input image (LIP-greyscale)]{\includegraphics[width=0.28\mycolumnwidth]{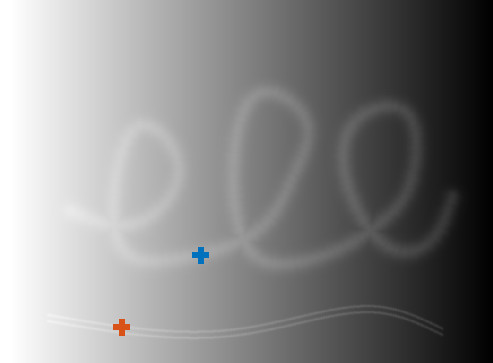}%
\label{fig:im_spiral_probe:1}}
\\
\subfloat[Probe $b$ and image surfaces at the red point]{\includegraphics[width=0.28\mycolumnwidth]{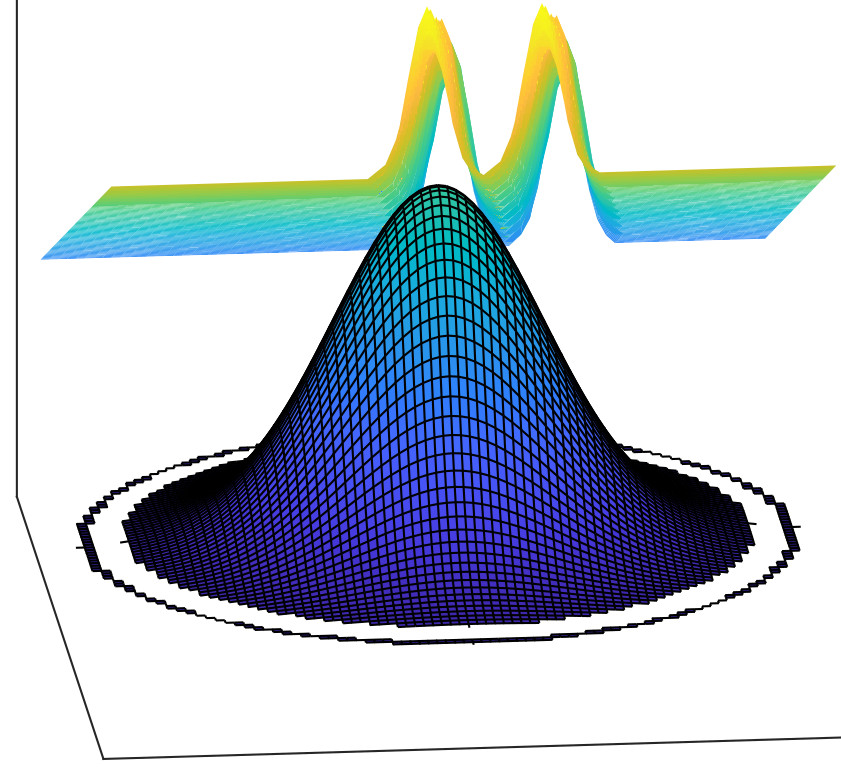}%
\label{fig:im_spiral_probe:2}}
\hfil
\subfloat[Probe $b$ and image surfaces at the blue point]{\includegraphics[width=0.28\mycolumnwidth]{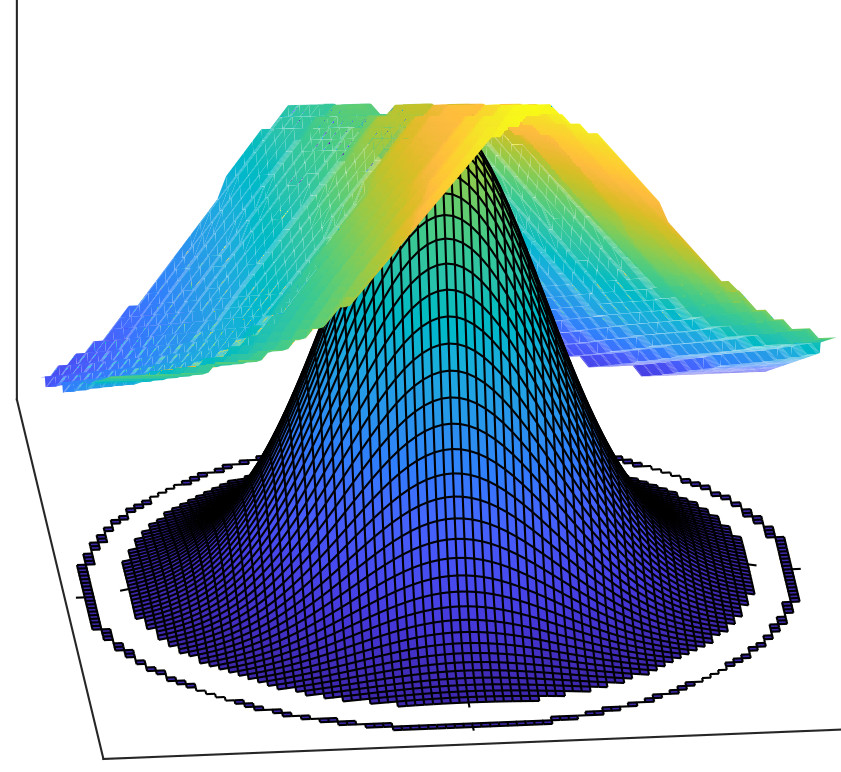}%
\label{fig:im_spiral_probe:3}}
\hfil
\subfloat[LIP-opening by the probe $b$]{\includegraphics[width=0.28\mycolumnwidth]{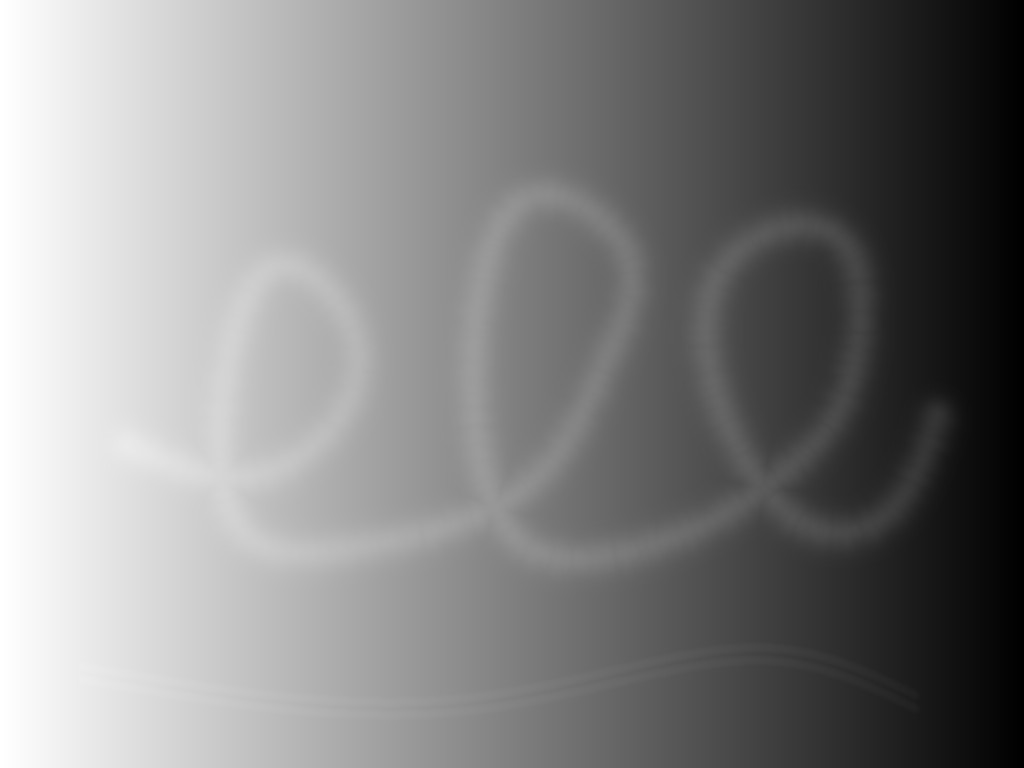}%
\label{fig:im_spiral_probe:4}}
\\
\subfloat[Probe $b_r$ and image surfaces at the red point]{\includegraphics[width=0.28\mycolumnwidth]{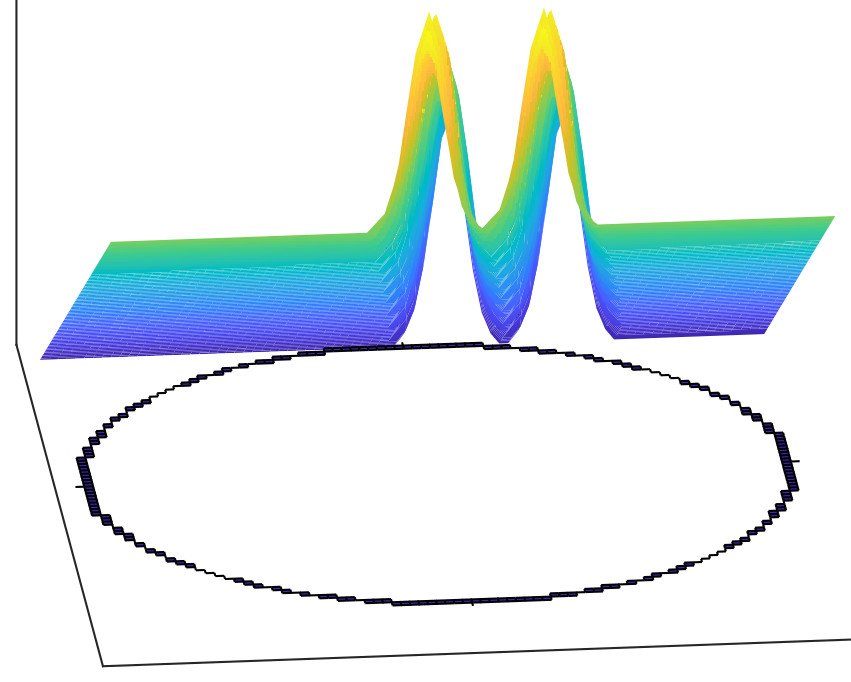}%
\label{fig:im_spiral_probe:5}}
\hfil
\subfloat[Probe $b_r$ and image surfaces at the blue point]{\includegraphics[width=0.28\mycolumnwidth]{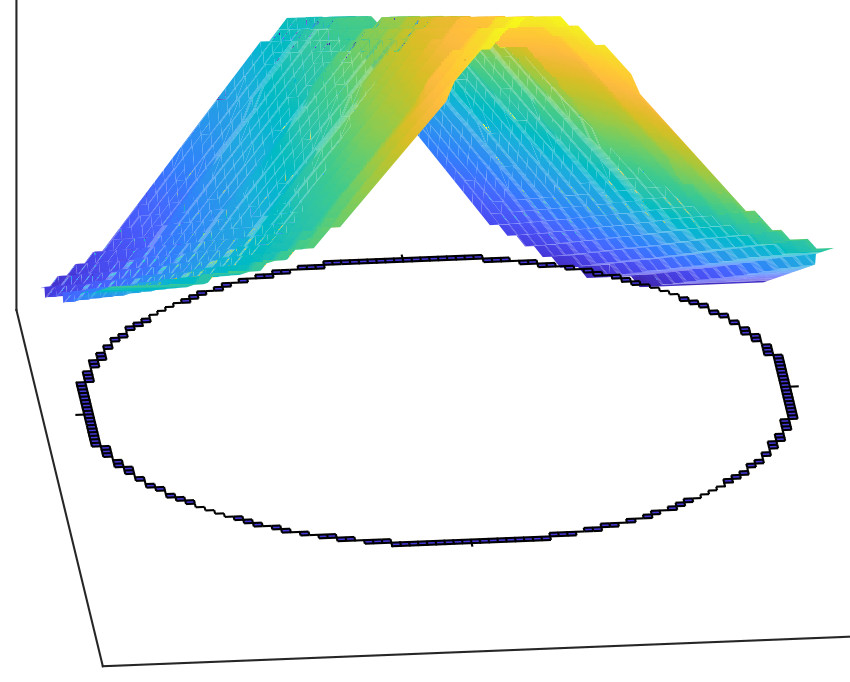}%
\label{fig:im_spiral_probe:6}}
\hfil
\subfloat[LIP-opening by the probe $b_r$]{\includegraphics[width=0.28\mycolumnwidth]{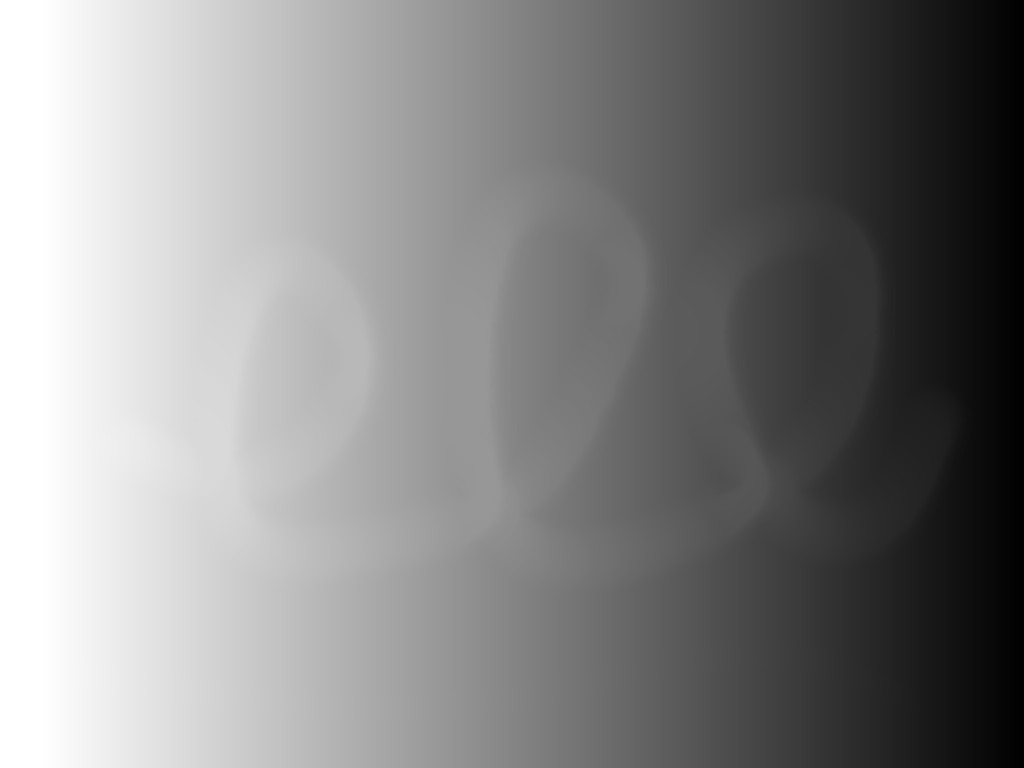}%
\label{fig:im_spiral_probe:7}}\\
\subfloat[LMM $G_b^{\protect \LP}(f)$]{\includegraphics[width=0.45\mycolumnwidth]{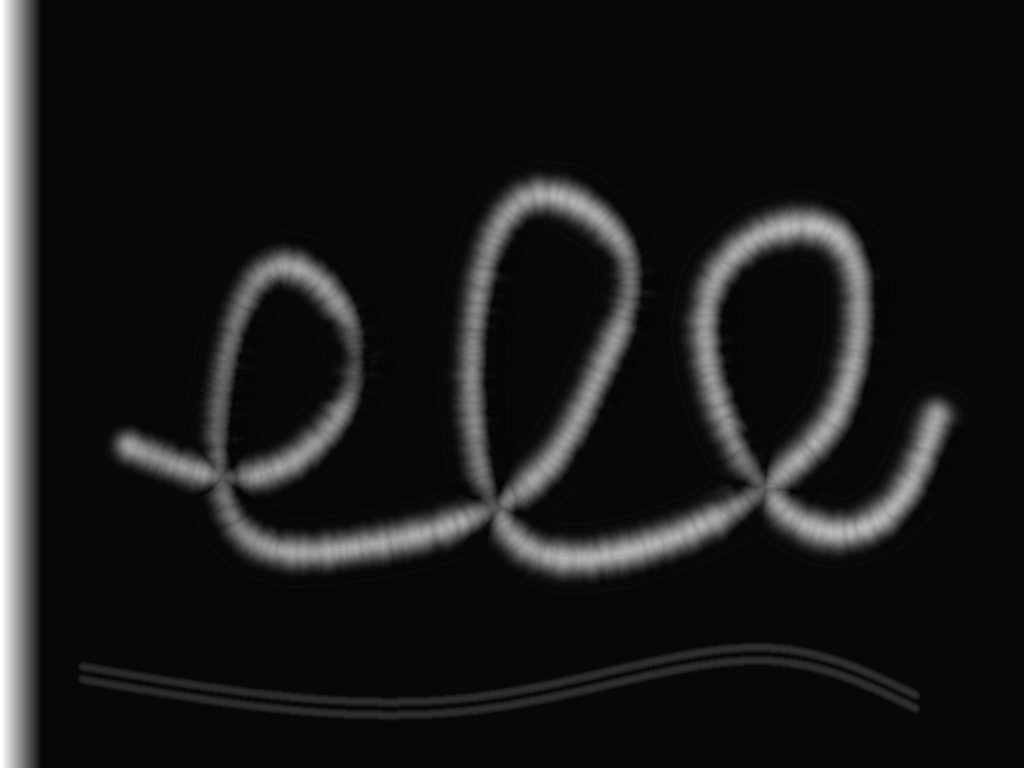}%
\label{fig:im_spiral_probe:8}}
\hfil
\subfloat[MM $G_b (f)$]{\includegraphics[width=0.45\mycolumnwidth]{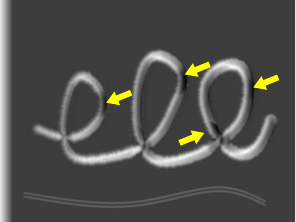}%
\label{fig:im_spiral_probe:9}}

\caption{(a) Selection of two points in the image: the red point is in the curves and the blue point is in the spiral.  The image of \figurename~\ref{fig_intro:input} is shown in the LIP-greyscale.
Intensities of the probe $b$ and the image $f$ represented as a surface: (b) in the curves, for the red point, and (c)  in the spiral, for the blue point.
(d) LIP-opening $\gamma_b^{\protect \LP}(f)$ of the image $f$ by the probe $b$ composed of a Gaussian and a ring.
Surfaces of the probe $b_r$ and the image $f$: (e) in the curves, for the red point, and (f) in the spiral, for the blue point.
(g) LIP-opening $\gamma_{b_r}^{\protect \LP}(f)$ of the image $f$ by the probe $b_r$ composed of a ring.
Results (h) of the LMM operator $G_b^{\protect \LP}(f)$ and
(i) of the classical MM operator $G_b (f) = \gamma_b (f) - \gamma_{b_r} (f)$. The yellow arrows indicate the contrast changes caused by the lighting drift.}
\label{fig:im_spiral_probe}
\end{figure}

In the same way as in section~\ref{ssec:RobOp:LipDiff_LipEro}, the LIP-difference $\LM$ between two operations of LMM can be robust to lighting variations. For example, \figurename~\ref{fig_intro:Ldiff_Lopen} illustrates the robustness to a lighting drift, of the operator $G^{\LP}_b$ defined by: 
\begin{align}
	G^{\LP}_b (f) &= \gamma_{b}^{\LP}(f) \LM \gamma_{b_r}^{\LP}(f). \label{eq:LIPdiff_LIPopenings}
\end{align}
It is equal to the LIP-difference between two logarithmic openings $\gamma_{b}^{\LP}$ and $\gamma_{b_r}^{\LP}$ by two different probes $b$ and  $b_r$. $b$ is a Gaussian-shape probe $g$ surrounded by a ring (\figurename~\ref{fig_intro:probe}) and $b_r$ is a ring-shape probe. 
In both probes $b$ and $b_r$, the ring is composed of the points lying in the interval $\left[39,41\right]$ pixels with an altitude of $h=4.1$ grey levels in the usual grey-scale. The Gaussian-shape probe is 
defined by $g(x)=a \exp{\left(\frac{-\left\|x\right\|_2^2}{2\sigma^2}\right)} + h$,  if $\left\|x\right\|_2\leq 3\sigma$, where $x \in \mathbb{Z}^2$. It has the following parameters: an amplitude $a=-0.8$ grey levels, a standard deviation $\sigma=12$ pixels, a height $h=4.1$ grey levels in the usual grey-scale. 
In \figurename~\ref{fig_intro:probe}, \figurename~\ref{fig:im_spiral_probe:2}, \figurename~\ref{fig:im_spiral_probe:3}, \figurename~\ref{fig:im_spiral_probe:5} and \figurename~\ref{fig:im_spiral_probe:6}, the probes $b$ and $b_r$ are represented in the LIP-greyscale by using the following transform $b_{LIP} = (M-1) - b$, where $M=256$.

The operator $G^{\LP}$ possesses the following property, the proof of which is in Appendix~\ref{app:property:LIPdiff_LIPopenings_rob_LIPadd_property:LIPdiff_LIPopenings_rob_LIPadd}.

\begin{property}
The operator $G^{\LP}_b : D \mapsto \left]-\infty,M\right[$ is insensitive to the LIP-addition of any constant $c \in \left]-\infty,M\right[$ to the function $f : D \mapsto \left]-\infty,M\right[\>$: 
\begin{equation*}
	G^{\LP}_b (f \LP c) = G^{\LP}_b (f). \label{eq:LIPdiff_LIPopenings_rob_LIPadd}
\end{equation*}
\label{property:LIPdiff_LIPopenings_rob_LIPadd}
\end{property}

As the openings are performed by using the probe $b$ defined on a local domain $D_b \subset D$, the operator $G^{\LP}_b$ is locally (and globally) insensitive to the addition of any constant in this domain $D_b$. Such a local domain $D_b$ corresponds to a sliding window around any pixel $x$ of the image domain $D$. In \figurename~\ref{fig_intro:input}, the image $f$  presents a lighting drift caused by the LIP-addition of a linear function (i.e. a plane). As a plane (and several other lighting drifts) can be approximated by a piecewise constant function, the operator $G^{\LP}_b$ is robust to such a lighting drift. In addition, the Gaussian shape of the probe allows to detect the spiral without detecting both close curves. 
\begin{inparaenum}[(i)]
\item Firstly, the width of the Gaussian at its top is larger than the width of each one of both curves, but it is smaller than the width of the spiral. When probing the image $f$, as the probe $b$  is composed of a Gaussian and a ring, it cannot enter into the curves (\figurename~\ref{fig:im_spiral_probe:2}), but it can enter into the spiral (\figurename~\ref{fig:im_spiral_probe:3}). As a consequence, the logarithmic opening $\gamma_{b}^{\LP}(f)$ of $f$ by the Gaussian and ring probe $b$ almost completely removes both curves but it keeps the spiral (\figurename~\ref{fig:im_spiral_probe:4}).
\item Secondly, the ring probe $b_r$ is larger than the widths of the spiral and of both curves. This prevents it from entering into them (\figurename~\ref{fig:im_spiral_probe:5} and ~\ref{fig:im_spiral_probe:6}). The logarithmic opening $\gamma_{b_r}^{\LP}(f)$ of $f$ by the ring $b_r$ strongly attenuates the spiral and removes both curves (\figurename~\ref{fig:im_spiral_probe:7}). 
\end{inparaenum}
In the resulting image $G_b^{\LP} (f)$ (\figurename~\ref{fig_intro:Ldiff_Lopen} or \ref{fig:im_spiral_probe:8}), the LMM operator $G_b^{\LP}$ extracts the spiral without the lighting drift and strongly attenuates the confounding curves. This image is obtained by the LIP-difference between both openings, $\gamma_{b}^{\LP}(f)$ (\figurename~\ref{fig:im_spiral_probe:4}) and $\gamma_{b_r}^{\LP}(f)$ (\figurename~\ref{fig:im_spiral_probe:7}), of the image $f$ by the probes $b$ and $b_r$, respectively.
However, the classical Mathematical Morphology operator $G_b (f) = \gamma_b (f) - \gamma_{b_r} (f)$ defined as the difference between the openings $\gamma_b$ and $\gamma_{b_r}$ (\figurename~\ref{fig:im_spiral_probe:9}), does not extract the spiral without keeping contrast changes caused by the lighting drift. A detection robust to lighting drifts -- with a physical cause and modelled by the LIP-addition law -- is therefore not possible by using classical MM operators, whereas it is possible by using LMM operators.

Other operators robust to lighting variations which are modelled by the LIP-additive law, can be defined as the LIP-difference between LMM operations, e.g.: the LIP-difference between logarithmic-closings or the LIP-difference between an image and its logarithmic-opening (see \eqref{eq:LIP:LMM:res_LIPopening}).


\subsection{Extensions of top-hat operators}
\label{ssec:RobOp:Extensions_top_hats}

Let us focus on the difference, or residue, of an image $f$ by its opening $\gamma_b(f)$ or its logarithmic-opening $\gamma^{\LP}_b(f)$. Both operators $R_b$ and $R^{\LP}_b$ are defined as follows:
\begin{align}
	R_b(f) &= f - \gamma_b(f), \label{eq:LIP:LMM:res_opening}\\
	R^{\LP}_b(f) &= f \LM \gamma^{\LP}_b(f). \label{eq:LIP:LMM:res_LIPopening}
\end{align}
When $b : D \mapsto \left]-\infty , M \right[$ is  a flat structuring element denoted by $b = B$, the operators $R_b$ and $R^{\LP}_b$ correspond to the top-hat operators $TH_B(f) = f - \gamma_B(f)$ and $TH^{\LP}_b(f) = f \LM \gamma_B(f)$, respectively. These top-hat operators are presented in section~\ref{sssec:rel_wrk:top_hat}. When $b$ is a non-flat structuring function, both operators $R_b$  and $R^{\LP}_b$ constitute extensions of the top-hat operators.
The operator $R_b$ is named the \textit{extended top-hat} and the operator $R^{\LP}_b$ the \textit{extended LIP-top-hat}.
However, only this latter operator $R^{\LP}_b$ possesses the following property, the proof of which is in Appendix~\ref{app:property:LIPresidue_LIPopening_rob_LIPadd}.
\begin{property}
The extended LIP-top-hat, $R^{\LP}_b$, is insensitive to the LIP-addition of any constant $c \in \left]-\infty,M\right[$ to the function $f : D \mapsto \left]-\infty,M\right[\>$: $R^{\LP}_b(f \LP c) = R^{\LP}_b(f)$.
\label{property:LIPresidue_LIPopening_rob_LIPadd}
\end{property}

\FloatBarrier

\begin{figure}[!thp]
\begin{center}
\subfloat[Bright image $\mathbf{f}$ ({\protect \qty[parse-numbers=false,per-mode = symbol]{1/40}{\second}})]{\includegraphics[width=0.3\mycolumnwidth]{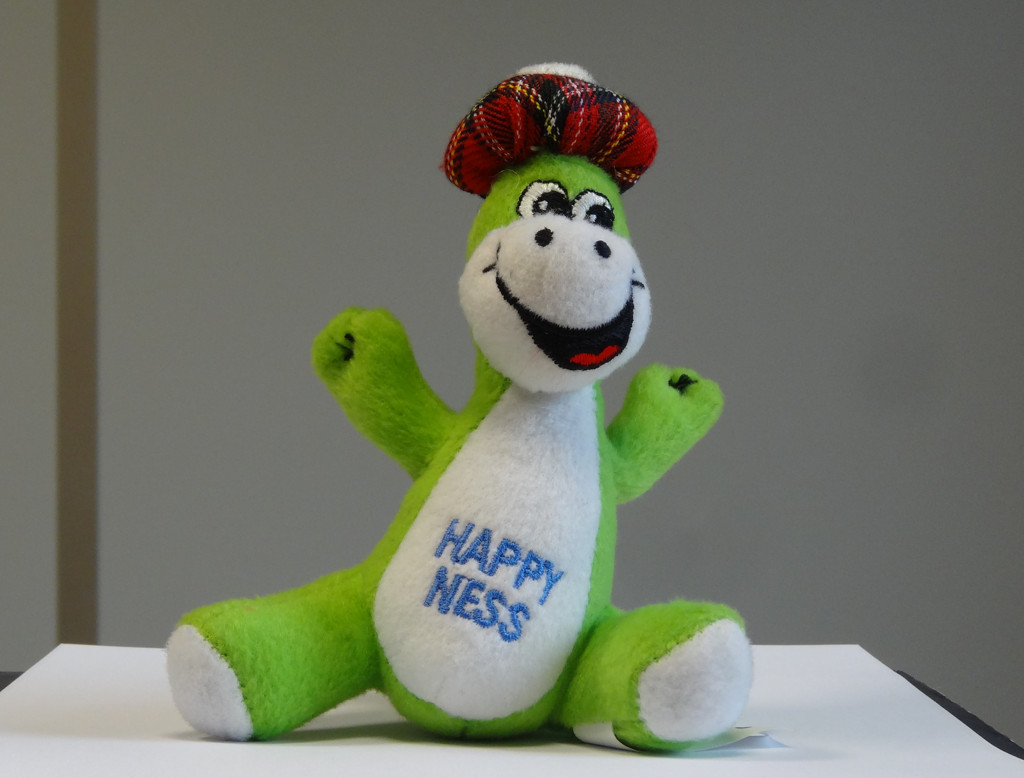}%
\label{fig:exptime:im_monstre:1}}
\hfil%
\subfloat[Dark image $\mathbf{f}^d$ ({\protect \qty[parse-numbers=false,per-mode = symbol]{1/800}{\second}})]{\includegraphics[width=0.3\mycolumnwidth]{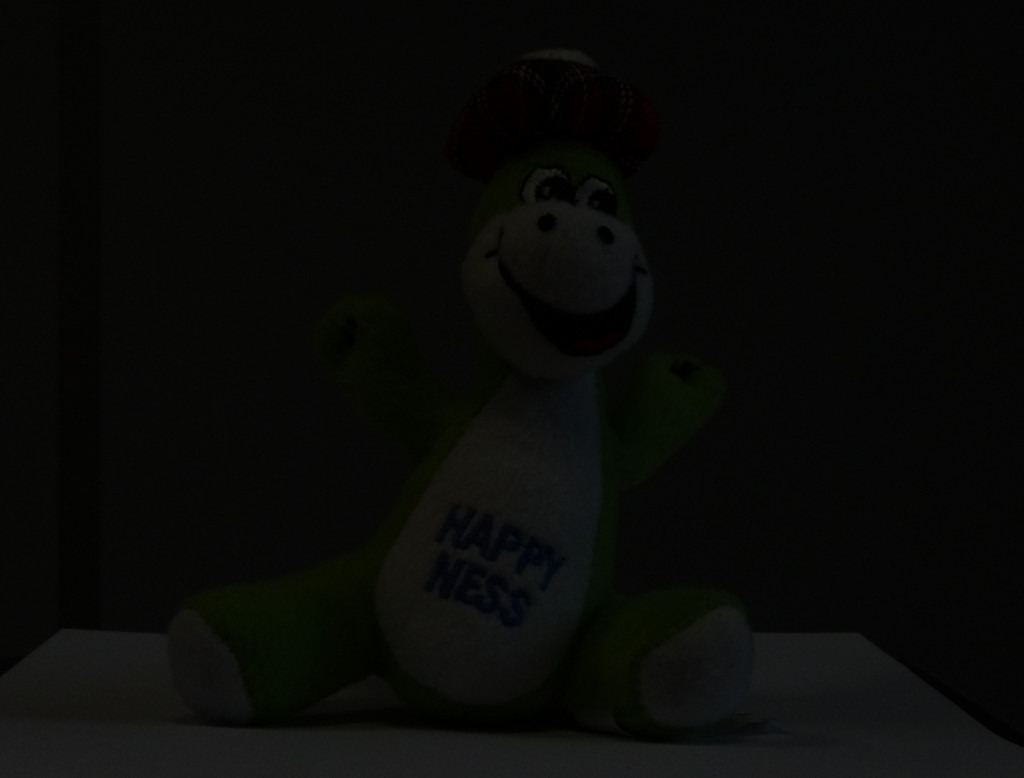}%
\label{fig:exptime:im_monstre:2}}
\hfil%
\phantom{\includegraphics[width=.3\mycolumnwidth]{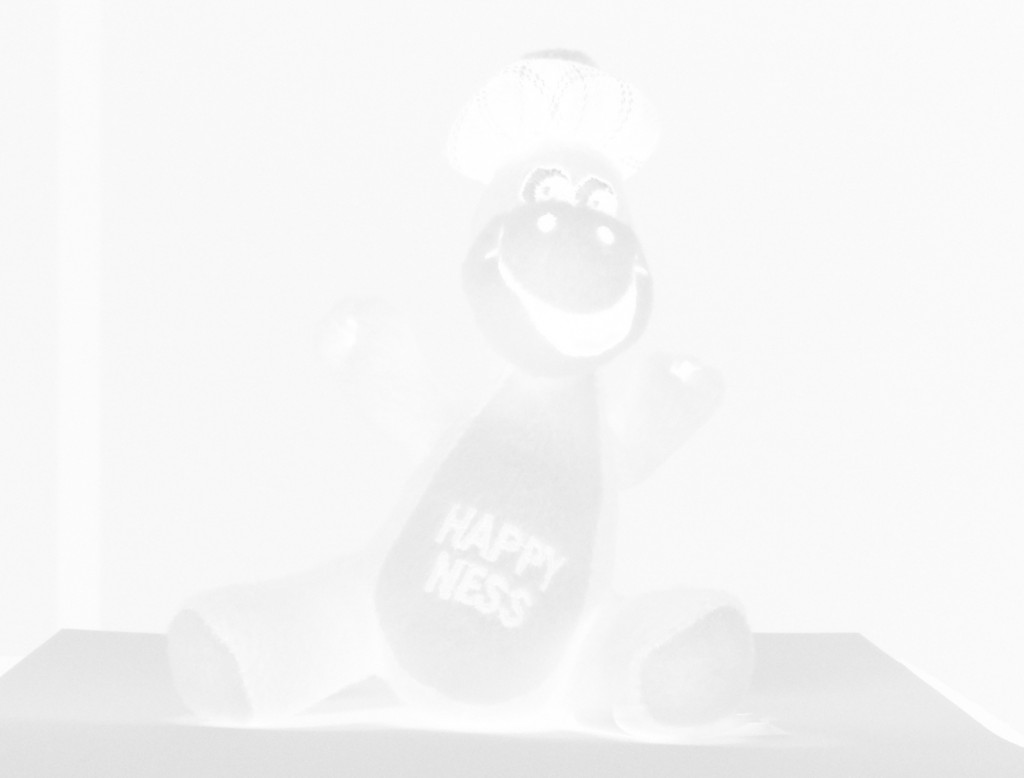}}
\\
\subfloat[$f$, luminance of $\mathbf{f}$ (LIP-greyscale)]{\includegraphics[width=0.3\mycolumnwidth]{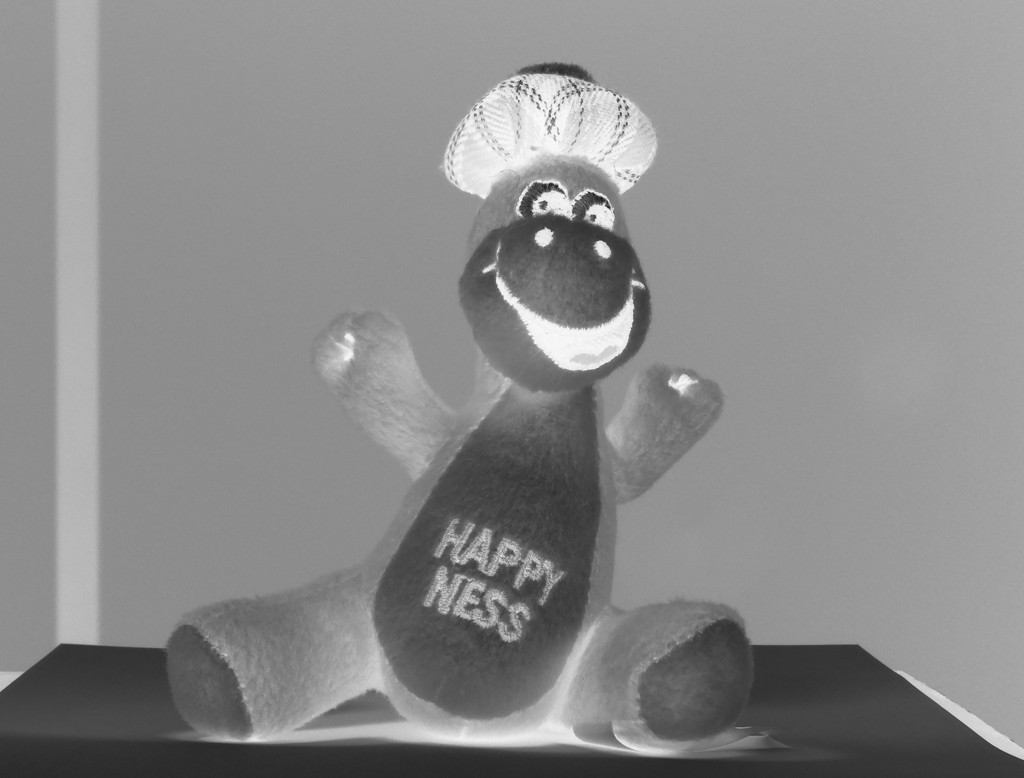}%
\label{fig:exptime:im_monstre:3}}%
\hfil%
\subfloat[$f^d$, luminance of $\mathbf{f}^d$ (LIP-greyscale)]{\includegraphics[width=0.3\mycolumnwidth]{Nessie_800ms_dark_comp}%
\label{fig:exptime:im_monstre:4}}%
\hfil%
\phantom{\includegraphics[width=.3\mycolumnwidth]{Nessie_800ms_dark_comp}}
\\
\subfloat[MM, $R_b(f)$, extended top-hat of $f$]{\includegraphics[width=0.3\mycolumnwidth]{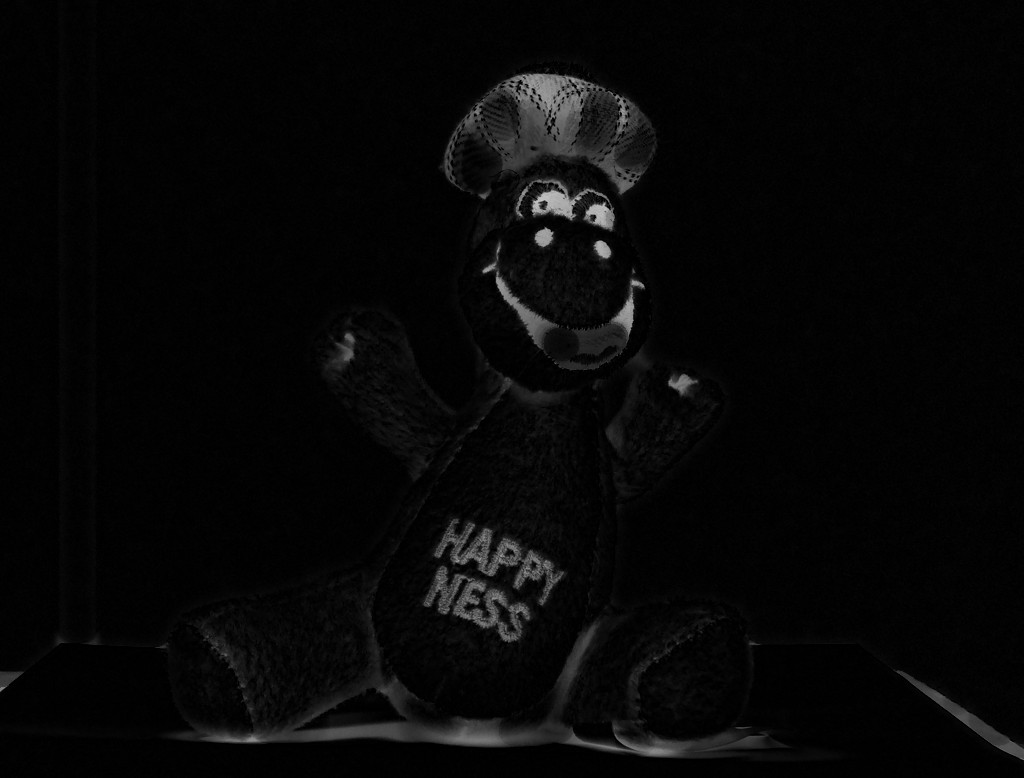}%
\label{fig:exptime:im_monstre:5}}
\hfil
\subfloat[MM, $R_b(f^d)$, extended top-hat of $f^d$.]
{\includegraphics[width=0.3\mycolumnwidth]{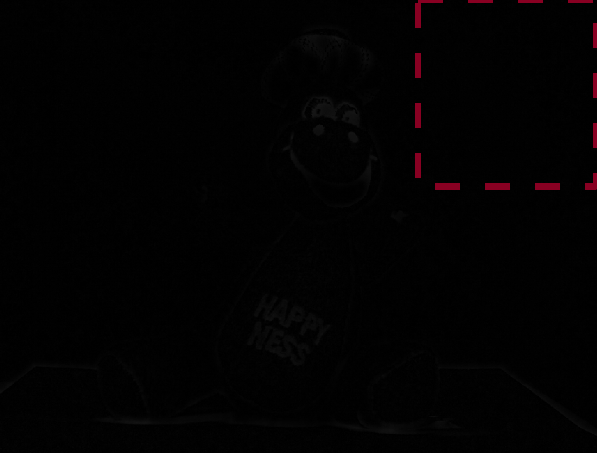}%
\label{fig:exptime:im_monstre:6}}
\hfil
\subfloat[Probe $b$ (LIP-greyscale)]{\includegraphics[width=0.3\mycolumnwidth]{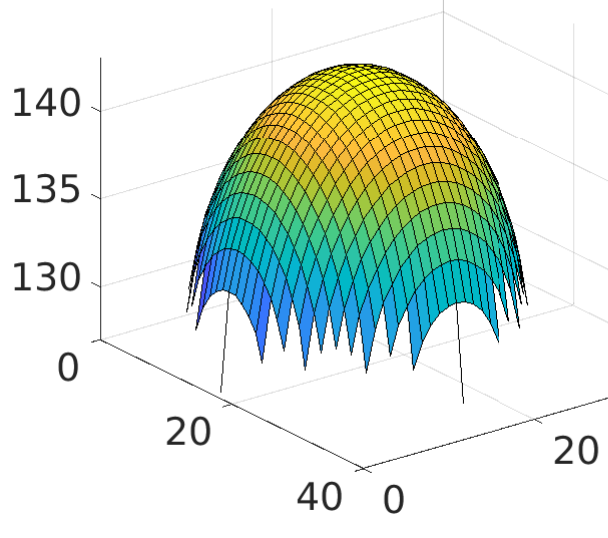}%
\label{fig:exptime:im_monstre:9}}
\\
\subfloat[LMM, $R^{\protect \LP}_b(f)$, extended LIP-top-hat of $f$]
{\includegraphics[width=0.3\mycolumnwidth]{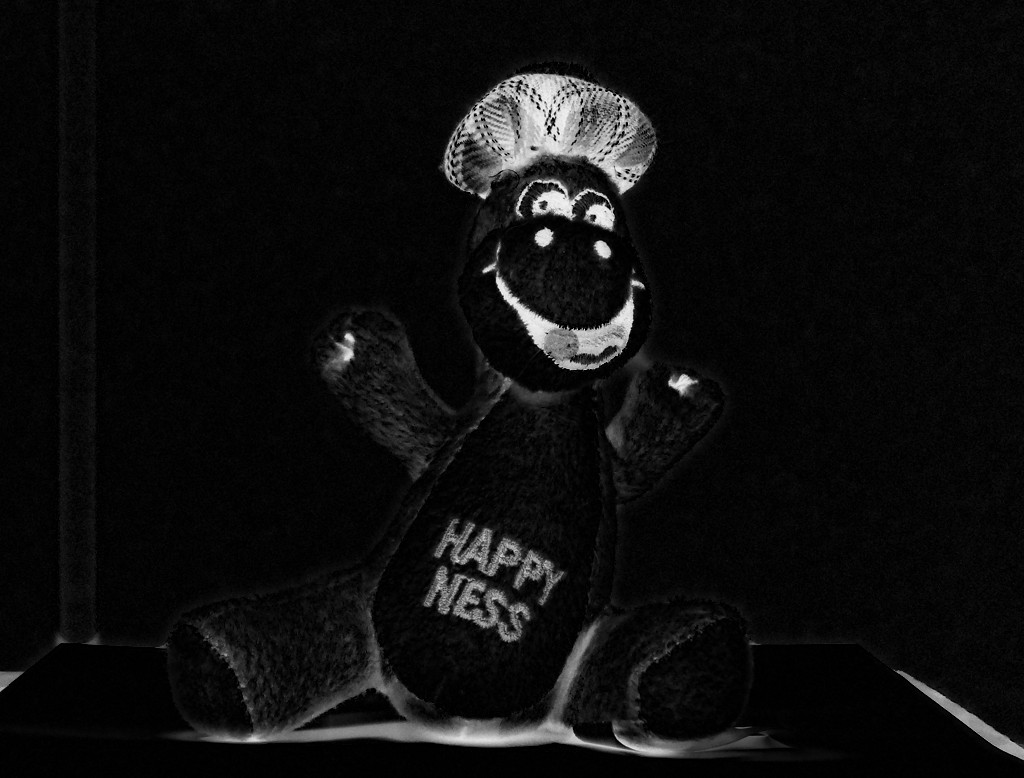}%
\label{fig:exptime:im_monstre:7}}
\hfil
\subfloat[LMM, $R^{\protect \LP}_b(f^d)$, extended LIP-top-hat of $f^d$]{\includegraphics[width=0.3\mycolumnwidth]{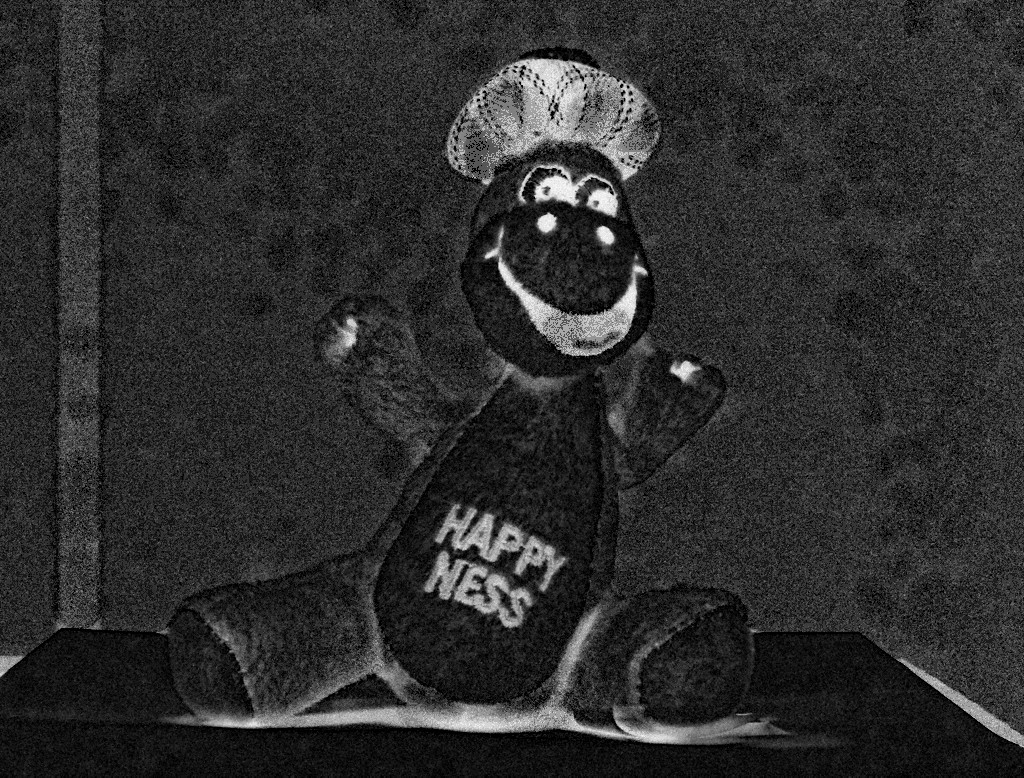}%
\label{fig:exptime:im_monstre:8}}
\hfil
\hspace{0.05\mycolumnwidth}
\subfloat[Zoom in (f)]{\includegraphics[width=0.2\mycolumnwidth]{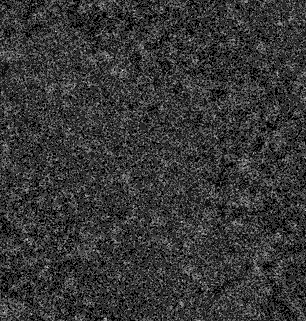}%
\label{fig:exptime:im_monstre:10}}
\phantom{\hspace{0.03\mycolumnwidth}}
\end{center}
\caption{Comparison of the robustness to camera exposure-times, between the operators of extended top-hat $R_b$ and extended LIP-top-hat $R^{\protect \LP}_b$. 
(a) $\mathbf{f}$ and 
(b) $\mathbf{f}^d$: colour images  captured with the camera exposure-times of \qty[parse-numbers=false,per-mode = symbol]{1/40}{\second} and \qty[parse-numbers=false,per-mode = symbol]{1/800}{\second}, respectively.
(c) $f$ and (d) $f^d$: luminance images (in the LIP-greyscale) of the colour images $\mathbf{f}$ and $\mathbf{f}^d$. 
(e) $R_b(f)$ and 
(f) $R_b(f^d)$: extended top-hat of $f$ and $f^d$ by the structuring function $b$. 
(g) The probe $b$ (in the LIP-greyscale) is made of a 16-pixel radius hemisphere, whose base is set at the grey value 127.
(h) $R^{\protect \LP}_b(f)$ and 
(i) $R^{\protect \LP}_b(f^d)$: extended LIP-top-hat of $f$ and $f^d$ by $b$.
(j) Zoom in and rescaling of the background ROI of $R_b(f^d)$, which is in the red rectangle of (f).
}
\label{fig:exptime:im_monstre}
\end{figure}

%
%

\section{Experiments and results}
\label{sec:Exp}



\subsection{Robustness to lighting variations caused by changes in the exposure-time of a camera}
\label{ssec:Exp:exptime}


\subsubsection{Extended top-hats}
\label{sssec:Exp:exptime:extTophat}

Let us focus on the extensions of top-hat operators $R_b$ and $R^{\LP}_b$ defined in section~\ref{ssec:RobOp:Extensions_top_hats}.
Only the extended LIP-top-hat operator $R^{\LP}_b$ is insensitive to the LIP-addition of any constant $c$ lying  in the interval $\left]-\infty, M\right[\>$.
As the LIP-addition of a constant models a change of the camera exposure-time or of the source intensity, the logarithmic operator $R^{\LP}_b$ is expected to have a low sensitivity to such changes.
In order to verify this assumption, an experiment has been conducted.
An image of the same scene is acquired with two significantly different exposure-times. The scene is composed of a soft toy monster named ``Nessie'', which is put down on a white support. (\figurename~\ref{fig:exptime:im_monstre}). The first colour image $\mathbf{f}$ (\figurename~\ref{fig:exptime:im_monstre:1}) is captured with a sufficient exposure-time of \qty[parse-numbers=false,per-mode = symbol]{1/40}{\second}. It is therefore bright and highly-contrasted. Its luminance is denoted by $f$  and it is represented in the LIP-greyscale: i.e., the inverted greyscale (\figurename~\ref{fig:exptime:im_monstre:3}). The second colour image $\mathbf{f}^d$ (\figurename~\ref{fig:exptime:im_monstre:2}) is captured with a too small exposure-time of \qty[parse-numbers=false,per-mode = symbol]{1/800}{\second}, which makes it dark and lowly-contrasted. Its luminance is denoted by $f^d$ (\figurename~\ref{fig:exptime:im_monstre:4}). 

In the LIP-greyscale composed of $M=256$ grey levels, the probe $b$ is a 16-pixel radius hemisphere whose base is set at a grey value of 127 (\figurename~\ref{fig:exptime:im_monstre:9}). In the LIP-greyscale, it is mathematically defined as follows: $b(x) = (M-1)- [-\sqrt{(\rho^2 - \left\|x\right\|_2^2)} + h ]$, where $\rho=16$ pixels is the diameter, $h=128=(M-1)-127$ grey levels is the minimal height, $x \in \Real^2$ is the probe coordinates lying within the interval $\left[-\rho,\rho\right]^2$. 
The probe is hemispherical because this shape is a canonical function in mathematical morphology, particularly in morphological scale-space theory and in the Slope Transform, where it plays a role analogous to that of the Gaussian function in the Fourier Transform \cite{Dorst1994,Maragos_1995,Jackway1996}.
The probe diameter, $\rho = 16$ pixels, was chosen to be larger than both the mouth and the letters on the soft toy. The minimum height, $h = 128$, corresponds to the midpoint of the grey scale. It is important to note that the amplitude of the probe translated by LIP-addition with the image value varies according to the local image intensity. It is precisely this LIP-addition that provides a physical justification for both the operator and its translated probe.

The extended top-hat operator $R_b$ is computed on both luminance images $f$ and $f^d$. The resulting images are denoted by $R_b(f)$ (\figurename~\ref{fig:exptime:im_monstre:5}) and $R_b(f^d)$ (\figurename~\ref{fig:exptime:im_monstre:6}). It can be noticed that the extended top-hat in the dark image $R_b(f^d)$ is much less contrasted than in the bright image $R_b(f)$. 
The extended top-hat operator $R_b$ is therefore sensitive to lighting changes caused by a variation of the camera exposure-time.

The other operator, the extended LIP-top-hat $R^{\LP}_b$ is then computed on both luminance images $f$ and $f^d$. The resulting images are denoted by $R^{\LP}_b(f)$ (\figurename~\ref{fig:exptime:im_monstre:7}) and $R^{\LP}_b(f^d)$ (\figurename~\ref{fig:exptime:im_monstre:8}). It can be noticed that both results, $R^{\LP}_b(f)$ and $R^{\LP}_b(f^d)$, are similar for the most-contrasted part of the scene (i.e, the foreground): the hat, the mouth of ``Nessie'', the letters on its body, the bottom of its body and the contour of the white support. For the background (i.e. the very low-contrasted parts of the scene), there exist differences between the results in the bright image $R^{\LP}_b(f)$ and in the dark image $R^{\LP}_b(f^d)$. They are due to the noise caused by the acquisition in the lowly-contrasted image $\mathbf{f}^d$ (\figurename~\ref{fig:exptime:im_monstre:2}). However, such a noise also exists in the background of the extended top-hat of the dark image $R_b(f^d)$ (\figurename~\ref{fig:exptime:im_monstre:6}), although it is hidden by some important amplitudes in $R_b(f^d)$. Indeed, when zooming in the background part of 
$R_b(f^d)$ (\figurename~\ref{fig:exptime:im_monstre:6}) and rescaling its amplitude, 
the noise can be observed (\figurename~\ref{fig:exptime:im_monstre:10}). It is similar to the one observed in the extended LIP-top-hat of the dark image $R^{\LP}_b(f^d)$ (\figurename~\ref{fig:exptime:im_monstre:8}).
In very low-contrasted parts of a dark image, the extended LIP-top-hat operator $R^{\LP}_b$ enhances the noise caused by the acquisition; i.e., when very few photons are captured by the camera sensor. However, in the most contrasted parts, this extended LIP-top-hat operator $R^{\LP}_b$ has the same amplitude in the bright image $R^{\LP}_b(f)$ (\figurename~\ref{fig:exptime:im_monstre:7}) as in the dark image $R^{\LP}_b(f^d)$ (\figurename~\ref{fig:exptime:im_monstre:8}).
As a consequence, the extended LIP-top-hat operator $R^{\LP}_b$ is much more robust than the extended top-hat operator $R_b$, to strong lighting variations caused by changes of camera exposure-time.

\subsubsection{Map of Asplund distances with a tolerance to extrema}
\label{sssec:Exp:exptime:AsplTol}

\begin{figure}[!t]
\centering
\subfloat[$\mathbf{f}_{sta}$ ({\protect \qty[parse-numbers=false,per-mode = symbol]{1/13}{\second}})]{\includegraphics[width=0.28\mycolumnwidth]{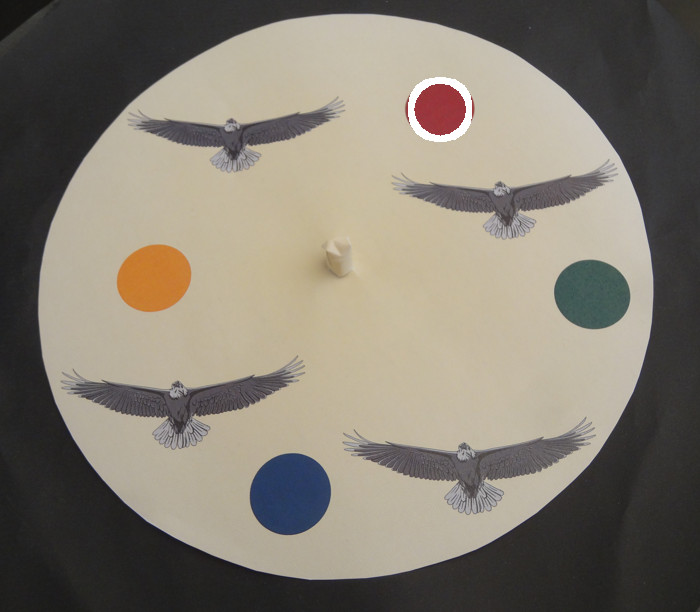}%
\label{fig:exptime:im_disk:AsplundTol:1}}
\hfil
\subfloat[$\mathbf{f}_{mov,bl}$ ({\protect \qty[parse-numbers=false,per-mode = symbol]{1/13}{\second}})]{\includegraphics[width=0.28\mycolumnwidth]{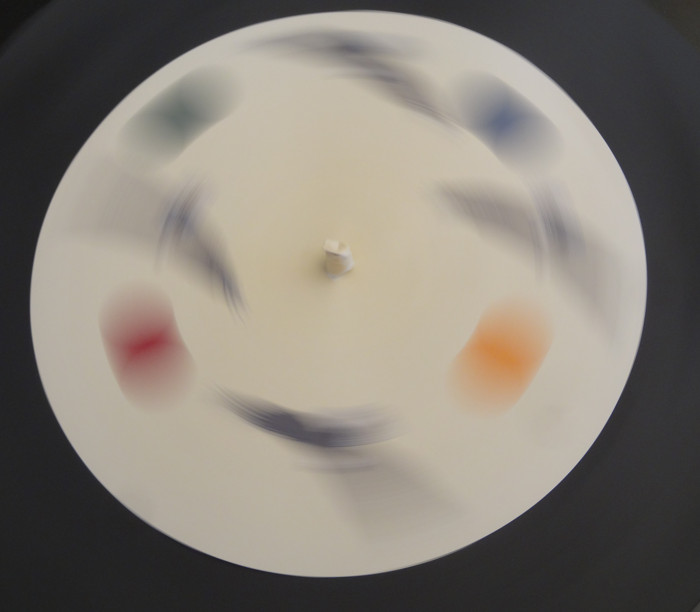}%
\label{fig:exptime:im_disk:AsplundTol:2}}\\
\hspace{0.28\mycolumnwidth}
\hfil
\subfloat[$\mathbf{f}_{mov,dk}$ ({\protect \qty[parse-numbers=false,per-mode = symbol]{1/160}{\second}})]{\includegraphics[width=0.28\mycolumnwidth]{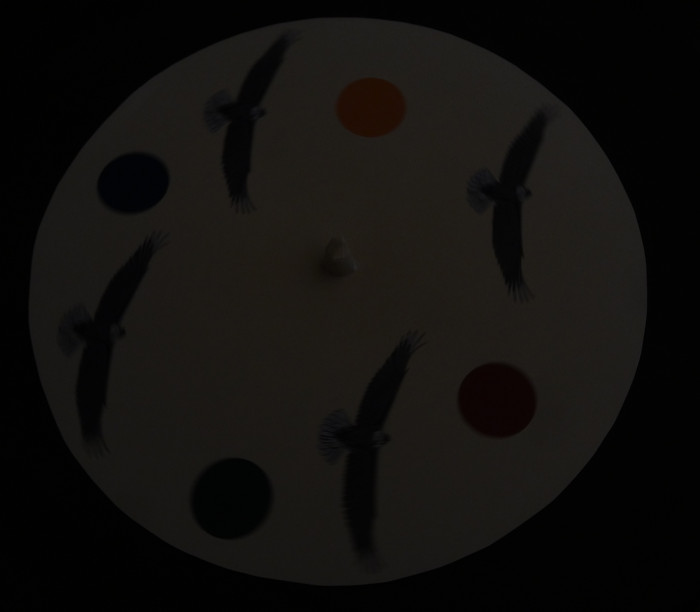}%
\label{fig:exptime:im_disk:AsplundTol:3}}
\\
\subfloat[MM, $A_{b,p}(f_{sta})$]{\includegraphics[width=0.28\mycolumnwidth]{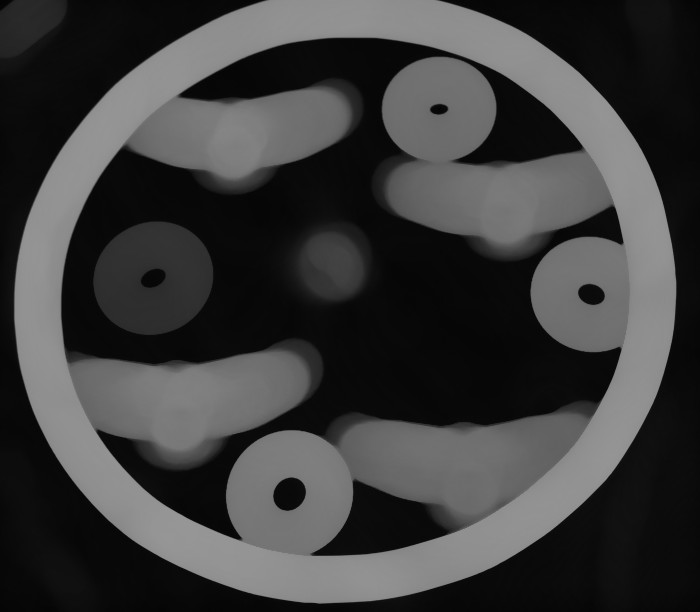}%
\label{fig:exptime:im_disk:AsplundTol:4}}
\hfil
\subfloat[MM, $A_{b,p}(f_{mov,bl})$]{\includegraphics[width=0.28\mycolumnwidth]{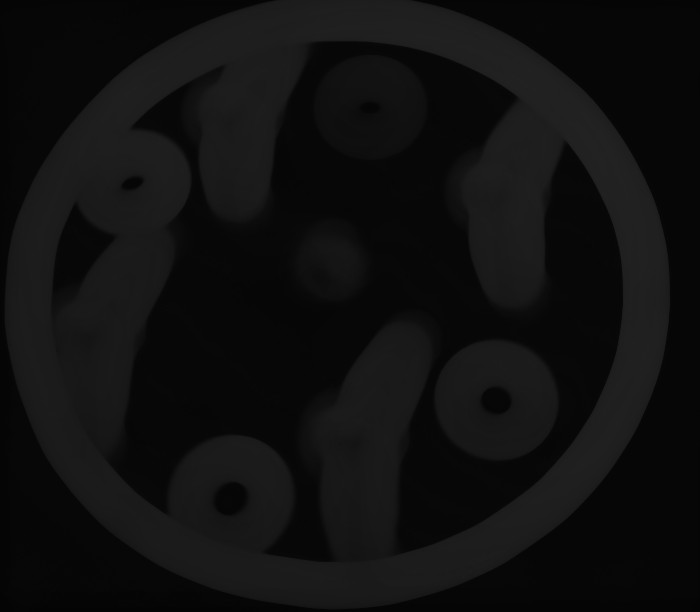}%
\label{fig:exptime:im_disk:AsplundTol:5}}
\\
\subfloat[LMM, $Asp_{b,p}^{\protect \LP}(f_{sta})$]{\includegraphics[width=0.28\mycolumnwidth]{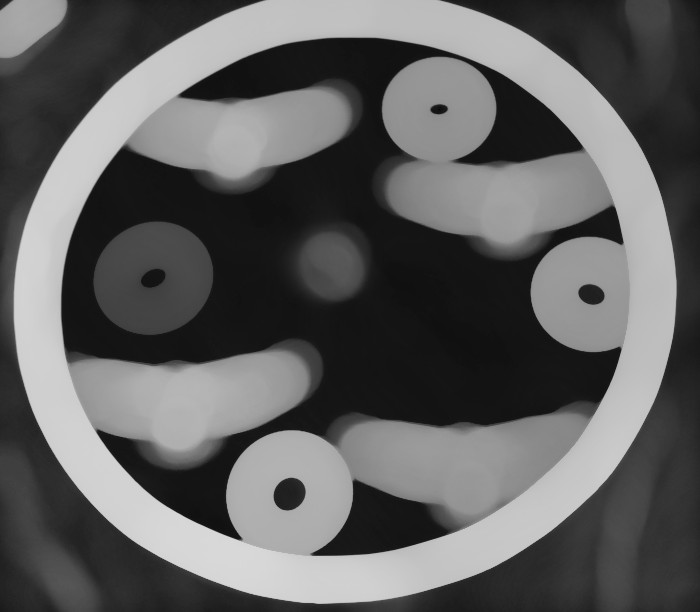}%
\label{fig:exptime:im_disk:AsplundTol:6}}
\hfil
\subfloat[LMM, $Asp_{b,p}^{\protect \LP}(f_{mov,bl})$]{\includegraphics[width=0.28\mycolumnwidth]{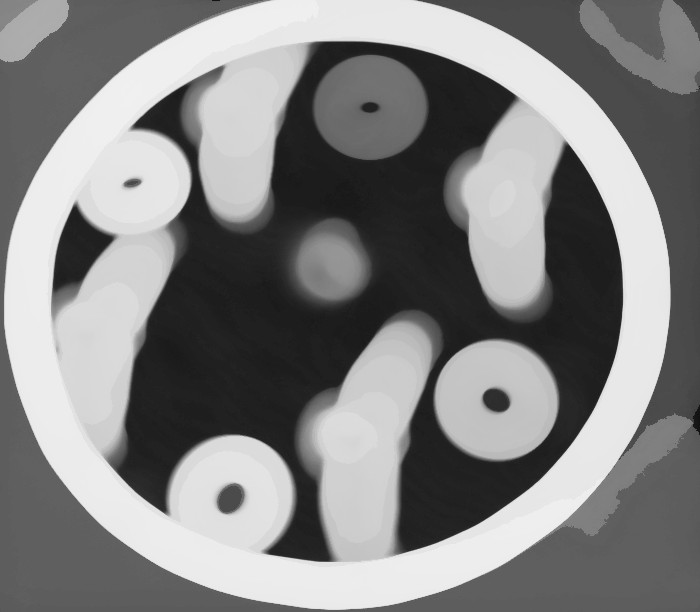}%
\label{fig:exptime:im_disk:AsplundTol:7}}
\caption{
(a)~Colour image $\mathbf{f}_{sta}$ of a static white disk captured with a camera exposure-time of \qty[parse-numbers=false,per-mode = symbol]{1/13}{\second}. The colour probe function $\mathbf{b}$ is delineated in white and converted into a luminance probe $b$.
(b)~Colour image $\mathbf{f}_{mov,bl}$ of the same disk in rotation and captured with a camera exposure-time of \qty[parse-numbers=false,per-mode = symbol]{1/13}{\second}. 
(c)~Colour image $\mathbf{f}_{mov,dk}$ of the white disk in rotation captured with a camera exposure-time of \qty[parse-numbers=false,per-mode = symbol]{1/160}{\second}.
(d)~In usual MM, equivalent maps of Asplund distances applied to the luminance of the static disk image, $A_{b,p}(f_{sta})$, and (e) to the luminance of the moving disk image, $A_{b,p}(f_{mov,dk})$.
(f)~Maps of LIP-additive Asplund distances, with a tolerance $p$, applied to the luminance of the static disk image, $Asp_{b,p}^{\protect \LP}(f_{sta})$, and (g)~to the luminance of the moving disk image, $Asp^{\protect \LP}_{b,p}(f_{mov,dk})$. For each map, the tolerance parameter $p$ is set to \SI{95}{\percent}.
}
\label{fig:exptime:im_disk:AsplundTol}
\end{figure}

The map of LIP-additive Asplund distances with a tolerance to extrema, $Asp^{\LP}_{b,p}$, defined in~\eqref{eq:map_AsAddtol_LMM}, is expected to be robust to strong lighting variations caused by different camera exposure-times. 
In order to verify this assumption, an experiment is performed with images of a moving object. In those images, the blur effect caused by the movement can be avoided by decreasing the camera exposure-time. However, this is done at the detriment of the image contrast.
In \figurename~\ref{fig:exptime:im_disk:AsplundTol}, a white disk with patterns is mounted on a turn table of a record player. The patterns include four small coloured disks and confounding shapes (i.e., the eagles).
Firstly, in \figurename~\ref{fig:exptime:im_disk:AsplundTol:1}, an image $\mathbf{f}_{sta}$ of the static white disk is captured with an appropriate camera exposure-time of \qty[parse-numbers=false,per-mode = symbol]{1/13}{\second}. This makes the image well-contrasted.
A well-contrasted probe $\mathbf{b}$ is manually selected inside the red disk within this image. This probe is delineated by a white curve in \figurename~\ref{fig:exptime:im_disk:AsplundTol:1} and is converted into a grey-level luminance image probe, $b$.
Then, the record player is started up at a speed of \SI{45}{revolutions/\minute}. With the same camera exposure-time as the one of $\mathbf{f}_{sta}$, a first image $\mathbf{f}_{mov,bl}$ is captured (\figurename~\ref{fig:exptime:im_disk:AsplundTol:2}). It is correctly contrasted but blurred which makes it useless to detect the coloured disks. In order to suppress the blur effect, the camera exposure-time is decreased to \qty[parse-numbers=false,per-mode = symbol]{1/160}{\second} and a second image $\mathbf{f}_{mov,dk}$ is captured (\figurename~\ref{fig:exptime:im_disk:AsplundTol:3}). This second image is not blurred but is darker than the image of the static disk $\mathbf{f}_{sta}$ (\figurename~\ref{fig:exptime:im_disk:AsplundTol:2}).
In the luminance images, $f_{sta}$ and $f_{mov,dk}$, of those two differently contrasted images $\mathbf{f}_{sta}$ (\figurename~\ref{fig:exptime:im_disk:AsplundTol:1}) and $\mathbf{f}_{mov,dk}$ (\figurename~\ref{fig:exptime:im_disk:AsplundTol:3}), the map of LIP-additive Asplund distances with a tolerance to extrema $Asp^{\LP}_{b,p}$ is compared to its equivalent in classical functional MM, $A_{b,p}$\>, defined as follows:
\begin{align}
	A_{b,p} (f) &= \vartheta_{- \overline{b},n_1} (f) - \zeta_{b,n_2}(f). \label{eq:map_AsAddtol_MM}%
\end{align}
The parameters $n_1$ and $n_2$ are defined as in Proposition~\ref{prop:link_AsAddtol_LMM}. In classical MM, the map $A_{b,p}(f_{sta})$ of the static disk image (\figurename~\ref{fig:exptime:im_disk:AsplundTol:4}) appears brighter than the map $A_{b,p}(f_{mov,dk})$ of the moving disk image (\figurename~\ref{fig:exptime:im_disk:AsplundTol:5}). However, in LMM, the maps of LIP-additive Asplund distance with tolerance, computed from the the static disk image $Asp^{\LP}_{b,p}(f_{sta})$ (\figurename~\ref{fig:exptime:im_disk:AsplundTol:6}) and from the moving disk image $Asp^{\LP}_{b,p}(f_{mov,bl})$ (\figurename~\ref{fig:exptime:im_disk:AsplundTol:7}), respectively, are similarly contrasted. Unlike the classical MM maps, the LMM maps\pagebreak[5] 
are robust to strong illumination variations caused by different camera exposure times.
\FloatBarrier


\subsection{Robustness to non-uniform lighting variations in images}
\label{ssec:Exp:loclight}

In order to test the robustness to non-uniform lighting variations of a segmentation task, a LMM approach \cite{Noyel2020} is compared to other state-of-the-art methods \cite{Liu2022,Kamran2021,Zhou2021}. The segmentation task consists of extracting vessels in eye fundus images coming from the test set of the DRIVE dataset \cite{Staal2004}. However, the testing set is composed of \qty{20} colour eye fundus images which are well contrasted and do not present any significant non-uniform lighting variations. As a consequence, such variations were previously added to those images.

\subsubsection{Adding lighting variations to the images}
\label{sssec:Exp:loclight:descr}

\begin{figure}[!t]
\centering
\subfloat[$\mathbf{f}$]{\includegraphics[width=0.4\mycolumnwidth]{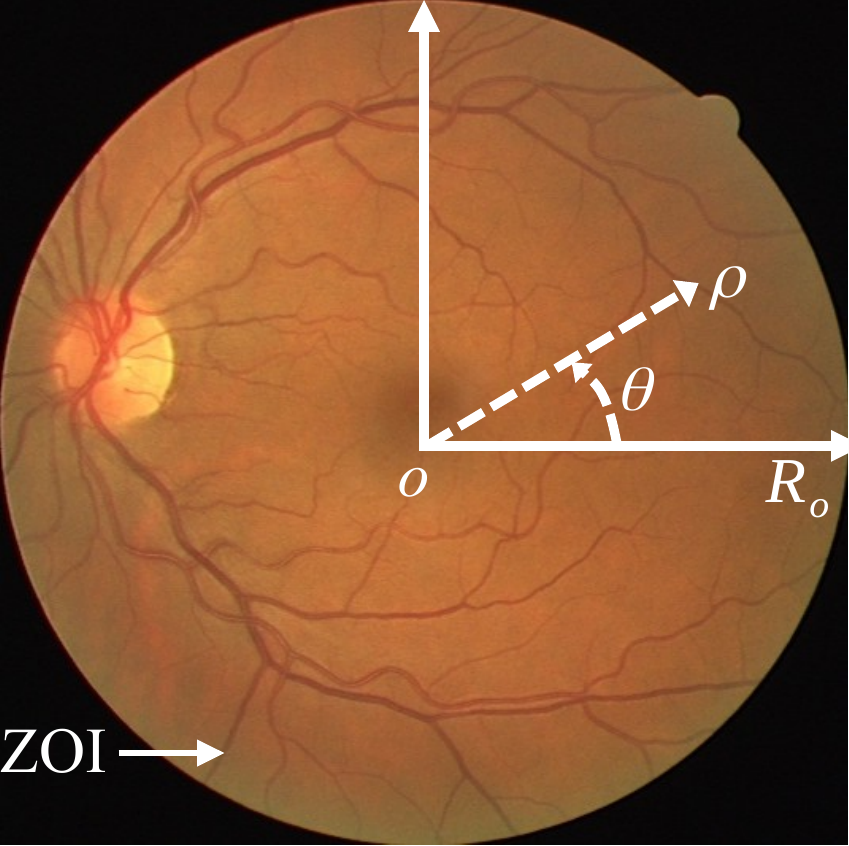}%
\label{fig:Exp:loclight:descr:1}}
\hfil
\subfloat[$c_{dk}$]{\includegraphics[width=0.49\mycolumnwidth,clip,trim=0cm 0cm 0cm 0cm]{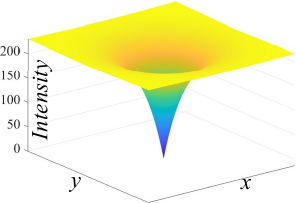}%
\label{fig:Exp:loclight:descr:2}}
\\
\subfloat[$\mathbf{f}^{dk}$]{\includegraphics[width=0.4\mycolumnwidth,clip,trim=0cm 0cm 0cm 0cm]{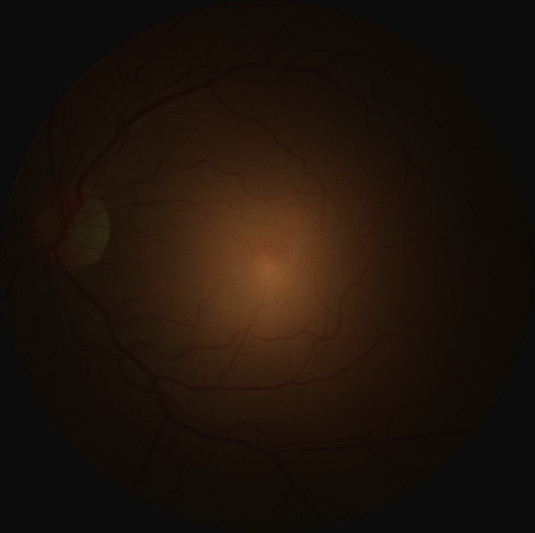}%
\label{fig:Exp:loclight:descr:3}}

\caption{(a)~Colour eye fundus image $\mathbf{f}$, its ZOI and the polar coordinates $(\rho,\theta)$.
(b)~Surface of the darkening function $c_{dk}$ in the LIP-greyscale.
(c)~Darkened image $\mathbf{f}^{dk}$.}
\label{fig:Exp:loclight:descr}
\end{figure}

In a colour image $\mathbf{f}=(f_R, f_G, f_B)$, where $\mathbf{f} : D \mapsto \left[0,M\right[^3$, the non-uniform lighting variations are generated by LIP-adding the same darkening function $c_{dk} : D \mapsto \left]-\infty,M\right[$ to each of the three image components: $f_R$, $f_G$ and $f_B$. The darkening function is a 2D increasing function whose origin is located at the centre of the Zone of Interest (ZOI). The ZOI is assimilated to a circle of radius $R_o$ and centre $o=(x_o,y_o)$, where $o \in D \subset \Real^2$ (\figurename~\ref{fig:Exp:loclight:descr:1}). Let $\rho \in \left[0,+\infty\right[$ and $\theta \in \left]-\pi,\pi\right]$ be the polar coordinates of the pixels from the circle centre $o$. 
The darkening function $c_{dk}$ is defined by (\figurename~\ref{fig:Exp:loclight:descr:2}):
$c_{dk}(\rho,\theta) = I_0 \left[1 -  \exp{\left( \frac{-\rho}{R_o/4} \right)} \right]$. 
The intensity value $I_0 =$ \qty{230} is chosen so that the image is strongly darkened in its external part. 
The darkened image $\mathbf{f}^{dk}=(f^{dk}_R, f^{dk}_G, f^{dk}_B)$ is then defined for each of its components $f^{dk}_i$ as follows:
$f^{dk}_i = (M-1)-\lfloor \left(M-1-f_i\right) \LP c_{dk} \rfloor$,
where $\lfloor x \rfloor$ is the floor function of the value $x \in \Real^+$. The floor function allows to save the darkened images in png or tif format in order to use them with different segmentation methods. 
Those darken images have a brighter area in their centre than elsewhere (\figurename~\ref{fig:Exp:loclight:descr:3}).

\subsubsection{Vessel segmentation method based on LMM}
\label{sssec:Exp:loclight:LMM}

\begin{figure}[!t]
\centering
\hfil
\subfloat[Probe domain]{\includegraphics[width=0.32\mycolumnwidth]{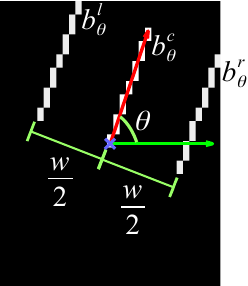}%
\label{fig:Exp:loclight:LMM:probe:1}}
\hfil
\subfloat[Probe intensity]{\includegraphics[width=0.38\mycolumnwidth,clip,trim=0cm 0cm 0cm 0cm]{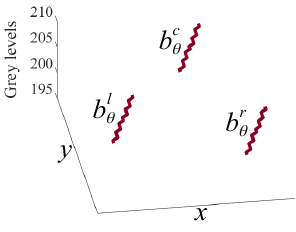}%
\label{fig:Exp:loclight:LMM:probe:2}}
\hfil
\caption{(a)~2D probe $b$ with an orientation $\theta$ and a width $w$. 
(b)~The central segment $b^c_{\theta}$ has a higher intensity than both others ones $b^l_{\theta}$ and $b^r_{\theta}$.}
\label{fig:Exp:loclight:LMM:probe}
\end{figure}

In \cite{Noyel2020,Noyel2021}, a method based on Logarithmic Mathematical Morphology was introduced to segment vessels in eye fundus images. The colour images $\mathbf{f}=(f_R, f_G, f_B)$ are converted to greyscale images with the LIP-greyscale thanks to the equation $f = M-1-(\qty{0.299}f_R + \qty{0.587} f_G + \qty{0.114} f_B)$.
In this inverted greyscale, the vessels appear as brighter than their surroundings. As in section~\ref{ssec:RobOp:LipDiff_LipEro}, a bump detector is defined. It is based on a 2D probe $b_{\theta} : D_{b_{\theta}} \mapsto \left[0,M\right[$ composed of 3 parallel segments in the orientation $\theta$ and with the same length (\figurename~\ref{fig:Exp:loclight:LMM:probe:1}). The probe origin is chosen as one of the extremities of the central segment $b^c_{\theta}$. Its intensity is greater than the one of the left and right segments $b^l_{\theta}$ and $b^r_{\theta}$ (\figurename~\ref{fig:Exp:loclight:LMM:probe:2}). These two segments are equidistant of the central one and the width of the probe is $w$.
The left and right detectors $E (b^l_{\theta}, f)$ and $E (b^r_{\theta}, f)$ of \eqref{eq:LAC_er_left} and \eqref{eq:LAC_er_right} are now defined with the $k^{th}$ minimum $\wedge^k$, for any $x\in D$, as follows:
\begin{align}
	E^k(b^l_{\theta}, f)(x) &= \wedge^k_{ h \in D_{b^l_{\theta}} }{ \{ f(x+h) \LM b^l_{\theta}(h) \} \LM \grave{c}_{\protect b_{\theta},k}(f)(x) }  \nonumber \\
	&= \zeta^{\LP}_{b^l_{\theta},k} (f) (x) \LM \grave{c}_{\protect b_{\theta},k}(f)(x)  \label{eq:LIP:seg:yeux:map_LAC_er_left2}\\		
	E^k(b^r_{\theta}, f)(x) &= \wedge^k_{ h \in D_{b^r_{\theta}} }{ \{ f(x+h) \LM b^r_{\theta}(h) \} \LM \grave{c}_{\protect b_{\theta},k}(f)(x) } \nonumber \\
	&= \zeta^{\LP}_{b^r_{\theta},k} (f) (x) \LM \grave{c}_{\protect b_{\theta},k}(f)(x).  \label{eq:LIP:seg:yeux:map_LAC_er_right2}	
\end{align}                                                                                                              
The map $\grave{c}_{\protect b_{\theta},k}(f)$ is defined as the pointwise infimum $\bigwedge$ of the maps $c_{b^c_{\theta}}(f)$, $c_{b^l_{\theta},k}(f)$ and $c_{b^r_{\theta},k}(f)$ for each segment $b^c_{\theta}$, $b^l_{\theta}$ and $b^r_{\theta}$ of the probe:\linebreak
$\grave{c}_{b_{\theta},k}(f) = \bigwedge{\{\>  c_{b^c_{\theta}}(f) , \bigwedge{[ c_{b^l_{\theta},k}(f) , c_{b^r_{\theta},k}(f) ]} \} }
	= \bigwedge{\{\>  \varepsilon^{\LP}_{b^c_{\theta}}(f) , \bigwedge{[ \zeta^{\LP}_{b^l_{\theta},k}(f) , \zeta^{\LP}_{b^r_{\theta},k}(f) ]} \} }$.

\begin{figure}[!t]
\centering
\hfil
\subfloat[$e^k_{b}(f)$]{\includegraphics[width=0.4\mycolumnwidth]{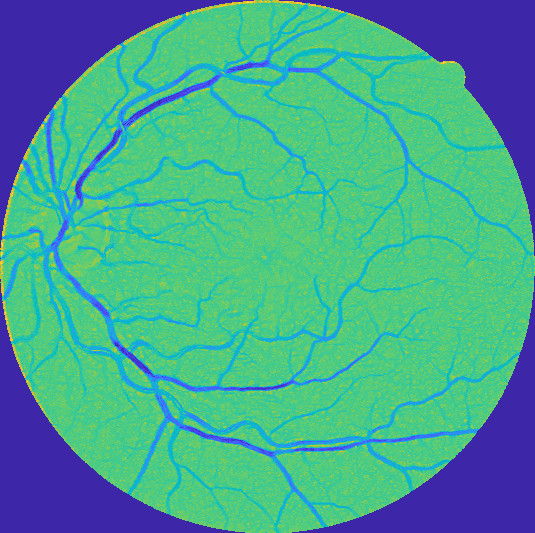}%
\label{fig:Exp:loclight:LMM:seg:1}}
\hfil
\subfloat[Segmentation]{\includegraphics[width=0.4\mycolumnwidth,clip,trim=0cm 0cm 0cm 0cm]{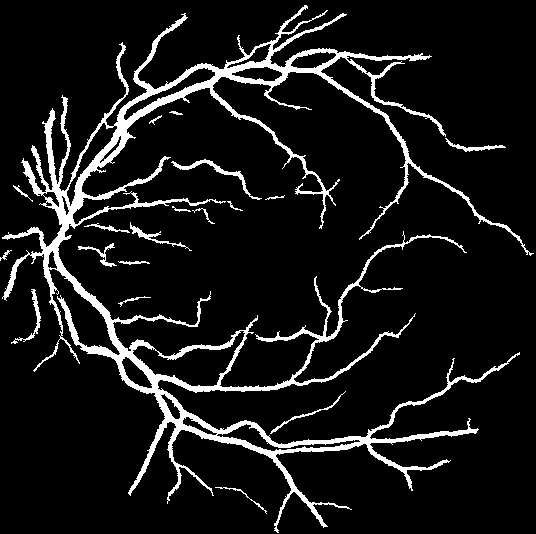}%
\label{fig:Exp:loclight:LMM:seg:2}}
\caption{(a)~Map of vesselness $e^k_{b}(f)$.
(b)~Vessel segmentation.}
\label{fig:Exp:loclight:LMM:seg}
\end{figure}

As the central segment $b^c_{\theta}$ must fully enter into the vessel relief, the infimum must be extracted and therefore in the previous equation, the map $c_{{\protect b^c_{\theta}}} (f) = \wedge \{ f(x+h) \LM b^c_{\theta}(h) , h \in b^c_{\theta} \} = \varepsilon^{\LP}_{b^c_{\theta}}(f)$ is used (see \eqref{eq:lower_map_add_2} and \eqref{eq:mglb_LMM}). 
However, in order to reduce the effects of noise for the left and right segments, $b^l_{\theta}$ and  $b^r_{\theta}$, the maps with the $k^{th}$ minimum, $c_{{\protect b^l_{\theta},k}}(f) = \wedge^k \{ f(x+h) \LM b^l_{\theta}(h), h \in b^l_{\theta} \} = \zeta^{\LP}_{b^l_{\theta},k}(f)$ and $c_{{\protect b^r_{\theta},k}}(f) = \zeta^{\LP}_{b^r_{\theta},k}(f)$, are used (see \eqref{eq:mglb_AsAddtol_LMM}). 
As in \eqref{eq:LIP:LMM:detector_one_dir}, the bump detector map in orientation $\theta$, $E^k (b_{\theta}, f)$, is defined by:
\begin{align}
	E^k (b_{\theta}, f) &= \bigvee{\{ E^k (b^l_{\theta} , f) , E^k (b^r_{\theta}, f) \}}. \label{eq:LIP:seg:yeux:detector_one_or}
\end{align}
The bump detector map is expressed as the point-wise infimum of the maps $E^k (b_{\theta}, f)$ in all the orientations $\theta \in \Theta$:
\begin{align}
	E^k (b, f) &= \bigwedge{\{ E^k (b_{\theta}, f) \mid \theta \in \Theta \}}. \label{eq:LIP:seg:yeux:detector_tol_1probe}
\end{align}
As the vessel detection is a multi-scale problem, $I$ different probes,  $\{b_i\}_{i \in [\![1 \ldots I]\!]}$, of width $\{w_i\}_i$ and length $\{l_i\}_i$ will be used.
The bump detector maps $E^k (b_i, f)$ for the probes $b_i$ are
then combined by point-wise infimum:
\begin{align}
	e^k_{b}(f) &= \bigwedge{\{ E^k (b_i, f) \mid i \in [\![1 \ldots I]\!] \}}. \label{eq:LIP:seg:yeux:map_detector_tol_3probes}
\end{align}        
In the \textit{map of vesselness} $e^k_{b}(f)$ (\figurename~\ref{fig:Exp:loclight:LMM:seg:1}), the vessels appear as valleys. They are extracted by a threshold such that \qty{12}{\percent} of the ZOI area are considered as vessels (\figurename~\ref{fig:Exp:loclight:LMM:seg:2}). $I=3$ probes with \qty{18} orientations $\theta$ between \qty{0}{\degree} and \qty{360}{\degree} are chosen. The width $w$ of the probe was chosen in order to be slightly greater than the largest vessel diameter. All the parameters and the experiments used to estimate them are given in \cite{Noyel2020}.
Similarly to property~\ref{property:lr_detectors_rob_LIPadd}, the map of vesselness $e^k_{b}(f)$ \eqref{eq:LIP:seg:yeux:map_detector_tol_3probes}
is insensitive to the LIP-addition (or the LIP-subtraction) of any constant $c \in \left]-\infty,M\right[$ to (or from) an image $f$, as shown by the following equation, the proof of which can be found in Appendix \ref{app:eq:LMM_vessels_rob_LIPadd}:
\begin{align}
	e^k_{b}(f \LP c) 			&= e^k_{b}(f). \label{eq:LMM_vessels_rob_LIPadd}
\end{align}
The map of vesselness $e^k_{b}(f)$ is therefore robust to uniform variations  of light intensity or of camera exposure-time in an image $f$. Indeed, such variations are modelled by the LIP-addition (or the LIP-subtraction) of a constant $c \in \left]-\infty,M\right[$\>. The vessel segmentation by this LMM approach is therefore robust to such uniform lighting variations.

\subsubsection{Comparison of the LMM approach with other methods}
\label{sssec:Exp:loclight:comp}

The robustness to non-uniform lighting variations will be tested for the previous approach and state-of-the-art approaches. As explained in section \ref{sssec:Exp:loclight:descr}, a non-uniform lighting variation was generated by LIP-adding to the images a function $c_{dk} : D \mapsto \left]-\infty,M\right[$ which varies across the domain $D$, in place of a constant $c$. The 20 test images of the dataset were darkened.

For comparison purposes, the three best methods of vessel segmentation were selected in the DRIVE dataset using the ranking given in \cite{DriveBenchmark2023}. They are named FR-UNet \cite{Liu2022}, RV-GAN \cite{Kamran2021} and SGL (Study Group Learning) \cite{Zhou2021} and are based on Deep Learning architectures. FR-UNET is based on a multi-resolution U-Net architecture \cite{Ronneberger2015}. RV-GAN is composed of a multi-scale Generative Adversarial Network \cite{Goodfellow2014}. SGL is based on a U-Net consisting of an image enhancement module and a segmentation module. Those three methods have been tested on the test set using the pre-trained weights given by their authors.
In table~\ref{tab:Exp:loclight:comp}, the performance of each method was evaluated by several indicators for the initial and the darkened images of the test set. The indicators are the mean values over the test set of the Area Under ROC Curve (AUC), Accuracy (Acc), Sensitivity (Se) and Specificity (Sp). Each indicator was computed for each image and the mean value was taken over the test set.
The same groundtruth coming from the DRIVE dataset was used. The same program in MATLAB\textsuperscript{\tiny\textregistered} language was used to estimate those indicators.

\begin{figure}[!t]
\vspace{-1em}
\centering
\subfloat[LMM, ini.]{\includegraphics[width=0.3\mycolumnwidth]{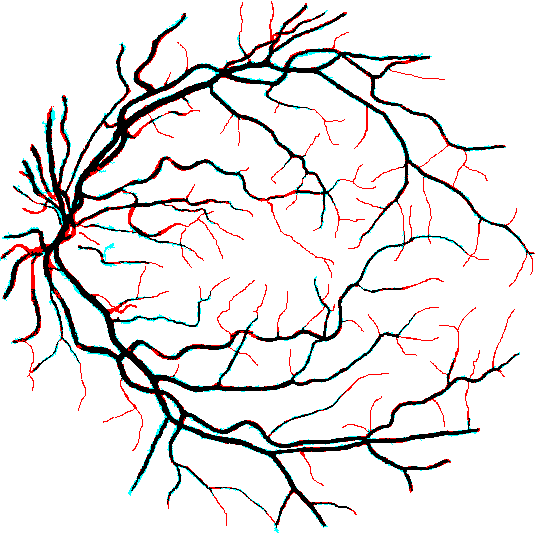}%
\label{fig:Exp:loclight:comp:1}}
\hfil
\subfloat[LMM, dark]{\includegraphics[width=0.3\mycolumnwidth]{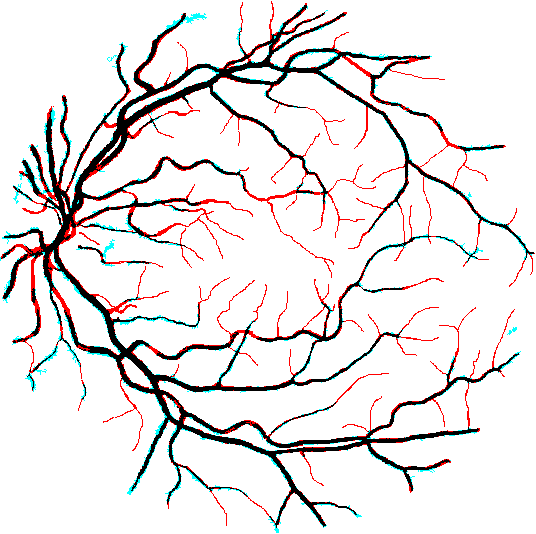}%
\label{fig:Exp:loclight:comp:2}}
\\
\subfloat[FR-UNet, ini.]{\includegraphics[width=0.3\mycolumnwidth]{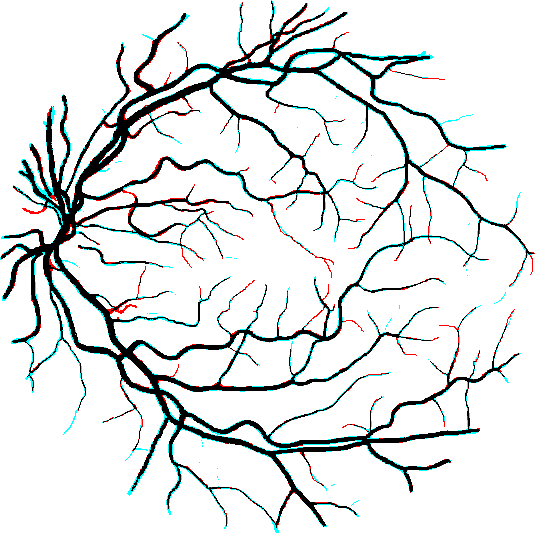}%
\label{fig:Exp:loclight:comp:3}}
\hfil
\subfloat[FR-UNet, dark]{\includegraphics[width=0.3\mycolumnwidth]{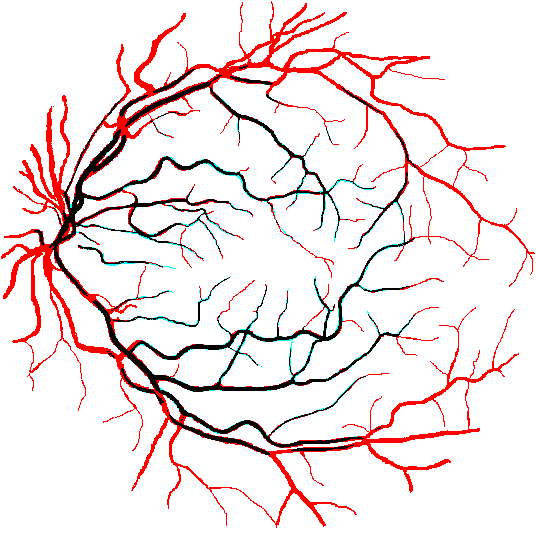}%
\label{fig:Exp:loclight:comp:4}}
\\
\subfloat[SGL, ini.]{\includegraphics[width=0.3\mycolumnwidth]{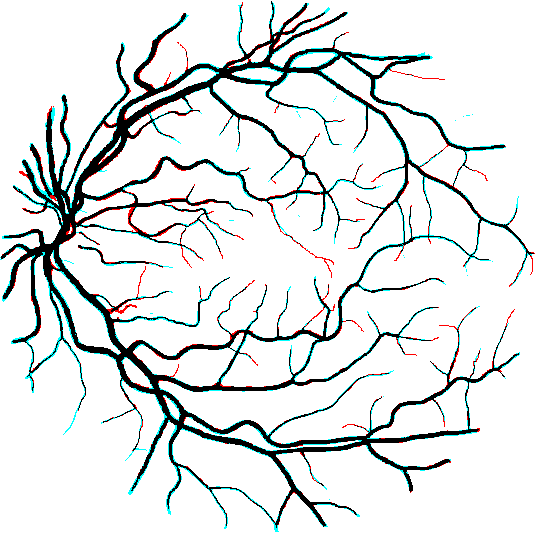}%
\label{fig:Exp:loclight:comp:5}}
\hfil
\subfloat[SGL, dark]{\includegraphics[width=0.3\mycolumnwidth]{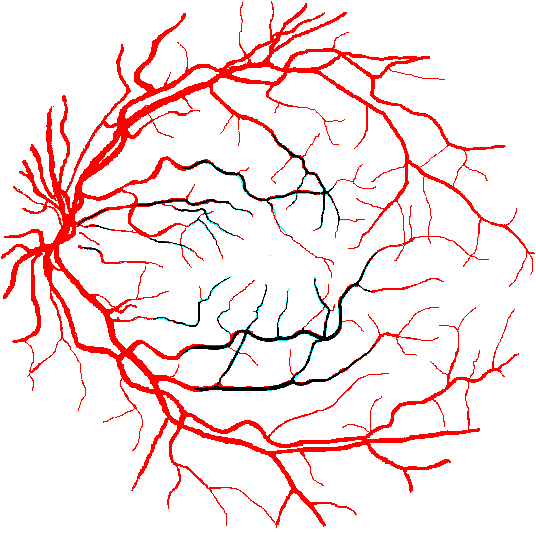}%
\label{fig:Exp:loclight:comp:6}}
\\
\subfloat[RV-GAN, ini.]{\includegraphics[width=0.3\mycolumnwidth]{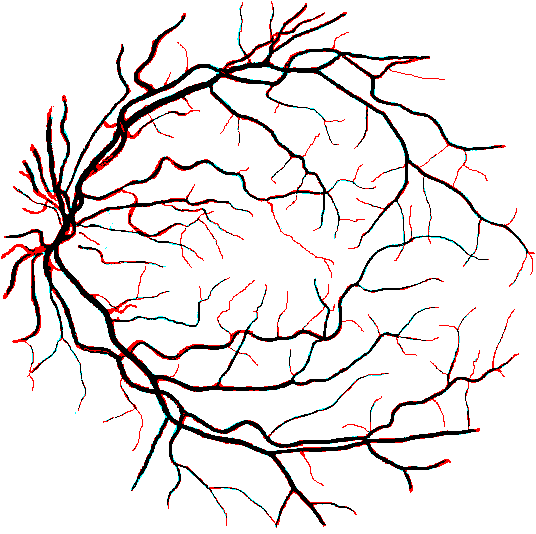}%
\label{fig:Exp:loclight:comp:7}}
\hfil
\subfloat[RV-GAN, dark]{\includegraphics[width=0.3\mycolumnwidth]{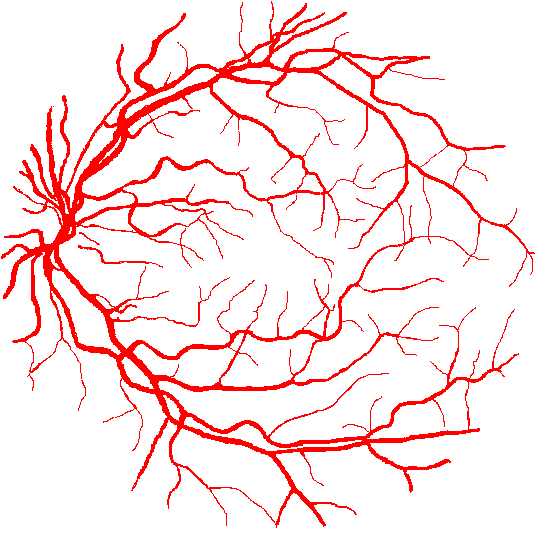}%
\label{fig:Exp:loclight:comp:8}}
\caption{Segmentation comparison with the groundtruth. Black pixels
are true positives, white pixels are true negatives, cyan pixels
are false positives and red pixels are false negatives.
(a) and (b)~LMM segmentations for the initial $\mathbf{f}$ and dark $\mathbf{f}^{dk}$ images of \figurename~\ref{fig:Exp:loclight:descr}.
(c) and (d)~FR-UNet segmentations.
(e) and (f)~SGL segmentations.
(g) and (h)~RV-GAN segmentations.}
\label{fig:Exp:loclight:comp}
\end{figure}

The relative difference between the AUC for the initial images and the darkened ones has been computed. The LMM approach obtains the smallest relative difference between AUC with \qty{2.51}{\percent}. This is better than the other methods: FR-UNet (\qty{8.48}{\percent}), SGL (\qty{27.95}{\percent}) and RV-GAN (\qty{31.56}{\percent}).
In \figurename~\ref{fig:Exp:loclight:comp}, one can notice that the LMM segmentations are much more similar between the initial test image (\figurename~\ref{fig:Exp:loclight:comp:1}) and its darkened version (\figurename~\ref{fig:Exp:loclight:comp:2}) than the segmentations with the other approaches. Indeed, in the initial images, the FR-UNet (\figurename~\ref{fig:Exp:loclight:comp:3}), the SGL (\figurename~\ref{fig:Exp:loclight:comp:5}) and the RV-GAN (\figurename~\ref{fig:Exp:loclight:comp:7}) methods obtains good segmentation results. However, in the darken image, the FR-UNet (\figurename~\ref{fig:Exp:loclight:comp:4}) and the SGL (\figurename~\ref{fig:Exp:loclight:comp:6}) methods only segment the vessels in the brightest part located in the image centre. The RV-GAN approach does not segment the vessels (\figurename~\ref{fig:Exp:loclight:comp:8}).
As a consequence, the LMM approach has a better robustness to non-uniform lighting variations than the other state-of-the-art approaches. This is caused by the LIP-differences between LMM operations (see \eqref{eq:LIP:seg:yeux:map_LAC_er_left2} and \eqref{eq:LIP:seg:yeux:map_LAC_er_right2}) which are used in this approach. 


\begin{table}[!t]
\centering
\caption{Comparison of the segmentation methods on the test DRIVE dataset with initial and darkened images. Average values 
of the Area Under ROC Curve (AUC), Accuracy (Acc), Sensitivity (Se) and Specificity (Sp). Absolute Relative difference between the AUC for the initial and the dark images (R. Diff).}
\begin{tabular}{@{}lc
S[table-number-alignment = center,table-format=1.4]
S[table-number-alignment = center,table-format=1.4]
S[table-number-alignment = center,table-format=1.4]
S[table-number-alignment = center,table-format=1.4]
@{ }c}
\hline
Method						&Images	& \mcc{AUC} & \mcc{Acc} & \mcc{Se} 	& \mcc{Sp} & \mcc{R. Diff.}\\
\hline
\multirow{2}{4em}{LMM}		& initial & 0.9434 	& 0.9625 	& 0.7370 	& 0.9844 & \multirow{2}*{\tablenum[table-number-alignment = center,table-format=2.2]{2.51}{\,\si{\percent}}} \\
				  			& dark    & 0.9197 	& 0.9547 	& 0.6664 	& 0.9826 & \\
\hline
\multirow{2}{4em}{FR-UNet} 	& initial & 0.9889 	& 0.9705 	& 0.8356 	& 0.9837 & \multirow{2}*{\tablenum[table-number-alignment = center,table-format=2.2]{8.48}{\,\si{\percent}}} \\
			 			    & dark 	  & 0.9051 	& 0.9405	& 0.3683 	& 0.9954 &  \\				  			
\hline
\multirow{2}{4em}{SGL}		& initial & 0.9882 	& 0.9704 	& 0.8376 	& 0.9834 & \multirow{2}*{\tablenum[table-number-alignment = center,table-format=2.2]{27.95}{\,\si{\percent}}} \\
				  			& dark    & 0.7120 	& 0.9218 	& 0.1269 	& 0.9982 &  \\			 			    
\hline
\multirow{2}{4em}{RV-GAN}	& initial & 0.9864 	& 0.9650 	& 0.6373 	& 0.9967 & \multirow{2}*{\tablenum[table-number-alignment = center,table-format=2.2]{31.56}{\,\si{\percent}}} \\
			 				& dark	  & 0.6751 	& 0.9124 	& 0.0000 	& 1.0000 &  \\				  			
\hline
\end{tabular}
\label{tab:Exp:loclight:comp}
\end{table}

\FloatBarrier

\section{Conclusion}\label{sec:conclusion}

A framework named \textit{Logarithmic Mathematical Morphology} (LMM) has been presented. It allows to define Mathematical Morphology operations for images and functions with an upper bound value $M$ by using the Logarithmic Image Processing (LIP) vector space and its additive law $\LP$. The sum $f \LP b$ between two functions $f$ and $b$ with an upper bound value $M$ is smaller than this upper bound value. The amplitude of the sum $f \LP b$, i.e. the structuring function $b$ which is translated by the image $f$, varies according to the intensity of the function $f$ and in a way which is physically justified. Such a physical property comes from the LIP model which is defined thanks to the transmittance law and which is coherent with the human vision. The LMM framework allows the definition of morphological operators which are robust to lighting variations modelled by the LIP-additive law $\LP$. Those variations correspond to a change of light intensity or of camera exposure-time. Experiments have shown that those operators are robust to such uniform lighting variations and perform better than usual morphological operations defined with the usual additive law $+$. With such non-uniform lighting variations, a LMM approach for vessel segmentation in eye-fundus images is more robust than three state-of-the-art methods based on deep-learning, namely FR-UNet, SGL and RVGAN. LMM framework paves the way for the definition of morphological operators and neural nets \cite{Noyel2022} allowing a robust analysis of images acquired in uncontrolled lighting variations. Such variations occur in numerous practical applications (outdoor scenes, industry, medicine, remote-sensing, etc.). This paper does not study other types of lighting variation caused by factors other than changes in light intensity or camera exposure time. However, with lighting variations caused by changes in opacity or absorption and modelled by the LIP-multiplicative law \cite{Jourlin2014,Jourlin2016}, new morphological operators can be defined using this law. This would require further study.

\backmatter

%

\bmhead{Acknowledgements}

The author sincerely thanks Prof. Michel Jourlin for his careful rereading of the manuscript and the reviewers for their constructive comments, which led to the logarithmic mathematical morphology framework being centred on the isomorphism ($\xi$) and to a simplified presentation of the proofs based on properties \LIPPref{LIP:P1}--\LIPPref{LIP:P9}.

\section*{Declarations}

\begin{itemize}
\item Funding
No funding was received for conducting this study.

\item Conflict of interest/Competing interests 
The author has no competing interests to declare that are relevant to the content of this article.

\item Ethics approval and consent to participate
All medical data are coming from publicly available databases.

\item Consent for publication
The author gives his consent for publication.

\item Data availability 
The data are coming from publicly available databases.

\item Materials availability
Not applicable.

\item Code availability 
The code is publicly available at the following address:\\ \url{https://github.com/guillaumenoyel/LogMM}.

\item Author contribution
The unique author has contributed to everything in this manuscript: Conceptualisation; Methodology; Formal analysis and investigation; Writing - original draft preparation; Writing - review and editing.
\end{itemize}

\begin{appendices}

\section{Mathematical Morphology (p. \pageref{sssec:back:MM})}
\label{sec:app:MM}

\subsection{Lattice of functions with real value}
\label{ssec:app:MM:real_fct}

The set of functions $\Realb^D$ defined on a domain $D$ with values in $\Realb$ is a complete lattice with the order $\leq$. The infimum and the supremum are defined for any family $\mathscr{X} \subset \Realb^D$ by $\left(\wedge \mathscr{X}\right)(x) = \wedge_{\Realb} \left\{ f(x):f \in \mathscr{X}, \> x\in D \right\}$ and $\left(\vee \mathscr{X}\right)(x) = \vee_{\Realb} \left\{ f(x):f \in \mathscr{X}, \> x\in D  \right\}$, respectively. The least and greatest elements, $O$ and $I$, are the constant functions equal to $O(x)=-\infty$ and $I(x)=+\infty$, for all $x \in D$, respectively.

\subsection{Details on Definition~\ref{pre:def_dilation_erosion} about erosion and dilation}
\label{ssec:app:MM:def_ero_dil}

\begin{remark}
In Definition~\ref{def:erosion_dilation}, the same symbol $\wedge$ is used for the supremum of both $\Lscr_1$ and $\Lscr_2$. The same applies to the infimum, represented by the symbol $\vee$.
\end{remark}
The definitions of these mappings apply even to the empty subset $\emptyset$ of $\Lscr_1$ or $\Lscr_2$ because of the relations $O = \vee \emptyset$ and $I = \wedge \emptyset$, where $O$ and $I$ are the least and greatest elements, respectively \cite{Heijmans1990}. We have therefore: $\varepsilon(I)=I$ and $\delta(O)=O$.

%
%

\section{Detailed remark \ref{rem:link_mult_add_morpho} (p.~\pageref{rem:link_mult_add_morpho})}
\label{sec:app:detailed_remark_log_iso}

Remark \ref{rem:link_mult_add_morpho} concerning the logarithmic mapping of grey values between morphological adjunctions is detailed below.

Let us recall the definition of the erosion and the dilation with a
multiplicative structuring function \cite{Heijmans1990,Heijmans1994}.
Let $f \in (\overline{\Real}^{+})^{D}$ be a function, and $b \in (\overline{\Real}^{+})^{D_b}$ a ``multiplicative'' structuring function. The dilation $\dot{\oplus} b$ and the erosion $\dot{\ominus} b$ with a multiplicative structuring function $b$ are defined as follows:
\begin{align}
 \left(f \dot{\oplus} b\right) (x) &= \vee_{h \in D_b} \left\{ f(x - h) . b(h) \right\}  \label{eq:dilation_funct_mult}\\
\left(f \dot{\ominus} b\right) (x) &= \wedge_{h \in D_b} \left\{ f(x + h) / b(h) \right\}, \label{eq:erosion_funct_mult}%
\end{align} with the convention that $f(x-h).b(h)=0$, when $f(x-h)=0$ or $b(h) =0$ and that $f(x+h)/b(h)= + \infty$, when $f(x+h) = + \infty$ or $b(h) = 0$. The symbols $\dot{\oplus}$ and $\dot{\ominus}$ represent the extension to multiplicative structuring functions of Minkowski operations between sets \citep{Serra1982}.
The erosion and dilation with a multiplicative structuring function are respectively linked to the erosion \eqref{eq:erode_funct} and dilation \eqref{eq:dilate_funct} with and additive structuring function by the following relations \citep{Heijmans1990,Heijmans1994}:
\begin{align*}
	f \dot{\oplus} b  &=  \exp{ \left(\ln f \oplus \ln b \right)},\\ 
	f \dot{\ominus} b &=  \exp{ \left(\ln f \ominus \ln b \right)}.
\end{align*}
In this case, the isomorphism is the usual logarithm $\ln : (\overline{\Real}^{+})^{D} \mapsto \overline{\Real}^{D} $ and its inverse, the exponential function $\exp : \overline{\Real}^{D} \mapsto (\overline{\Real}^{+})^{D}$.

The dilation $\dot{\oplus} b$ in \eqref{eq:dilation_funct_mult}, can be expressed as a supremum of conjunctions, which can be for instance a product. This is a link with fuzzy sets \citep[Chap. 6]{Najman2013}.

%
%





%
%

\section{Map of Asplund distances: link between double probing and logarithmic-opening and closing (p.~\pageref{link_double_prob_LIPopen_close})}
\label{app:link_double_prob_LIPopen_close}

%

Let us study the location of the supremum or infimum of the translated probes $b \in \Fcurvb_M$ such that they are in contact with the function $f$ from below or from above, respectively.
In a similar way to that used by Heijmans et al. in \cite{Heijmans1990}, the mapping 
$\tau_{h,v}$ of horizontal and vertical translations on $\Fcurvb_M$ is defined by $(\tau_{h,v}(f))(x) = f(x-h) \LP v$, where $h \in D$ and $v \in \left]-\infty,M\right[$\>. 
For the lower probes, the supremum of the translated probes $b$ which are in contact with the function $f$ from below is the supremum of the translated probes $\tau_{h,v}(b)$ which are less or equal to the function $f$. It is equal to $\Gamma^{\LP}_b(f) = \sup{\{ \tau_{h,v}(b)\mid h \in D, v \in \left]-\infty,M\right[, \tau_{h,v}(b) \leq f \}}$. As demonstrated in \cite{Ronse1991}, $\Gamma^{\LP}_b(f)$ is an opening. 

Similarly, for the upper probes, the infimum of the translated probes $b$ which are in contact with the function $f$ from above is equal to\linebreak $\Phi^{\LP}_b(f) = \inf{\{ \tau_{h,v}(b)\mid h \in D, v \in \left]-\infty,M\right[, \tau_{h,v}(b) \geq f \}}$. As demonstrated in \cite{Ronse1991}, $\Phi^{\LP}_b(f)$ is a closing. 

\begin{proposition}
	The supremum of the translated probes $b$ which are in contact with the function $f$ from below, $\Gamma^{\LP}_b(f)$, is equal to the logarithmic-opening of $f$, $\gamma_{b}^{\LP} (f)$:
\begin{align}
	\Gamma_{b}^{\LP} (f)(x) &= \gamma_{b}^{\LP} (f)(x). \label{eq:probe_ctc_above}
\end{align}

The infimum of the translated probes $b$ which are in contact with the function $f$ from above, $\Phi_{b}^{\LP} (f)$, is equal to the logarithmic-closing of $f$, $\varphi_{\LM \bar{b}}^{\LP} (f)$:
\begin{align}
	\Phi_{b}^{\LP} (f)(x) &=  \varphi_{\LM \bar{b}}^{\LP} (f)(x). \label{eq:probe_ctc_below}
\end{align}

	\label{prop:contact_probes}
\end{proposition}

\begin{figure}[!t]
\centering
\subfloat[Image and probe]{\includegraphics[width=0.49\mycolumnwidth]{Signal_and_probe_mapAspAdd_thin_curve}%
\label{fig:signal_mapAspAdd_2:sgn}}
\hfil
\subfloat[Maps]{\includegraphics[width=0.49\mycolumnwidth]{Signal_and_mapAspAdd_thin_curve}%
\label{fig:signal_mapAspAdd_2:map}}
\\
\subfloat[Logarithmic-opening]{\includegraphics[width=0.49\mycolumnwidth]{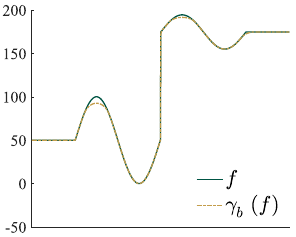}%
\label{fig:signal_mapAspAdd_2:open}}
\hfil
\subfloat[Logarithmic-closing]{\includegraphics[width=0.49\mycolumnwidth]{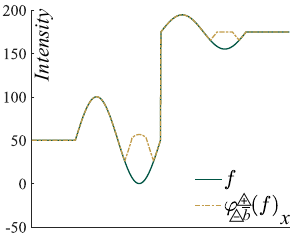}%
\label{fig:signal_mapAspAdd_2:close}}
\caption{(a) An image $f$ is analysed by a probe $b$ from above and below. (b) The \textit{mlub} $c_{1_b} (f)$, the \textit{mglb} $c_{2_b} (f)$ and the map of Asplund distances $Asp_b^{\protect \LP} (f)$ between the image and the probe. (c) The supremum of the probe $b$ such that the probe is in contact with the function $f$ from below is the logarithmic-opening $\gamma_{b}^{\protect \LP}$ of the function $f$. (d) The infimum of the probe $b$ such that the probe is in contact with the function $f$ from above is the logarithmic-closing $\varphi_{\protect \LM \bar{b}}^{\protect \LP} (f)$ of the function $f$ by the structuring function $\protect \LM \bar{b}$.}
\label{fig:signal_mapAspAdd_2}
\end{figure}

\figurename~\ref{fig:signal_mapAspAdd_2:open} and \ref{fig:signal_mapAspAdd_2:close} illustrate the supremum or infimum of the translated probes $b$ which are in contact with the function $f$ from below or from above, respectively. In \figurename~\ref{fig:signal_mapAspAdd_2:open}, one can notice that
the opening $\gamma_{b}^{\LP} (f)$ of $f$ is very close the image $f$. In \figurename~\ref{fig:signal_mapAspAdd_2:close}, the closing $\varphi_{\LM \bar{b}}^{\LP} (f)$ of $f$ is similar to the image $f$ with the exception of both valleys of $f$, where the structuring function $\LM \bar{b}$ is very dissimilar to $f$.  

\begin{proof}[Proof of proposition \ref{prop:contact_probes}]

Let $f$ and $b$ be a function and a probe of $\Fcurvb_M$, respectively. For $h \in D$ and $v \in -]\infty,M[$\>, the mapping $\tau_{h,v}$ of horizontal and vertical translations on $\Fcurvb_M$ is defined by $(\tau_{h,v}(f))(x) = f(x-h) \LP v$.
The supremum of the translated probes $b$ which are in contact with the function $f$ from below is equal to:
\begin{align*}
	\Gamma_{b}^{\LP} (f)(x) &= \sup{\{ \tau_{h,v}(b)(x)\mid h \in D, v \in \left]-\infty,M\right[, 
	\tau_{h,v}(b) \leq f \}}\\
	&= \sup{\{ b(x-h) \LP v\mid h \in D, v \in \left]-\infty,M\right[, 
	x' \in D,\> b(x'-h) \LP v \leq f(x') \}} \\
	&= \vee_{h \in D}{ \{b(x-h) \LP v(h)\mid v(h) =
	\sup_{x' \in D}{ \{v , b(x'-h) \LP v \leq f(x') \}} \} }  \\
	&=  \vee_{h \in D}{ \{b(x-h) \LP v(h)\mid v(h) = 
	\sup_{x' \in D}{ \{v , b(x') \LP v \leq f(h+x') \}} \} }\\
	&=  \vee_{h \in D}{ \{b(x-h) \LP c_{2_b} f(h) \}}\quad  \text{(from \eqref{eq:lower_map_add_2})}\\
	&=  \vee_{h \in D}{ \{b(x-h) \LP \varepsilon_b^{\LP}(f)(h) \}} \quad \text{(from Prop. \ref{prop:link_AsAdd_LMM})} \\
	&=  \vee_{h \in D}{ \{\varepsilon_b^{\LP}(f)(h) \LP b(x-h)\}}   \qquad \LIPPref{LIP:P2}\\
	&=  \vee_{h \in D}{ \{\varepsilon_b^{\LP}(f)(x-h) \LP b(h)\}}  \\
	&=  \delta_{b}^{\LP} {\varepsilon_b^{\LP}(f)}(x) \quad \text{(from \eqref{eq:LIP-dilation})}\\
	&=  \gamma_{b}^{\LP} (f)(x). \quad \text{(from \eqref{eq:LIP-opening})}
\end{align*}
Such a relation is an opening one.

The infimum of the translated probes $b$ which are in contact with the function $f$ from above is equal to:\label{demo:eq:probe_ctc_below}
\begin{align*}
	\Phi_{b}^{\LP} (f)(x) &=  \inf{\{ \tau_{h,v}(b)(x)\mid h \in D, v \in \left]-\infty,M\right[, 
	\tau_{h,v}(b) \geq f \}}\\
	&= \wedge_{h \in D_b}{ \{b(x-h) \LP v(h)\mid v(h) = 
	\inf_{x' \in D_b}{ \{v , b(x'-h) \LP v \geq f(x') \}} \} }\\
	&=  \wedge_{h \in D_b}{ \{b(x-h) \LP c_{1_b} f(h) \}} \quad  \text{(from \eqref{eq:upper_map_add_2})}  \\
	&=  \wedge_{h \in D_b}{ \{b(x-h) \LP \delta_{\LM \overline{b}}^{\LP} (f)(h) \}} \quad \text{(from Prop. \ref{prop:link_AsAdd_LMM})} \\
	&=  \wedge_{-h \in D_b}{ \{\delta_{\LM \overline{b}}^{\LP} (f)(x+h) \LP b(-h)\}}  \qquad \LIPPref{LIP:P2}\\
	&=  \wedge_{h \in D_{\bar{b}}}{ \{\delta_{\LM \overline{b}}^{\LP} (f)(x+h) \LM (\LM \bar{b}(h))\}}   \qquad \LIPPref{LIP:P1}\\
	&=  \varepsilon_{\LM \bar{b}}^{\LP} {\delta_{\LM \bar{b}}^{\LP} (f)}(x) \quad \text{(from \eqref{eq:LIP-erosion})}\\
	&=  \varphi_{\LM \bar{b}}^{\LP} (f)(x) \quad \text{(from \eqref{eq:LIP-closing})}.
\end{align*}
Such a relation is a closing one. 
\end{proof}

\section{Proofs of properties \ref{property:lr_detectors_LMM} (p.~\pageref{property:lr_detectors_LMM}) and \ref{property:lr_detectors_rob_LIPadd} (p.~\pageref{property:lr_detectors_rob_LIPadd})}
\label{app:property:lr_detectors_LMM_property:lr_detectors_rob_LIPadd}

\begin{proof}[Proof of property \ref{property:lr_detectors_LMM}]
Let $f \in \Fcurvb_M$ be a function and $b^l$, $b^r \in \Fcurvb_M$ be structuring functions.
From~\eqref{eq:LIP-erosion}, we have $\forall x \in D$, $\wedge_{h \in D_b}{ \{ f(x+h) \LM b^l(h) \} } = \varepsilon_{b^l}^{\LP}(f)(x) $ and from~\eqref{eq:mglb_LMM}, $c_{2_b}(f) = \varepsilon_b^{\LP}(f)$.
From~\eqref{eq:LAC_er_left}, we deduce that:
\begin{align*}
	E (b^l, f)(x) &= \wedge_{h \in D_b}{ \{ f(x+h) \LM b^l(h) \} } \LM c_{2_b}(f)(x)\\
	&= \varepsilon_{b^l}^{\LP}(f)(x) \LM \varepsilon_b^{\LP}(f)(x)\text{.}
\end{align*}
In a similar way, we have $E (b^r, f)(x) = \varepsilon_{b^r}^{\LP}(f)(x) \LM \varepsilon_b^{\LP}(f)(x)$.
\end{proof}

\begin{proof}[Proof of property \ref{property:lr_detectors_rob_LIPadd}]\label{proof:property:lr_detectors_rob_LIPadd}
From~\eqref{eq:LIP-erosion}, we have $\forall c \in \left]-\infty,M\right[$ and $\forall x \in D$:
\begin{align}
	\varepsilon_{b}^{\LP}(f\LP c)(x) 
	&= \xi^{-1} \left( \varepsilon_{\xi(b)} \left[\xi(f \LP c)\right] \right) \qquad \text{(from~\eqref{eq:LIP_ero_prop})}\nonumber\\
	&= \xi^{-1} \left( \varepsilon_{\xi(b)} \left[\xi(f) + \xi(c)\right] \right) \qquad \text{$\xi$ is an isomorphism}\nonumber\\
	&= \xi^{-1} \left( \varepsilon_{\xi(b)} \left[\xi(f)\right] + \xi(c) \right) \qquad \text{$\varepsilon_{\xi(b)}$ is invariant to a constant addition}\nonumber\\
	&= \varepsilon_{b}^{\LP}(f)(x) \LP c \qquad \text{(from~\eqref{eq:LIP_ero_prop})}\text{.}\label{eq:LIPerosion_linearity}
\end{align}

From~\eqref{eq:LIP:LMM:LAC_er_left2}, one deduces that:
\begin{align*}
	E (b^l, f \LP c) &= \varepsilon_{b^l}^{\LP}(f\LP c) \LM \varepsilon_b^{\LP}(f \LP c)\\
	&= \left(\varepsilon_{b^l}^{\LP}(f) \LP c\right) \LM \left(\varepsilon_b^{\LP}(f)\LP c\right)\\
	&= \varepsilon_{b^l}^{\LP}(f) \LM \varepsilon_b^{\LP}(f)  \qquad \LIPPref{LIP:P7}\\
	&= E (b^l, f).
\end{align*}
In a similar way, we obtain $E (b^r, f \LP c) = E (b^r, f)$.
As a result, from~\eqref{eq:LIP:LMM:detector_one_dir}, we have\newline $E (b, f \LP c) = E (b, f)$.
\end{proof}

%
%

\section{Proof of property \ref{property:LIPdiff_LIPopenings_rob_LIPadd} (p.~\pageref{property:LIPdiff_LIPopenings_rob_LIPadd})}
\label{app:property:LIPdiff_LIPopenings_rob_LIPadd_property:LIPdiff_LIPopenings_rob_LIPadd}

\begin{proof}[Proof of property \ref{property:LIPdiff_LIPopenings_rob_LIPadd}]\label{proof:property:LIPdiff_LIPopenings_rob_LIPadd}
Let $f \in \Fcurvb_M$ be a function and $b$, $b_r \in \Fcurvb_M$ be structuring functions.
Similarly as in~\eqref{eq:LIPerosion_linearity}, we have $\forall c \in \left]-\infty,M\right[$ and $\forall x \in D$:
\begin{align}
	\delta_{b}^{\LP}(f\LP c)(x)
	&= \delta_{b}^{\LP}(f)(x) \LP c \text{.}\label{eq:LIPdilation_linearity}
\end{align}

From \eqref{eq:LIP-opening}, \eqref{eq:LIPerosion_linearity} and \eqref{eq:LIPdilation_linearity}, we have $\forall c \in \left]-\infty,M\right[\>$:
\begin{align}
	\gamma_{b}^{\LP}(f \LP c) &= \delta_b^{\LP} \left[ \varepsilon_b^{\LP}(f \LP c)\right]\nonumber\\
	&= \delta_b^{\LP} \left[\varepsilon_b^{\LP}(f)\right] \LP c \nonumber\\
	&= \gamma_{b}^{\LP}(f) \LP c \text{.}\label{eq:LIPopening_linearity}
\end{align}
From \eqref{eq:LIPdiff_LIPopenings}, one deduces $\forall c \in \left]-\infty,M\right[$ that:
\begin{align*}
	G^{\LP}_b (f \LP c) &= \gamma_{b}^{\LP}(f \LP c) \LM \gamma_{b_r}^{\LP}(f \LP c)\\
	&= \left[\gamma_{b}^{\LP}(f) \LP c\right] \LM \left[\gamma_{b_r}^{\LP}(f) \LP c\right]\\
	&= \gamma_{b}^{\LP}(f) \LM \gamma_{b_r}^{\LP}(f) \qquad \LIPPref{LIP:P7}\\
	&= G_{b}^{\LP}(f) \text{.}
\end{align*}
\end{proof}

%
%

\section{Proof of property \ref{property:LIPresidue_LIPopening_rob_LIPadd} (p.~\pageref{property:LIPresidue_LIPopening_rob_LIPadd})}
\label{app:property:LIPresidue_LIPopening_rob_LIPadd}

\begin{proof}[Proof of property \ref{property:LIPresidue_LIPopening_rob_LIPadd}]
Let $f \in \Fcurvb_M$ be a function and $b$, $b_r \in \Fcurvb_M$ be structuring functions.
From \eqref{eq:LIP:LMM:res_LIPopening} and \eqref{eq:LIPopening_linearity}, we have $\forall c \in \left]-\infty,M\right[\>$:
\begin{align*}
	R^{\LP}_b(f \LP c) &= \left[f \LP c\right] \LM \gamma^{\LP}_b(f \LP c)\\
	&= \left[f \LP c\right] \LM \left[\gamma^{\LP}_b(f) \LP c\right]\\
	&= f \LM \gamma^{\LP}_b(f) \qquad \LIPPref{LIP:P7}\\
	&= R^{\LP}_b(f) \text{.}
\end{align*}
\end{proof}

%
%

\section{Proof of proposition \ref{prop:link_AsAddtol_LMM} (p.~\pageref{prop:link_AsAddtol_LMM})}
\label{app:prop:link_AsAddtol_LMM}

\begin{proof}[Proof of proposition \ref{prop:link_AsAddtol_LMM}]
First of all, let us remind the definition of the LIP-additive Asplund metric with tolerance which was introduced in \cite{Noyel2019b}.

\begin{definition}[LIP-additive Asplund metric with tolerance]
Let $(1-p)$ be a percentage of points of $D$ to be discarded and $D'$ the set of these discarded points. The LIP-additive Asplund metric with tolerance between two functions $f$ and $g \in \Fcurv_M$ is defined by:
\begin{align}
d^{\LP}_{asp,p} (f,g) &= c_{1,p} \LM c_{2,p}. \label{eq:dAsAdd_tol}%
\end{align}
The constants $c_{1,p}$ and $c_{2,p}$ are equal to: 
\begin{align*}
	c_{1,p} &= \inf \{c, \forall x \in D, \gamma^{\LP}_{(f_{|D \setminus D'},g_{|D \setminus D'})}(x) \leq c \},\\
	c_{2,p} &= \sup \{ c , \forall x \in D, c \leq \gamma^{\LP}_{(f_{|D \setminus D'},g_{|D \setminus D'})}(x) \}. 
\end{align*}
A percentage $(1-p)/2$ of the points $x\in D$ with the greatest, respectively lowest contrast values $\gamma^{\LP}_{(f,g)}(x) = f(x) \LM g(x)$ are discarded.
\label{def:dAsAdd_tol}
\end{definition}

\begin{remark}
In practice, for a function $f : D \mapsto \left]\infty,M\right[$ defined on a discrete grid $D$, e.g. $D \subset \Zint^n$, the number of points to be suppressed are selected as follows. 
Let $\#D$ be the cardinal of $D$.
The number of points $n_{supr}$ to be suppressed is equal to $n_{suppr} = round[(1-p)\#D]$, where $round$ is the rounding operator of any real number to its nearest integer. For the constant $c_{1,p}$ and $c_{2,p}$, the number of points to be suppressed are respectively equal to:
\begin{align}
n_{1} &= round(n_{suppr}/2),\label{eq:n1_mapAsp}\\
n_{2} &= n_{suppr} - n_{1},\label{eq:n2_mapAsp}
\end{align}
\end{remark}

The map of LIP-additive Asplund distances with tolerance (to noise extrema) is defined as follows.

\begin{definition}[Map of LIP-additive Asplund distances with tolerance \cite{Noyel2019b}]
Let  $f \in \Fcurv_M$  be a grey-level image and $b : D_{b} \mapsto \left]-\infty,M\right[$ a probe. Let $(1-p)$ be a percentage of points of $D_b$ to be discarded. The map of Asplund distances $Asp_{b}^{\LP}: \Fcurv_M \mapsto \Fcurv_M$ with a tolerance (to extrema) $p$ is defined by:
\begin{align}
Asp_{b,p}^{\LP}(f)(x) &= d^{\LP}_{asp,p} (f_{|D_b(x)},b).\label{eq:map_LIPaddAspdisttol}
\end{align}
\label{def:map_LIPaddAspdisttol}
\end{definition}
For each point $x \in D$, the distance $d^{\LP}_{asp,p} (f_{\left|D_b(x)\right.},b)$ is computed in the neighbourhood $D_b(x)$ centred in $x$ and the template $b$ is acting like a structuring function. $f_{\left|D_b(x)\right.}$ is the restriction of $f$ to $D_b(x)$.  $Asp_{b,p}^{\LP} (f): D \mapsto \Fcurv_M$ is the map of Asplund distances between the image $f$ and the probe $b$.

In order to establish the link with MM, we will express the map of Asplund distances with neighbourhood operations.
From \eqref{eq:dAsAdd_tol}, for each $x\in D$, the map expression becomes 
\begin{align*}
d^{\LP}_{asp,p} (f_{\left|D_b(x)\right.},b) &= c_{1_b,p}(f)(x) \LIPminus c_{2_b,p}(f)(x),
\end{align*}
where 
\begin{align}
	c_{1_b,p}(f)(x) &=  \inf \left\{c, \forall h \in D_b, 
					\quad \gamma^{\LP}_{(f_{|D_b(x) \setminus D'_b(x)},b_{|D_b \setminus D'_b(x)})}(h) \leq c \right\} ,\quad \label{eq:upper_map_add_tol}\\
	c_{2_b,p} (f)(x) &= \sup \left\{ c , \forall h \in D_b, 
					\quad c \leq \gamma^{\LP}_{(f_{|D_b(x) \setminus D'_b(x)},b_{|D_b \setminus D'_b(x)})}(h) \right\}.\quad\label{eq:lower_map_add_tol}%
\end{align}
This leads to the following definition.

\begin{definition}[LIP-additive maps of the least upper and of the greatest lower bounds with tolerance]
Let $f \in \Fcurvb_M$ be a function and $b : D_{b} \mapsto \left]\infty,M\right[$ a probe.
For any $x \in D$, let $(1-p)$ be a percentage of points of the neighbourhood $D_b(x)$ to be discarded and $D'_b(x)$ the set of these discarded points. The map of the least upper bounds (mlub) $c_{1_b,p}: \Fcurvb_M  \mapsto \Fcurvb_M$ and the map of the greatest lower bounds (mglb) $c_{2_{b,p}}: \Fcurvb_M \mapsto \Fcurvb_M$, of $f$ by $b$ are defined in \eqref{eq:upper_map_add_tol} and \eqref{eq:lower_map_add_tol}, respectively. 
\end{definition}

The number of points to be suppressed $n_1$ and $n_2$ for the mlub $c_{1_b,p}$ and for the mglb $c_{2_b,p}$ are equal to
$n_1 = round( n_{suppr} /2)$, and $n_2 = n_{suppr} - n_1$, respectively, where $n_{suppr} = round[(1-p)\#D_b]$ and $D_b$ is the cardinal of $\#D_b$.

The $k^{th}$ LIP-minimum-rank filter $\zeta_{b,k}^{\LP}: \Fcurvb_M^D \mapsto \Fcurvb_M^D$ and the $k^{th}$ LIP-maximum-rank filter $\vartheta_{b,k}^{\LP}: \Fcurvb_M^D \mapsto \Fcurvb_M^D$ are defined in \eqref{eq:_LIP_erotol} and in \eqref{eq:_LIP_diltol}, respectively.
From \eqref{eq:upper_map_add_tol}, one deduces, for any $x \in D$, that:
\begin{align*}
	c_{1_b,p}(f)(x) &= \bigvee \left\{ \gamma^{\LP}_{(f_{|D_b(x) \setminus D'_b(x)},b_{|D_b \setminus D'_b(x)})}(h) , h \in D_b \right\}\\
	&= \bigvee^{n_1} \left\{ \gamma^{\LP}_{(f_{|D_b(x)},b)}(h), h \in D_b \right\}\\
	&= \bigvee^{n_1} \left\{ f(x+h) \LM b(h) , h \in D_b \right\}\\
	&= \bigvee^{n_1} \left\{ f(x-h) \LM b(-h) , -h \in D_b \right\} \\
	&= \bigvee^{n_1} \left\{ f(x-h) \LP (\LM \overline{b}(h)) , h \in D_{\overline{b}} \right\}  \qquad \LIPPref{LIP:P5}\\
	&= \vartheta_{\LM \overline{b},n_1}^{\LP} (f)(x)\hspace{2cm}\text{(from \eqref{eq:_LIP_diltol})}.
\end{align*}
Similarly, from \eqref{eq:lower_map_add_tol}, one deduces, for any $x \in D$, that:
\begin{align*}
	c_{2_b,p} (f)(x) &= \bigwedge \left\{ \gamma^{\LP}_{(f_{|D_b(x) \setminus D'_b(x)},b_{|D_b \setminus D'_b(x)})}(h) , h \in D_b \right\}\\
	&= \bigwedge^{n_2} \left\{ f(x+h) \LM b(h) , h \in D_b \right\}\\
	&= \zeta_{b,n_2}^{\LP}(f)(x)\hspace{2cm}\text{(from \eqref{eq:_LIP_erotol})}.
\end{align*}

As a consequence, proposition \ref{prop:link_AsAddtol_LMM} (p.~\pageref{prop:link_AsAddtol_LMM}) holds.

\end{proof}

%
%

\section{Proof of equation~\eqref{eq:LMM_vessels_rob_LIPadd} (p.~\pageref{eq:LMM_vessels_rob_LIPadd})}
\label{app:eq:LMM_vessels_rob_LIPadd}

\begin{proof}[Proof of equation~\eqref{eq:LMM_vessels_rob_LIPadd}]

Let $f \in \Fcurvb_M$ be a function and $b$, $b_{\theta}$, $b^l_{\theta}$, $b^c_{\theta}$ $b^r_{\theta} \in \Fcurvb_M$ be structuring functions.

From~\eqref{eq:_erotol}, we have $\forall c \in \left]-\infty,M\right[$, $\zeta_{b,k}(f+ c) = \zeta_{b,k}(f) + c$.  By the isomorphism $\xi$, from~\eqref{eq:_LIP_erotol}, we have:
\begin{align}
	\zeta_{b,k}^{\LP}(f\LP c) &= \zeta_{b,k}^{\LP}(f) \LP c \text{.}\label{eq:LIPrankerosion_linearity}
\end{align}


From~\eqref{eq:LIPerosion_linearity},~\eqref{eq:LIPrankerosion_linearity}, one deduces that:
\begin{align}
	\grave{c}_{b_{\theta},k}(f \LP c) 
	&= \bigwedge{\left\{\>  \varepsilon^{\LP}_{b^c_{\theta}}(f \LP c) , 
	\bigwedge{[ \zeta^{\LP}_{b^l_{\theta},k}(f \LP c) , \zeta^{\LP}_{b^r_{\theta},k}(f \LP c) ]} \>\right\} } &\nonumber\\
	&= \bigwedge{\left\{\>  \varepsilon^{\LP}_{b^c_{\theta}}(f)\LP c , 
	\bigwedge{[ \zeta^{\LP}_{b^l_{\theta},k}(f)\LP c , \zeta^{\LP}_{b^r_{\theta},k}(f)\LP c ]} \>\right\} } &\nonumber\\
	&= \bigwedge{\left\{\>  \varepsilon^{\LP}_{b^c_{\theta}}(f)\LP c , 
	\bigwedge{[ \zeta^{\LP}_{b^l_{\theta},k}(f) , \zeta^{\LP}_{b^r_{\theta},k}(f) ] \LP c } \>\right\} } & \LIPPref{LIP:P10} \nonumber\\		
	&= \bigwedge{\left\{\>  \varepsilon^{\LP}_{b^c_{\theta}}(f) , 
	\bigwedge{[ \zeta^{\LP}_{b^l_{\theta},k}(f) , \zeta^{\LP}_{b^r_{\theta},k}(f) ] } \>\right\}  \LP c }.  &\label{eq:LIP:seg:yeux:er_tol_3seg_linearity}		
\end{align}  

From~\eqref{eq:LIP:seg:yeux:map_LAC_er_left2}, \eqref{eq:LIPrankerosion_linearity} and \eqref{eq:LIP:seg:yeux:er_tol_3seg_linearity}, we have:
\begin{align}
	E^k(b^l_{\theta}, f \LP c)
	&= \zeta^{\LP}_{b^l_{\theta},k} (f \LP c) \LM \grave{c}_{\protect b_{\theta},k}(f \LP c) \nonumber\\
	&= \left[\zeta^{\LP}_{b^l_{\theta},k} (f) \LP c \right] \LM \left[\grave{c}_{\protect b_{\theta},k}(f) \LP c \right] \nonumber\\
	&= \zeta^{\LP}_{b^l_{\theta},k} (f) \LM \grave{c}_{\protect b_{\theta},k}(f)  \qquad \LIPPref{LIP:P7}\nonumber\\
	&= E^k(b^l_{\theta}, f ).\label{eq:LIP:seg:yeux:map_LAC_er_left_linearity}
\end{align}  
Similarly, from~\eqref{eq:LIP:seg:yeux:map_LAC_er_right2}, we have:
\begin{align}	
	E^k(b^r_{\theta}, f \LP c) &= E^k(b^r_{\theta}, f ). \label{eq:LIP:seg:yeux:map_LAC_er_right_linearity}	
\end{align}                                                                                                              


From~\eqref{eq:LIP:seg:yeux:detector_one_or}, \eqref{eq:LIP:seg:yeux:map_LAC_er_left_linearity} and \eqref{eq:LIP:seg:yeux:map_LAC_er_right_linearity}, we have:
\begin{align}
	E^k (b_{\theta}, f \LP c) &= \bigvee{\{ E^k (b^l_{\theta} , f \LP c) , E^k (b^r_{\theta}, f \LP c) \}}\nonumber\\
	 &= \bigvee{\{ E^k (b^l_{\theta} , f ) , E^k (b^r_{\theta}, f ) \}}.\nonumber\\ 
	 &= E^k (b_{\theta}, f)\label{eq:LIP:seg:yeux:detector_one_or_linearity}
\end{align}
As a consequence from~\eqref{eq:LIP:seg:yeux:detector_tol_1probe}, we have:
\begin{align}
	E^k (b, f \LP c) &= \bigwedge{\{ E^k (b_{\theta}, f \LP c) \mid \theta \in \Theta \}}\nonumber\\
	&= \bigwedge{\{ E^k (b_{\theta}, f ) \mid \theta \in \Theta \}}\nonumber\\
	&= E^k (b, f).\label{eq:LIP:seg:yeux:detector_tol_1probe_linearity}
\end{align}

Finally, from \eqref{eq:LIP:seg:yeux:map_detector_tol_3probes}, we have :
\begin{align}
	e^k_{b}(f \LP c) &= \bigwedge{\{ E^k (b_i, f \LP c) \mid i \in [\![1 \ldots I]\!] \}}\nonumber\\
	&= \bigwedge{\{ E^k (b_i, f) \mid i \in [\![1 \ldots I]\!] \}}\nonumber\\
	&= e^k_{b}(f). \label{eq:LIP:seg:yeux:map_detector_tol_3probes_linearity}
\end{align} 
\end{proof}

%




\end{appendices}


\bibliography{refs}


\begin{thebibliography}{61}
\ifx \bisbn   \undefined \def \bisbn  #1{ISBN #1}\fi
\ifx \binits  \undefined \def \binits#1{#1}\fi
\ifx \bauthor  \undefined \def \bauthor#1{#1}\fi
\ifx \batitle  \undefined \def \batitle#1{#1}\fi
\ifx \bjtitle  \undefined \def \bjtitle#1{#1}\fi
\ifx \bvolume  \undefined \def \bvolume#1{\textbf{#1}}\fi
\ifx \byear  \undefined \def \byear#1{#1}\fi
\ifx \bissue  \undefined \def \bissue#1{#1}\fi
\ifx \bfpage  \undefined \def \bfpage#1{#1}\fi
\ifx \blpage  \undefined \def \blpage #1{#1}\fi
\ifx \burl  \undefined \def \burl#1{\textsf{#1}}\fi
\ifx \doiurl  \undefined \def \doiurl#1{\url{https://doi.org/#1}}\fi
\ifx \betal  \undefined \def \betal{\textit{et al.}}\fi
\ifx \binstitute  \undefined \def \binstitute#1{#1}\fi
\ifx \binstitutionaled  \undefined \def \binstitutionaled#1{#1}\fi
\ifx \bctitle  \undefined \def \bctitle#1{#1}\fi
\ifx \beditor  \undefined \def \beditor#1{#1}\fi
\ifx \bpublisher  \undefined \def \bpublisher#1{#1}\fi
\ifx \bbtitle  \undefined \def \bbtitle#1{#1}\fi
\ifx \bedition  \undefined \def \bedition#1{#1}\fi
\ifx \bseriesno  \undefined \def \bseriesno#1{#1}\fi
\ifx \blocation  \undefined \def \blocation#1{#1}\fi
\ifx \bsertitle  \undefined \def \bsertitle#1{#1}\fi
\ifx \bsnm \undefined \def \bsnm#1{#1}\fi
\ifx \bsuffix \undefined \def \bsuffix#1{#1}\fi
\ifx \bparticle \undefined \def \bparticle#1{#1}\fi
\ifx \barticle \undefined \def \barticle#1{#1}\fi
\bibcommenthead
\ifx \bconfdate \undefined \def \bconfdate #1{#1}\fi
\ifx \botherref \undefined \def \botherref #1{#1}\fi
\ifx \url \undefined \def \url#1{\textsf{#1}}\fi
\ifx \bchapter \undefined \def \bchapter#1{#1}\fi
\ifx \bbook \undefined \def \bbook#1{#1}\fi
\ifx \bcomment \undefined \def \bcomment#1{#1}\fi
\ifx \oauthor \undefined \def \oauthor#1{#1}\fi
\ifx \citeauthoryear \undefined \def \citeauthoryear#1{#1}\fi
\ifx \endbibitem  \undefined \def \endbibitem {}\fi
\ifx \bconflocation  \undefined \def \bconflocation#1{#1}\fi
\ifx \arxivurl  \undefined \def \arxivurl#1{\textsf{#1}}\fi
\csname PreBibitemsHook\endcsname

\bibitem[\protect\citeauthoryear{Matheron}{1975}]{Matheron1975}
\begin{bbook}
\bauthor{\bsnm{Matheron}, \binits{G.}}:
\bbtitle{Random Sets and Integral Geometry}.
\bsertitle{Wiley ser. in probability and math. statis.}
\bpublisher{Wiley},
\blocation{New York, NY, USA}
(\byear{1975})
\end{bbook}
\endbibitem

\bibitem[\protect\citeauthoryear{Serra}{1982}]{Serra1982}
\begin{bbook}
\bauthor{\bsnm{Serra}, \binits{J.}}:
\bbtitle{Image Analysis and Mathematical Morphology}
vol. \bseriesno{1}.
\bpublisher{Academic},
\blocation{Orlando, FL, USA}
(\byear{1982}).
\burl{\url{https://arks.org/ark:/13960/s2643qqkkmh}}
\end{bbook}
\endbibitem

\bibitem[\protect\citeauthoryear{{Sternberg}}{1979}]{Sternberg1979}
\begin{bchapter}
\bauthor{\bsnm{{Sternberg}}, \binits{S.R.}}:
\bctitle{Parallel architectures for image processing}.
In: \bbtitle{COMPSAC 79. Proc. Comput. Softw. and The IEEE Comput. Soc.'s Third
  Int. Appl. Conf.},
pp. \bfpage{712}--\blpage{717}
(\byear{1979}).
\doiurl{10.1109/CMPSAC.1979.762586}
\end{bchapter}
\endbibitem

\bibitem[\protect\citeauthoryear{Sternberg}{1986}]{Sternberg1986}
\begin{barticle}
\bauthor{\bsnm{Sternberg}, \binits{S.R.}}:
\batitle{Grayscale morphology}.
\bjtitle{Computer Vision, Graphics, and Image Processing}
\bvolume{35}(\bissue{3}),
\bfpage{333}--\blpage{355}
(\byear{1986})
\doiurl{10.1016/0734-189X(86)90004-6}
\end{barticle}
\endbibitem

\bibitem[\protect\citeauthoryear{{Maragos}}{1989}]{Maragos_1989}
\begin{barticle}
\bauthor{\bsnm{{Maragos}}, \binits{P.}}:
\batitle{A representation theory for morphological image and signal
  processing}.
\bjtitle{{IEEE} Trans. Pattern Anal. Mach. Intell.}
\bvolume{11}(\bissue{6}),
\bfpage{586}--\blpage{599}
(\byear{1989})
\doiurl{10.1109/34.24793}
\end{barticle}
\endbibitem

\bibitem[\protect\citeauthoryear{{Heijmans}}{1991}]{Heijmans1991}
\begin{barticle}
\bauthor{\bsnm{{Heijmans}}, \binits{H.J.A.M.}}:
\batitle{Theoretical aspects of gray-level morphology}.
\bjtitle{{IEEE} Trans. Pattern Anal. Mach. Intell.}
\bvolume{13}(\bissue{6}),
\bfpage{568}--\blpage{582}
(\byear{1991})
\doiurl{10.1109/34.87343}
\end{barticle}
\endbibitem

\bibitem[\protect\citeauthoryear{Heijmans and Ronse}{1990}]{Heijmans1990}
\begin{barticle}
\bauthor{\bsnm{Heijmans}, \binits{H.J.A.M.}},
\bauthor{\bsnm{Ronse}, \binits{C.}}:
\batitle{The algebraic basis of mathematical morphology {I}. {D}ilations and
  erosions}.
\bjtitle{Comput. Vision Graphics and Image Process.}
\bvolume{50}(\bissue{3}),
\bfpage{245}--\blpage{295}
(\byear{1990})
\doiurl{10.1016/0734-189X(90)90148-O}
\end{barticle}
\endbibitem

\bibitem[\protect\citeauthoryear{Heijmans}{1994}]{Heijmans1994}
\begin{bbook}
\bauthor{\bsnm{Heijmans}, \binits{H.J.A.M.}}:
\bbtitle{Morphological Image Operators}.
\bsertitle{Adv. Imag. Electron Phys.: Suppl.},
vol. \bseriesno{25}.
\bpublisher{Academic},
\blocation{San Diego, CA, USA}
(\byear{1994}).
\burl{\url{https://arks.org/ark:/13960/s2kr7r0gj4z}}
\end{bbook}
\endbibitem

\bibitem[\protect\citeauthoryear{Jourlin and Pinoli}{1988}]{Jourlin1988}
\begin{barticle}
\bauthor{\bsnm{Jourlin}, \binits{M.}},
\bauthor{\bsnm{Pinoli}, \binits{J.}}:
\batitle{A model for logarithmic image-processing}.
\bjtitle{J. Microsc.}
\bvolume{149}(\bissue{1}),
\bfpage{21}--\blpage{35}
(\byear{1988})
\doiurl{10.1111/j.1365-2818.1988.tb04559.x}
\end{barticle}
\endbibitem

\bibitem[\protect\citeauthoryear{Brailean et~al.}{1991}]{Brailean1991}
\begin{bchapter}
\bauthor{\bsnm{Brailean}, \binits{J.C.}},
\bauthor{\bsnm{Sullivan}, \binits{B.}},
\bauthor{\bsnm{Chen}, \binits{C.T.}},
\bauthor{\bsnm{Giger}, \binits{M.L.}}:
\bctitle{Evaluating the \uppercase{EM} algorithm for image processing using a
  human visual fidelity criterion}.
In: \bbtitle{IEEE Int. Conf. Acoustics, Speech, Signal Process.},
pp. \bfpage{2957}--\blpage{29604}
(\byear{1991}).
\doiurl{10.1109/ICASSP.1991.151023}
\end{bchapter}
\endbibitem

\bibitem[\protect\citeauthoryear{Sun et~al.}{2012}]{Sun2012}
\begin{barticle}
\bauthor{\bsnm{Sun}, \binits{J.Z.}},
\bauthor{\bsnm{Wang}, \binits{G.I.}},
\bauthor{\bsnm{Goyal}, \binits{V.K.}},
\bauthor{\bsnm{Varshney}, \binits{L.R.}}:
\batitle{A framework for bayesian optimality of psychophysical laws}.
\bjtitle{Journal Math Psychology}
\bvolume{56}(\bissue{6}),
\bfpage{495}--\blpage{501}
(\byear{2012})
\doiurl{10.1016/j.jmp.2012.08.002}
\end{barticle}
\endbibitem

\bibitem[\protect\citeauthoryear{Varshney and Sun}{2013}]{Varshney2013}
\begin{barticle}
\bauthor{\bsnm{Varshney}, \binits{L.R.}},
\bauthor{\bsnm{Sun}, \binits{J.Z.}}:
\batitle{Why do we perceive logarithmically?}
\bjtitle{Significance}
\bvolume{10}(\bissue{1}),
\bfpage{28}--\blpage{31}
(\byear{2013})
\doiurl{10.1111/j.1740-9713.2013.00636.x}
\end{barticle}
\endbibitem

\bibitem[\protect\citeauthoryear{Jourlin}{2016}]{Jourlin2016_chap1}
\begin{bchapter}
\bauthor{\bsnm{Jourlin}, \binits{M.}}:
\bctitle{{Chapter One - Gray-Level LIP Model. Notations, Recalls, and First
  Applications}}.
In: \beditor{\bsnm{Jourlin}, \binits{M.}} (ed.)
\bbtitle{Logarithmic Image Processing: Theory and Applications}.
\bsertitle{Adv. Imag. Electron Phys.},
vol. \bseriesno{195},
pp. \bfpage{1}--\blpage{26}.
\bpublisher{Elsevier},
\blocation{New York, NY, USA}
(\byear{2016}).
\doiurl{10.1016/bs.aiep.2016.04.001}
\end{bchapter}
\endbibitem

\bibitem[\protect\citeauthoryear{Land}{1977}]{Land1977}
\begin{barticle}
\bauthor{\bsnm{Land}, \binits{E.H.}}:
\batitle{The retinex theory of color vision}.
\bjtitle{Sci. Am.}
\bvolume{237}(\bissue{6}),
\bfpage{108}--\blpage{129}
(\byear{1977})
\end{barticle}
\endbibitem

\bibitem[\protect\citeauthoryear{Kimmel et~al.}{2003}]{Kimmel2003}
\begin{barticle}
\bauthor{\bsnm{Kimmel}, \binits{R.}},
\bauthor{\bsnm{Elad}, \binits{M.}},
\bauthor{\bsnm{Shaked}, \binits{D.}},
\bauthor{\bsnm{Keshet}, \binits{R.}},
\bauthor{\bsnm{Sobel}, \binits{I.}}:
\batitle{{A Variational Framework for Retinex}}.
\bjtitle{Int. J. Comput. Vision}
\bvolume{52}(\bissue{1}),
\bfpage{7}--\blpage{23}
(\byear{2003})
\doiurl{10.1023/A:1022314423998}
\end{barticle}
\endbibitem

\bibitem[\protect\citeauthoryear{Roerdink}{2009}]{Roerdink2009}
\begin{bchapter}
\bauthor{\bsnm{Roerdink}, \binits{J.B.T.M.}}:
\bctitle{Adaptivity and group invariance in mathematical morphology}.
In: \bbtitle{2009 IEEE Int. Conf. on Image Process.},
pp. \bfpage{2253}--\blpage{2256}
(\byear{2009}).
\doiurl{10.1109/ICIP.2009.5413983}
\end{bchapter}
\endbibitem

\bibitem[\protect\citeauthoryear{Jourlin and Pinoli}{2001}]{Jourlin2001}
\begin{botherref}
\oauthor{\bsnm{Jourlin}, \binits{M.}},
\oauthor{\bsnm{Pinoli}, \binits{J.C.}}:
Logarithmic image processing: The mathematical and physical framework for the
  representation and processing of transmitted images.
Adv. Imag. Electron Phys.,
vol. 115,
pp. 129--196.
Elsevier
(2001).
\doiurl{10.1016/S1076-5670(01)80095-1}
\end{botherref}
\endbibitem

\bibitem[\protect\citeauthoryear{Jourlin}{2016}]{Jourlin2016}
\begin{bbook}
\bauthor{\bsnm{Jourlin}, \binits{M.}}:
\bbtitle{Logarithmic Image Processing: Theory and Applications}.
\bsertitle{Adv. Imag. Electron Phys.},
vol. \bseriesno{195}.
\bpublisher{Elsevier},
\blocation{New York, NY, USA}
(\byear{2016}).
\doiurl{10.1016/S1076-5670(16)30070-2}
\end{bbook}
\endbibitem

\bibitem[\protect\citeauthoryear{Noyel}{2019}]{Noyel2019a}
\begin{bchapter}
\bauthor{\bsnm{Noyel}, \binits{G.}}:
\bctitle{Logarithmic mathematical morphology: A new framework adaptive to
  illumination changes}.
In: \beditor{\bsnm{Vera-Rodriguez}, \binits{R.}},
\beditor{\bsnm{Fierrez}, \binits{J.}},
\beditor{\bsnm{Morales}, \binits{A.}} (eds.)
\bbtitle{Progress in Pattern Recognition, Image Analysis, Computer Vision, and
  Applications. 23rd Iberoamerican Congress, CIARP 2018.}
\bsertitle{Lect. Notes Comput. Sci.},
vol. \bseriesno{11401},
pp. \bfpage{453}--\blpage{461}.
\bpublisher{Springer},
\blocation{Cham, Switzerland}
(\byear{2019}).
\doiurl{10.1007/978-3-030-13469-3_53}
\end{bchapter}
\endbibitem

\bibitem[\protect\citeauthoryear{Noyel}{2021}]{Noyel2021}
\begin{botherref}
\oauthor{\bsnm{Noyel}, \binits{G.}}:
Morphological and logarithmic analysis of large image databases.
Habilitation dissertation,
{Universit{\'e} de Reims Champagne-Ardenne, France}
(June 2021).
\url{https://tel.archives-ouvertes.fr/tel-03343079}
\end{botherref}
\endbibitem

\bibitem[\protect\citeauthoryear{Noyel et~al.}{2022}]{Noyel2022}
\begin{bchapter}
\bauthor{\bsnm{Noyel}, \binits{G.}},
\bauthor{\bsnm{Barbier-Renard}, \binits{E.}},
\bauthor{\bsnm{Jourlin}, \binits{M.}},
\bauthor{\bsnm{Fournel}, \binits{T.}}:
\bctitle{Logarithmic morphological neural nets robust to lighting variations}.
In: \beditor{\bsnm{Baudrier}, \binits{{\'E}.}},
\beditor{\bsnm{Naegel}, \binits{B.}},
\beditor{\bsnm{Kr{\"a}henb{\"u}hl}, \binits{A.}},
\beditor{\bsnm{Tajine}, \binits{M.}} (eds.)
\bbtitle{Discrete Geometry and Mathematical Morphology. DGMM 2022}.
\bsertitle{Lect. Notes Comput. Sci.},
pp. \bfpage{462}--\blpage{474}.
\bpublisher{Springer},
\blocation{Cham, Switzerland}
(\byear{2022}).
\doiurl{10.1007/978-3-031-19897-7_36}
\end{bchapter}
\endbibitem

\bibitem[\protect\citeauthoryear{Meyer}{1979}]{Meyer1979}
\begin{barticle}
\bauthor{\bsnm{Meyer}, \binits{F.}}:
\batitle{Iterative image transformations for an automatic screening of cervical
  smears.}
\bjtitle{J. Histochemistry \& Cytochemistry}
\bvolume{27}(\bissue{1}),
\bfpage{128}--\blpage{135}
(\byear{1979})
\doiurl{10.1177/27.1.438499}
\end{barticle}
\endbibitem

\bibitem[\protect\citeauthoryear{Jourlin and Montard}{1997}]{Jourlin1997}
\begin{barticle}
\bauthor{\bsnm{Jourlin}, \binits{M.}},
\bauthor{\bsnm{Montard}, \binits{N.}}:
\batitle{A logarithmic version of the top-hat transform in connection with the
  {A}splund distance}.
\bjtitle{Acta Stereologica}
\bvolume{16},
\bfpage{201}--\blpage{208}
(\byear{1997})
\end{barticle}
\endbibitem

\bibitem[\protect\citeauthoryear{{Zaharescu}}{2007}]{Zaharescu2007}
\begin{bchapter}
\bauthor{\bsnm{{Zaharescu}}, \binits{E.}}:
\bctitle{Morphological enhancement of medical images in a logarithmic image
  environment}.
In: \bbtitle{2007 15th Int. Conf. Digit. Signal Process.},
pp. \bfpage{171}--\blpage{174}
(\byear{2007}).
\doiurl{10.1109/ICDSP.2007.4288546}
\end{bchapter}
\endbibitem

\bibitem[\protect\citeauthoryear{Beucher and Meyer}{1993}]{Beucher1993}
\begin{bchapter}
\bauthor{\bsnm{Beucher}, \binits{S.}},
\bauthor{\bsnm{Meyer}, \binits{F.}}:
\bctitle{The morphological approach to segmentation: The watershed
  transformation}.
In: \bbtitle{Math. Morphology in Image Process.},
pp. \bfpage{433}--\blpage{481}.
\bpublisher{Marcel Dekker},
\blocation{New York, NY, USA}
(\byear{1993}).
\bcomment{Chap. 12}.
\burl{\url{https://www.taylorfrancis.com/books/9781482277234/chapters/10.1201/9781482277234-12}}
\end{bchapter}
\endbibitem

\bibitem[\protect\citeauthoryear{Beucher}{1990}]{Beucher1990}
\begin{botherref}
\oauthor{\bsnm{Beucher}, \binits{S.}}:
{Image segmentation and mathematical morphology}.
Th\`ese,
{{E}cole Nat. Sup{\'e}rieure Mines Paris},
Fr.
(June 1990).
\url{https://pastel.archives-ouvertes.fr/tel-00108290}
\end{botherref}
\endbibitem

\bibitem[\protect\citeauthoryear{Lowe}{2004}]{Lowe2004}
\begin{barticle}
\bauthor{\bsnm{Lowe}, \binits{D.G.}}:
\batitle{Distinctive image features from scale-invariant keypoints}.
\bjtitle{Int. J. Comput. Vision}
\bvolume{60}(\bissue{2}),
\bfpage{91}--\blpage{110}
(\byear{2004})
\doiurl{10.1023/B:VISI.0000029664.99615.94}
\end{barticle}
\endbibitem

\bibitem[\protect\citeauthoryear{{Salembier} et~al.}{1998}]{Salembier1998}
\begin{barticle}
\bauthor{\bsnm{{Salembier}}, \binits{P.}},
\bauthor{\bsnm{{Oliveras}}, \binits{A.}},
\bauthor{\bsnm{{Garrido}}, \binits{L.}}:
\batitle{Antiextensive connected operators for image and sequence processing}.
\bjtitle{{IEEE} Trans. Image Process.}
\bvolume{7}(\bissue{4}),
\bfpage{555}--\blpage{570}
(\byear{1998})
\doiurl{10.1109/83.663500}
\end{barticle}
\endbibitem

\bibitem[\protect\citeauthoryear{{Monasse} and {Guichard}}{2000}]{Monasse2000}
\begin{barticle}
\bauthor{\bsnm{{Monasse}}, \binits{P.}},
\bauthor{\bsnm{{Guichard}}, \binits{F.}}:
\batitle{Fast computation of a contrast-invariant image representation}.
\bjtitle{{IEEE} Trans. Image Process.}
\bvolume{9}(\bissue{5}),
\bfpage{860}--\blpage{872}
(\byear{2000})
\doiurl{10.1109/83.841532}
\end{barticle}
\endbibitem

\bibitem[\protect\citeauthoryear{Passat et~al.}{2011}]{Passat2011}
\begin{barticle}
\bauthor{\bsnm{Passat}, \binits{N.}},
\bauthor{\bsnm{Naegel}, \binits{B.}},
\bauthor{\bsnm{Rousseau}, \binits{F.}},
\bauthor{\bsnm{Koob}, \binits{M.}},
\bauthor{\bsnm{Dietemann}, \binits{J.-L.}}:
\batitle{Interactive segmentation based on component-trees}.
\bjtitle{Pattern Recogn}
\bvolume{44}(\bissue{10}),
\bfpage{2539}--\blpage{2554}
(\byear{2011})
\doiurl{10.1016/j.patcog.2011.03.025}
\end{barticle}
\endbibitem

\bibitem[\protect\citeauthoryear{{Xu} et~al.}{2016}]{Xu2016}
\begin{barticle}
\bauthor{\bsnm{{Xu}}, \binits{Y.}},
\bauthor{\bsnm{{G}\'eraud}, \binits{T.}},
\bauthor{\bsnm{{Najman}}, \binits{L.}}:
\batitle{Connected filtering on tree-based shape-spaces}.
\bjtitle{{IEEE} Trans. Pattern Anal. Mach. Intell.}
\bvolume{38}(\bissue{6}),
\bfpage{1126}--\blpage{1140}
(\byear{2016})
\doiurl{10.1109/TPAMI.2015.2441070}
\end{barticle}
\endbibitem

\bibitem[\protect\citeauthoryear{Noyel and Jourlin}{2019}]{Noyel2019c}
\begin{barticle}
\bauthor{\bsnm{Noyel}, \binits{G.}},
\bauthor{\bsnm{Jourlin}, \binits{M.}}:
\batitle{Region homogeneity in the logarithmic image processing framework:
  Application to region growing algorithms}.
\bjtitle{Image Anal. \& Stereology}
\bvolume{38}(\bissue{1}),
\bfpage{43}--\blpage{52}
(\byear{2019})
\doiurl{10.5566/ias.2038}
\end{barticle}
\endbibitem

\bibitem[\protect\citeauthoryear{Birkhoff}{1967}]{Birkhoff1967}
\begin{bbook}
\bauthor{\bsnm{Birkhoff}, \binits{G.}}:
\bbtitle{Lattice Theory},
\bedition{3}rd edn.
\bsertitle{Amer. Math. Soc. Colloq. Publications},
vol. \bseriesno{25}.
\bpublisher{AMS},
\blocation{Providence, RI}
(\byear{1967}).
\burl{\url{https://arks.org/ark:/13960/s2m3mrc4bmm}}
\end{bbook}
\endbibitem

\bibitem[\protect\citeauthoryear{Banon and Barrera}{1993}]{Banon1993}
\begin{barticle}
\bauthor{\bsnm{Banon}, \binits{G.J.F.}},
\bauthor{\bsnm{Barrera}, \binits{J.}}:
\batitle{Decomposition of mappings between complete lattices by mathematical
  morphology, part {I}: {G}eneral lattices}.
\bjtitle{Signal Process.}
\bvolume{30}(\bissue{3}),
\bfpage{299}--\blpage{327}
(\byear{1993})
\doiurl{10.1016/0165-1684(93)90015-3}
\end{barticle}
\endbibitem

\bibitem[\protect\citeauthoryear{Serra}{1988}]{Serra1988}
\begin{bbook}
\bauthor{\bsnm{Serra}, \binits{J.}}:
\bbtitle{Image Analysis and Mathematical Morphology: Theoretical Advances}
vol. \bseriesno{2}.
\bpublisher{Academic},
\blocation{San Diego, CA, USA}
(\byear{1988}).
\doiurl{10.1111/j.1365-2818.1988.tb01425.x}
\end{bbook}
\endbibitem

\bibitem[\protect\citeauthoryear{Ronse and Heijmans}{1991}]{Ronse1991}
\begin{barticle}
\bauthor{\bsnm{Ronse}, \binits{C.}},
\bauthor{\bsnm{Heijmans}, \binits{H.J.A.M.}}:
\batitle{The algebraic basis of mathematical morphology: {II}. {O}penings and
  closings}.
\bjtitle{CVGIP: Image Understanding}
\bvolume{54}(\bissue{1}),
\bfpage{74}--\blpage{97}
(\byear{1991})
\doiurl{10.1016/1049-9660(91)90076-2}
\end{barticle}
\endbibitem

\bibitem[\protect\citeauthoryear{Soille}{2003}]{Soille2003}
\begin{bbook}
\bauthor{\bsnm{Soille}, \binits{P.}}:
\bbtitle{Morphological Image Analysis: Principles and Applications},
\bedition{2}nd edn.
\bpublisher{Springer},
\blocation{New York}
(\byear{2003}).
\doiurl{10.1007/978-3-662-05088-0}
\end{bbook}
\endbibitem

\bibitem[\protect\citeauthoryear{Najman and Talbot}{2013}]{Najman2013}
\begin{bbook}
\bauthor{\bsnm{Najman}, \binits{L.}},
\bauthor{\bsnm{Talbot}, \binits{H.}}:
\bbtitle{Mathematical Morphology: From Theory to Applications},
\bedition{1}st edn.
\bpublisher{Wiley},
\blocation{Hoboken, NJ, USA}
(\byear{2013}).
\doiurl{10.1002/9781118600788}
\end{bbook}
\endbibitem

\bibitem[\protect\citeauthoryear{{Meyer}}{2019a}]{Meyer2019_1}
\begin{bbook}
\bauthor{\bsnm{{Meyer}}, \binits{F.}}:
\bbtitle{Topographical Tools for Filtering and Segmentation 1: Watersheds on
  Node- or Edge-weighted Graphs}.
\bpublisher{Wiley},
\blocation{USA}
(\byear{2019}).
\doiurl{10.1002/9781119579519}
\end{bbook}
\endbibitem

\bibitem[\protect\citeauthoryear{{Meyer}}{2019b}]{Meyer2019_2}
\begin{bbook}
\bauthor{\bsnm{{Meyer}}, \binits{F.}}:
\bbtitle{Topographical Tools for Filtering and Segmentation 2: Flooding and
  Marker-based Segmentation on Node- or Edge-weighted Graphs}.
\bpublisher{Wiley},
\blocation{Hoboken, NJ, USA}
(\byear{2019}).
\doiurl{10.1002/9781119575139}
\end{bbook}
\endbibitem

\bibitem[\protect\citeauthoryear{Bouaynaya and Schonfeld}{2008}]{Schonfeld2008}
\begin{barticle}
\bauthor{\bsnm{Bouaynaya}, \binits{N.}},
\bauthor{\bsnm{Schonfeld}, \binits{D.}}:
\batitle{Theoretical foundations of spatially-variant mathematical morphology
  part {II}: Gray-level images}.
\bjtitle{{IEEE} Trans. Pattern Anal. Mach. Intell.}
\bvolume{30}(\bissue{5}),
\bfpage{837}--\blpage{850}
(\byear{2008})
\doiurl{10.1109/TPAMI.2007.70756}
\end{barticle}
\endbibitem

\bibitem[\protect\citeauthoryear{{Verd\'u-Monedero} et~al.}{2011}]{Angulo2011}
\begin{barticle}
\bauthor{\bsnm{{Verd\'u-Monedero}}, \binits{R.}},
\bauthor{\bsnm{{Angulo}}, \binits{J.}},
\bauthor{\bsnm{{Serra}}, \binits{J.}}:
\batitle{Anisotropic morphological filters with spatially-variant structuring
  elements based on image-dependent gradient fields}.
\bjtitle{{IEEE} Trans. Image Process.}
\bvolume{20}(\bissue{1}),
\bfpage{200}--\blpage{212}
(\byear{2011})
\doiurl{10.1109/TIP.2010.2056377}
\end{barticle}
\endbibitem

\bibitem[\protect\citeauthoryear{van~de Gronde and
  Roerdink}{2014}]{vanDeGronde2014}
\begin{barticle}
\bauthor{\bsnm{Gronde}, \binits{J.J.}},
\bauthor{\bsnm{Roerdink}, \binits{J.B.T.M.}}:
\batitle{Group-invariant colour morphology based on frames}.
\bjtitle{{IEEE} Trans. Image Process.}
\bvolume{23}(\bissue{3}),
\bfpage{1276}--\blpage{1288}
(\byear{2014})
\doiurl{10.1109/TIP.2014.2300816}
\end{barticle}
\endbibitem

\bibitem[\protect\citeauthoryear{{Merveille} et~al.}{2018}]{Merveille2018}
\begin{barticle}
\bauthor{\bsnm{{Merveille}}, \binits{O.}},
\bauthor{\bsnm{{Talbot}}, \binits{H.}},
\bauthor{\bsnm{{Najman}}, \binits{L.}},
\bauthor{\bsnm{{Passat}}, \binits{N.}}:
\batitle{Curvilinear structure analysis by ranking the orientation responses of
  path operators}.
\bjtitle{{IEEE} Trans. Pattern Anal. Mach. Intell.}
\bvolume{40}(\bissue{2}),
\bfpage{304}--\blpage{317}
(\byear{2018})
\doiurl{10.1109/TPAMI.2017.2672972}
\end{barticle}
\endbibitem

\bibitem[\protect\citeauthoryear{Jourlin and Pinoli}{1995}]{Jourlin1995}
\begin{barticle}
\bauthor{\bsnm{Jourlin}, \binits{M.}},
\bauthor{\bsnm{Pinoli}, \binits{J.-C.}}:
\batitle{Image dynamic range enhancement and stabilization in the context of
  the logarithmic image processing model}.
\bjtitle{Signal Process.}
\bvolume{41}(\bissue{2}),
\bfpage{225}--\blpage{237}
(\byear{1995})
\doiurl{10.1016/0165-1684(94)00102-6}
\end{barticle}
\endbibitem

\bibitem[\protect\citeauthoryear{{Maragos} and {Schafer}}{1987}]{Maragos1987}
\begin{barticle}
\bauthor{\bsnm{{Maragos}}, \binits{P.}},
\bauthor{\bsnm{{Schafer}}, \binits{R.}}:
\batitle{Morphological filters--part {II}: Their relations to median,
  order-statistic, and stack filters}.
\bjtitle{{IEEE} Trans. Acoust., Speech, Signal Process.}
\bvolume{35}(\bissue{8}),
\bfpage{1170}--\blpage{1184}
(\byear{1987})
\doiurl{10.1109/TASSP.1987.1165254}
\end{barticle}
\endbibitem

\bibitem[\protect\citeauthoryear{Noyel and Jourlin}{2017}]{Noyel2017a}
\begin{bchapter}
\bauthor{\bsnm{Noyel}, \binits{G.}},
\bauthor{\bsnm{Jourlin}, \binits{M.}}:
\bctitle{Double-sided probing by map of {A}splund's distances using logarithmic
  image processing in the framework of mathematical morphology}.
In: \bbtitle{Mathematical Morphology and Its Applications to Signal and Image
  Processing. ISMM 2017.}
\bsertitle{Lect. Notes Comput. Sci.},
vol. \bseriesno{10225},
pp. \bfpage{408}--\blpage{420}.
\bpublisher{Springer},
\blocation{Cham, Switzerland}
(\byear{2017}).
\doiurl{10.1007/978-3-319-57240-6_33}
\end{bchapter}
\endbibitem

\bibitem[\protect\citeauthoryear{Noyel and Jourlin}{2020}]{Noyel2019b}
\begin{barticle}
\bauthor{\bsnm{Noyel}, \binits{G.}},
\bauthor{\bsnm{Jourlin}, \binits{M.}}:
\batitle{Functional {A}splund metrics for pattern matching, robust to variable
  lighting conditions}.
\bjtitle{Image Analysis \& Stereology}
\bvolume{39}(\bissue{2}),
\bfpage{53}--\blpage{71}
(\byear{2020})
\doiurl{10.5566/ias.2292}
\end{barticle}
\endbibitem

\bibitem[\protect\citeauthoryear{Noyel}{2019}]{Noyel2019d}
\begin{bchapter}
\bauthor{\bsnm{Noyel}, \binits{G.}}:
\bctitle{A link between the multiplicative and additive functional {A}splund's
  metrics}.
In: \beditor{\bsnm{Burgeth}, \binits{B.}},
\beditor{\bsnm{Kleefeld}, \binits{A.}},
\beditor{\bsnm{Naegel}, \binits{B.}},
\beditor{\bsnm{Passat}, \binits{N.}},
\beditor{\bsnm{Perret}, \binits{B.}} (eds.)
\bbtitle{Mathematical Morphology and Its Applications to Signal and Image
  Processing. ISMM 2019.}
\bsertitle{Lect. Notes Comput. Sci.},
vol. \bseriesno{11564},
pp. \bfpage{41}--\blpage{53}.
\bpublisher{Springer},
\blocation{Cham, Switzerland}
(\byear{2019}).
\doiurl{10.1007/978-3-030-20867-7_4}
\end{bchapter}
\endbibitem

\bibitem[\protect\citeauthoryear{Dorst and {Van den
  Boomgaard}}{1994}]{Dorst1994}
\begin{barticle}
\bauthor{\bsnm{Dorst}, \binits{L.}},
\bauthor{\bsnm{{Van den Boomgaard}}, \binits{R.}}:
\batitle{Morphological signal processing and the slope transform}.
\bjtitle{Signal Process.}
\bvolume{38}(\bissue{1}),
\bfpage{79}--\blpage{98}
(\byear{1994})
\doiurl{10.1016/0165-1684(94)90058-2} .
\bcomment{Mathematical Morphology and its Applications to Signal Processing}
\end{barticle}
\endbibitem

\bibitem[\protect\citeauthoryear{Maragos}{1995}]{Maragos_1995}
\begin{barticle}
\bauthor{\bsnm{Maragos}, \binits{P.}}:
\batitle{Slope transforms: theory and application to nonlinear signal
  processing}.
\bjtitle{{IEEE} Trans. Signal Process.}
\bvolume{43}(\bissue{4}),
\bfpage{864}--\blpage{877}
(\byear{1995})
\doiurl{10.1109/78.376839}
\end{barticle}
\endbibitem

\bibitem[\protect\citeauthoryear{Jackway and Deriche}{1996}]{Jackway1996}
\begin{barticle}
\bauthor{\bsnm{Jackway}, \binits{P.T.}},
\bauthor{\bsnm{Deriche}, \binits{M.}}:
\batitle{Scale-space properties of the multiscale morphological
  dilation-erosion}.
\bjtitle{{IEEE} Trans. Pattern Anal. Mach. Intell.}
\bvolume{18}(\bissue{1}),
\bfpage{38}--\blpage{51}
(\byear{1996})
\doiurl{10.1109/34.476009}
\end{barticle}
\endbibitem

\bibitem[\protect\citeauthoryear{Noyel et~al.}{2020}]{Noyel2020}
\begin{bchapter}
\bauthor{\bsnm{Noyel}, \binits{G.}},
\bauthor{\bsnm{Vartin}, \binits{C.}},
\bauthor{\bsnm{Boyle}, \binits{P.}},
\bauthor{\bsnm{Kodjikian}, \binits{L.}}:
\bctitle{Retinal vessel segmentation by probing adaptive to lighting
  variations}.
In: \bbtitle{IEEE 17th Int. Sympo. Biomed. Imag. (ISBI)},
pp. \bfpage{1246}--\blpage{1249}
(\byear{2020}).
\doiurl{10.1109/ISBI45749.2020.9098332}
\end{bchapter}
\endbibitem

\bibitem[\protect\citeauthoryear{Liu et~al.}{2022}]{Liu2022}
\begin{barticle}
\bauthor{\bsnm{Liu}, \binits{W.}},
\bauthor{\bsnm{Yang}, \binits{H.}},
\bauthor{\bsnm{Tian}, \binits{T.}},
\bauthor{\bsnm{Cao}, \binits{Z.}},
\bauthor{\bsnm{Pan}, \binits{X.}},
\bauthor{\bsnm{Xu}, \binits{W.}},
\bauthor{\bsnm{Jin}, \binits{Y.}},
\bauthor{\bsnm{Gao}, \binits{F.}}:
\batitle{Full-resolution network and dual-threshold iteration for retinal
  vessel and coronary angiograph segmentation}.
\bjtitle{IEEE J. Biomed. Health Inform.}
\bvolume{26}(\bissue{9}),
\bfpage{4623}--\blpage{4634}
(\byear{2022})
\doiurl{10.1109/JBHI.2022.3188710}
\end{barticle}
\endbibitem

\bibitem[\protect\citeauthoryear{Kamran et~al.}{2021}]{Kamran2021}
\begin{bchapter}
\bauthor{\bsnm{Kamran}, \binits{S.A.}},
\bauthor{\bsnm{Hossain}, \binits{K.F.}},
\bauthor{\bsnm{Tavakkoli}, \binits{A.}},
\bauthor{\bsnm{Zuckerbrod}, \binits{S.L.}},
\bauthor{\bsnm{Sanders}, \binits{K.M.}},
\bauthor{\bsnm{Baker}, \binits{S.A.}}:
\bctitle{{RV-GAN}: Segmenting retinal vascular structure in fundus photographs
  using a novel multi-scale generative adversarial network}.
In: \beditor{\bsnm{Bruijne}, \binits{M.}},
\beditor{\bsnm{Cattin}, \binits{P.C.}},
\beditor{\bsnm{Cotin}, \binits{S.}},
\beditor{\bsnm{Padoy}, \binits{N.}},
\beditor{\bsnm{Speidel}, \binits{S.}},
\beditor{\bsnm{Zheng}, \binits{Y.}},
\beditor{\bsnm{Essert}, \binits{C.}} (eds.)
\bbtitle{Medical Image Computing and Computer Assisted Intervention -- MICCAI
  2021}.
\bsertitle{Lect. Notes Comput. Sci.},
vol. \bseriesno{12908},
pp. \bfpage{34}--\blpage{44}.
\bpublisher{Springer},
\blocation{Cham, Switzerland}
(\byear{2021}).
\doiurl{10.1007/978-3-030-87237-3_4}
\end{bchapter}
\endbibitem

\bibitem[\protect\citeauthoryear{Zhou et~al.}{2021}]{Zhou2021}
\begin{bchapter}
\bauthor{\bsnm{Zhou}, \binits{Y.}},
\bauthor{\bsnm{Yu}, \binits{H.}},
\bauthor{\bsnm{Shi}, \binits{H.}}:
\bctitle{Study group learning: Improving retinal vessel segmentation trained
  with noisy labels}.
In: \beditor{\bsnm{Bruijne}, \binits{M.}},
\beditor{\bsnm{Cattin}, \binits{P.C.}},
\beditor{\bsnm{Cotin}, \binits{S.}},
\beditor{\bsnm{Padoy}, \binits{N.}},
\beditor{\bsnm{Speidel}, \binits{S.}},
\beditor{\bsnm{Zheng}, \binits{Y.}},
\beditor{\bsnm{Essert}, \binits{C.}} (eds.)
\bbtitle{Medical Image Computing and Computer Assisted Intervention -- MICCAI
  2021}.
\bsertitle{Lect. Notes Comput. Sci.},
vol. \bseriesno{12901},
pp. \bfpage{57}--\blpage{67}.
\bpublisher{Springer},
\blocation{Cham, Switzerland}
(\byear{2021}).
\doiurl{10.1007/978-3-030-87193-2_6}
\end{bchapter}
\endbibitem

\bibitem[\protect\citeauthoryear{{Staal} et~al.}{2004}]{Staal2004}
\begin{barticle}
\bauthor{\bsnm{{Staal}}, \binits{J.}},
\bauthor{\bsnm{{Abramoff}}, \binits{M.D.}},
\bauthor{\bsnm{{Niemeijer}}, \binits{M.}},
\bauthor{\bsnm{{Viergever}}, \binits{M.A.}},
\bauthor{\bsnm{{van Ginneken}}, \binits{B.}}:
\batitle{Ridge-based vessel segmentation in color images of the retina}.
\bjtitle{{IEEE} Trans. Med. Imag.}
\bvolume{23}(\bissue{4}),
\bfpage{501}--\blpage{509}
(\byear{2004})
\doiurl{10.1109/TMI.2004.825627}
\end{barticle}
\endbibitem

\bibitem[\protect\citeauthoryear{{Papers with code}}{2023}]{DriveBenchmark2023}
\begin{botherref}
\oauthor{\bsnm{{Papers with code}}}:
Retinal Vessel Segmentation on {DRIVE}.
\url{https://web.archive.org/web/20221205193645/https://paperswithcode.com/task/retinal-vessel-segmentation}.
Accessed: 15 September 2023
(2023)
\end{botherref}
\endbibitem

\bibitem[\protect\citeauthoryear{Ronneberger et~al.}{2015}]{Ronneberger2015}
\begin{bchapter}
\bauthor{\bsnm{Ronneberger}, \binits{O.}},
\bauthor{\bsnm{Fischer}, \binits{P.}},
\bauthor{\bsnm{Brox}, \binits{T.}}:
\bctitle{U-net: Convolutional networks for biomedical image segmentation}.
In: \beditor{\bsnm{Navab}, \binits{N.}},
\beditor{\bsnm{Hornegger}, \binits{J.}},
\beditor{\bsnm{Wells}, \binits{W.M.}},
\beditor{\bsnm{Frangi}, \binits{A.F.}} (eds.)
\bbtitle{Medical Image Computing and Computer-Assisted Intervention -- MICCAI
  2015}.
\bsertitle{Lect. Notes Comput. Sci.},
vol. \bseriesno{9351},
pp. \bfpage{234}--\blpage{241}.
\bpublisher{Springer},
\blocation{Cham, Switzerland}
(\byear{2015}).
\doiurl{10.1007/978-3-319-24574-4_28}
\end{bchapter}
\endbibitem

\bibitem[\protect\citeauthoryear{Goodfellow et~al.}{2014}]{Goodfellow2014}
\begin{bchapter}
\bauthor{\bsnm{Goodfellow}, \binits{I.J.}},
\bauthor{\bsnm{Pouget-Abadie}, \binits{J.}},
\bauthor{\bsnm{Mirza}, \binits{M.}},
\bauthor{\bsnm{Xu}, \binits{B.}},
\bauthor{\bsnm{Warde-Farley}, \binits{D.}},
\bauthor{\bsnm{Ozair}, \binits{S.}},
\bauthor{\bsnm{Courville}, \binits{A.}},
\bauthor{\bsnm{Bengio}, \binits{Y.}}:
\bctitle{Generative adversarial nets}.
In: \bbtitle{NIPS'14},
pp. \bfpage{2672}--\blpage{2680}
(\byear{2014}).
\burl{\url{https://dl.acm.org/doi/10.5555/2969033.2969125}}
\end{bchapter}
\endbibitem

\bibitem[\protect\citeauthoryear{Jourlin et~al.}{2014}]{Jourlin2014}
\begin{barticle}
\bauthor{\bsnm{Jourlin}, \binits{M.}},
\bauthor{\bsnm{Couka}, \binits{E.}},
\bauthor{\bsnm{Abdallah}, \binits{B.}},
\bauthor{\bsnm{Corvo}, \binits{J.}},
\bauthor{\bsnm{Breugnot}, \binits{J.}}:
\batitle{Aspl{\"u}nd's metric defined in the {L}ogarithmic {I}mage {P}rocessing
  ({LIP}) framework: {A} new way to perform double-sided image probing for
  non-linear grayscale pattern matching}.
\bjtitle{Pattern Recognit.}
\bvolume{47}(\bissue{9}),
\bfpage{2908}--\blpage{2924}
(\byear{2014})
\doiurl{10.1016/j.patcog.2014.03.031}
\end{barticle}
\endbibitem

\end{thebibliography}

\end{document}